\documentclass[fleqn,usenatbib]{mnras}

\usepackage{newtxtext,newtxmath}

\usepackage[T1]{fontenc}

\DeclareRobustCommand{\VAN}[3]{#2}
\let\VANthebibliography\thebibliography
\def\thebibliography{\DeclareRobustCommand{\VAN}[3]{##3}\VANthebibliography}

\usepackage{graphicx}	
\usepackage{amsmath}	
\usepackage{xcolor}     
\usepackage{makecell}
\usepackage{comment}
\usepackage{multirow}
\usepackage{mathtools}

\def \paul{\color{black}}
\def \p{\color{black}}
\def \pp{\color{black}}

\newcommand{\teff}{T_{\rm eff}}
\newcommand{\logg}{\log g}
\newcommand{\mh}{\rm{[M/H]}}

\newcommand{\afe}{\rm{[\alpha/Fe]}}
\newcommand{\nbMdwarfs}{44}
\newcommand{\nbMdwarfsCommon}{28}
\newcommand{\kms}{\rm km\ s^{-1}}
\newcommand{\vb}{v_{\rm b}}

\title[Estimating atmospheric properties for SPIRou]{Estimating the atmospheric properties of \nbMdwarfs{} M dwarfs from SPIRou spectra}

\author[P. I. Cristofari et al.]{
P. I. Cristofari$^{1}$\thanks{E-mail: paul.cristofari@irap.omp.eu (IRAP)},
J.-F. Donati$^{1}$,
T. Masseron$^{2,3}$, 
P. Fouqué$^{1,4}$, 
C. Moutou$^{1}$,
A. Carmona$^{5}$,
\newauthor
E. Artigau$^{6}$,
E. Martioli$^{7,8}$,
G. Hébrard$^{7,9}$,
E. Gaidos$^{10}$,
X. Delfosse$^{5}$,
and the SLS consortium
\\
$^{1}$Univ. de Toulouse, CNRS, IRAP, 14 av. Belin, 31400 Toulouse, France\\
$^{2}$Instituto de Astrofísica de Canarias, E-38205 La Laguna, Tenerife, Spain\\
$^{3}$Departamento de Astrofísica, Universidad de La Laguna, E-38206 La Laguna, Tenerife, Spain\\
$^{4}$Canada-France-Hawaii Telescope, CNRS, Kamuela, HI 96743, USA \\
$^{5}$Univ. Grenoble Alpes, CNRS, IPAG, F-38000 Grenoble, France\\
$^{6}$Universit\'e de Montréal, D\'epartement de Physique, IREX, Montréal, QC H3C 3J7, Canada  \\
$^{7}$Institut d’Astrophysique de Paris, CNRS, UMR 7095, Sorbonne Universit\'e, 75014 Paris, France\\
$^{8}$Laborat\'orio National de Astrof\'isica, 37504-364 Itajub\'a, MG, Brazil\\
$^{9}$Observatoire de Haute Provence, France\\
$^{10}$Department of earth sciences, University of  Hawai'i at M\=anoa, Honolulu HI 96822, USA \\
}

\date{Accepted 2022 August 18. Received 2022 August 12; in original form 2022 June 17}

\pubyear{2015}

\begin{document}
\label{firstpage}
\pagerange{\pageref{firstpage}--\pageref{lastpage}}
\maketitle

\begin{abstract}
	We describe advances on a method designed to derive accurate parameters of M dwarfs. Our analysis consists in comparing high-resolution  infrared spectra acquired with the near-infrared spectro-polarimeter SPIRou to synthetic spectra computed from \texttt{MARCS} model atmospheres, in order to derive the effective temperature ($\teff$), surface gravity ($\logg$), metallicity ($\mh$) and alpha-enhancement ($\afe$) of \nbMdwarfs{} M dwarfs monitored within the SPIRou Legacy Survey (SLS). Relying on 12 of these stars,
	we calibrated our method by  refining our selection of well modelled stellar lines, and adjusted the line list parameters to improve the fit when necessary. Our retrieved $\teff$, $\logg$ and $\mh$ are in good agreement with literature values, with dispersions of the order of 50~K in $\teff$ and  0.1~dex in $\logg$ and $\mh$.  We report that fitting $\afe$ has an impact on the derivation of the other stellar parameters, motivating us to extend our fitting procedure to this additional parameter. We find that our retrieved $\afe$ are compatible with  those expected from empirical relations derived in other studies.
\end{abstract}





\section{Introduction}

M dwarfs are obvious targets of interest to look for exoplanets, especially  those located in the habitable zones of their host stars~\citep{Bonfils_2013, dressing_2013, gaidos_2016}, as they dominate the stellar population of the solar neighbourhood.
In order to accurately characterize these planets, and derive their masses and radii, it is essential to obtain reliable estimates of the fundamental parameters of the host stars. In particular, the effective temperature ($\teff$), surface gravity ($\logg$)  and overall metallicity ($\mh$) of M dwarfs must be determined as accurately as possible.

Several techniques have been developed to characterize atmospheric parameters of low-mass stars. Some rely on the adjustment of equivalent widths~\citep{rojas_2010, neves_2014, fouque_2018}. Others attempt to fit spectral energy distributions (SEDs) on low to mid-resolution spectra~\citep{mann_2013}. More recently, advances in spectral modelling and the advent of new high-resolution spectrographs in the near-infrared (NIR) domain allowed some authors to perform direct fits of synthetic spectra on high-resolution spectroscopic observations~\citep{passegger_2018, schweitzer_2019, marfil_2021}.

Of these techniques, the latter is presumably the best option to retrieve precise estimates of the atmospheric parameters by modelling individual spectral lines rather than integrated quantities such as equivalent width or bandpass fluxes.
To succeed, this approach however requires accurate high-resolution synthetic spectra on the one hand, and high-resolution and high signal-to-noise ratio (SNR) spectroscopic observations on the other hand. To this end, model atmospheres  of low-mass stars such as \texttt{MARCS}~\citep{gustafsson_2008}, \texttt{ATLAS}~\citep{kurucz_1970} or \texttt{PHOENIX}~\citep{allard_1995}  were developed and refined over the last few decades. While \texttt{PHOENIX} also performs the radiative transfer to produce synthetic spectra, other codes are used to compute emergent spectra from model atmospheres, such as \texttt{Turbospectrum}~\citep{plez_1998, plez_2012} or \texttt{SYNTHE}~\citep{kurucz_2005}, in the case of \texttt{MARCS} and \texttt{ATLAS} atmospheric models respectively. 
In parallel, instruments such as SPIRou~\citep[][]{donati_2020}, CARMENES~\citep{quirrenbach_2014}, iSHELL~\citep{Rayner_2016}, IRD~\citep{kotani_2018} or HPF~\citep{hpf_instrument} have provided the community with high-quality and high-resolution spectra in the NIR domain.

For M dwarfs in the NIR domain, the modelling of stellar spectra is particularly challenging because of the high density of atomic and molecular lines, forming deep absorption bands. Furthermore, telluric features, extremely abundant in the NIR domain, often blend with stellar lines and forces one to carry out extra processing steps to extract the stellar spectrum. In spite of these challenges, the NIR domain remains an abundant source of information, particularly for M dwarfs that are brighter in the NIR than in the optical.

In this paper, we pursue the work  initiated in~\citet[hereafter C22]{cristofari_2022} with the ultimate goal of providing the community with accurate stellar parameters  for most M dwarfs observed with SPIRou. Over 70  of them have been  monitored with this instrument in the context of the SPIRou Legacy Survey~\citep[SLS, ][]{donati_2020}, an  ongoing observation program for which 310 nights were allocated on the 3.6-m Canada-France-Hawaii Telescope (CFHT). M dwarfs within the SLS are typically monitored tens of times  over successive seasons, allowing us to produce high-quality median spectra for our analysis~(C22), which we call "template spectra" in the following. In this work,  we focus on the \nbMdwarfs{}  M dwarfs  that were most intensively observed with SPIRou.

In contrast with C22, we focus in this paper on \texttt{MARCS} model atmospheres to derive stellar parameters, and bring several improvements to our method. More specifically, we extend our tools to constrain the abundance of alpha elements (O, Ne, Mg, Si, S, Ar, Ca, and Ti) for the studied targets, and demonstrate the importance of considering the alpha-enhancement parameter ($\afe$) when modelling spectra of M dwarfs.

In Sec.~\ref{sec:observations} we introduce the selected targets and the processing steps undertaken to produce template spectra from SPIRou observations. We recall the main steps of our analysis in Sec.~\ref{sec:method} along with the implemented improvements. We then discuss the impact of $\afe$ on the retrieved parameters in Sec.~\ref{sec:alpha},  outline the modifications brought to the parameters of some of the atomic lines used in our work (see Sec.~\ref{sec:line_selection}), and present the results of our analysis of \nbMdwarfs{} M dwarfs in Sec.~\ref{sec:results}. We conclude and discuss the results of our work in Sec.~\ref{sec:conclusions}.



\begin{table}
	\caption{Number of spectra, visits and typical SNR of the collected observations.}
	\label{tab:observations}
	\begin{tabular}{cccc}
		\hline
		Star & Nb. spectra & Nb. epochs & Med. SNR [SNR range] \\ 
		\hline
		Gl 338B & 124 & 31 & 250 [150 - 300] \\ 
		Gl 410 & 472 & 112 & 130 [50 - 150] \\ 
		Gl 846 & 792 & 194 & 160 [50 - 230] \\ 
		Gl 205 & 593 & 143 & 290 [50 - 350] \\ 
		Gl 880 & 634 & 155 & 200 [70 - 250] \\ 
		Gl 514 & 740 & 152 & 160 [50 - 280] \\ 
		Gl 382 & 238 & 59 & 150 [50 - 220] \\ 
		Gl 412A & 884 & 148 & 180 [60 - 350] \\ 
		Gl 15A & 1040 & 198 & 280 [60 - 360] \\ 
		Gl 411 & 592 & 143 & 360 [200 - 440] \\ 
		Gl 752A & 523 & 129 & 170 [50 - 230] \\ 
		Gl 48 & 786 & 195 & 130 [60 - 150] \\ 
		Gl 617B & 546 & 133 & 120 [50 - 150] \\ 
		Gl 480 & 283 & 70 & 110 [60 - 120] \\ 
		Gl 436 & 188 & 38 & 150 [70 - 220] \\ 
		Gl 849 & 771 & 189 & 120 [50 - 140] \\ 
		Gl 408 & 495 & 117 & 140 [50 - 170] \\ 
		Gl 687 & 898 & 214 & 200 [60 - 240] \\ 
		Gl 725A & 889 & 213 & 210 [50 - 260] \\ 
		Gl 317 & 108 & 27 & 100 [70 - 130] \\ 
		Gl 251 & 749 & 175 & 140 [50 - 170] \\ 
		GJ 4063 & 784 & 190 & 100 [50 - 120] \\ 
		Gl 581 & 124 & 31 & 120 [60 - 150] \\ 
		Gl 725B & 855 & 211 & 160 [70 - 200] \\ 
		PM J09553-2715 & 172 & 43 & 110 [80 - 140] \\ 
		Gl 876 & 369 & 88 & 160 [70 - 220] \\ 
		GJ 1012 & 522 & 129 & 100 [50 - 120] \\ 
		GJ 4333 & 734 & 181 & 100 [50 - 120] \\ 
		Gl 445 & 171 & 43 & 110 [50 - 140] \\ 
		GJ 1148 & 399 & 98 & 100 [50 - 110] \\ 
		PM J08402+3127 & 462 & 115 & 100 [50 - 110] \\ 
		GJ 3378 & 725 & 179 & 100 [50 - 130] \\ 
		GJ 1105 & 515 & 128 & 100 [50 - 130] \\ 
		Gl 699 & 950 & 231 & 200 [60 - 240] \\ 
		Gl 169.1A & 673 & 165 & 100 [50 - 130] \\ 
		PM J21463+3813 & 718 & 177 & 100 [50 - 120] \\ 
		Gl 15B & 755 & 188 & 100 [50 - 120] \\ 
		GJ 1289 & 812 & 202 & 100 [50 - 110] \\ 
		Gl 447 & 180 & 45 & 120 [60 - 170] \\ 
		GJ 1151 & 568 & 141 & 100 [50 - 120] \\ 
		GJ 1103 & 254 & 62 & 100 [50 - 110] \\ 
		Gl 905 & 484 & 117 & 110 [50 - 130] \\ 
		GJ 1002 & 524 & 130 & 100 [60 - 120] \\ 
		GJ 1286 & 438 & 113 & 100 [50 - 120] \\
		\hline
	\end{tabular}
\end{table}

\begin{table*}
	\caption{Parameters derived by~M15 for {\paul 12 calibration stars used in this study.} $\logg$ values are computed from reported masses and radii.}
	\label{tab:ref_stars}
	\begin{tabular}{cccccccc}
		\hline
		Star & Spectral type & $\teff$ & $\mh$ & Radius & Mass & $\logg$ \\
		\hline 
		Gl 846 & M0.5V & $3848$ $\pm$ $60$ & $0.02$ $\pm$ $0.08$ &$0.546 \pm 0.019$ &$0.590 \pm 0.059$ &$4.74 \pm 0.05$ &\\ 
		Gl 880 & M1.5V & $3720$ $\pm$ $60$ & $0.21$ $\pm$ $0.08$ &$0.549 \pm 0.018$ &$0.574 \pm 0.057$ &$4.72 \pm 0.05$ &\\ 
		Gl 15A & M2V & $3603$ $\pm$ $60$ & $-0.30$ $\pm$ $0.08$ &$0.388 \pm 0.013$ &$0.398 \pm 0.040$ &$4.86 \pm 0.05$ &\\ 
		Gl 411 & M2V & $3563$ $\pm$ $60$ & $-0.38$ $\pm$ $0.08$ &$0.389 \pm 0.013$ &$0.386 \pm 0.039$ &$4.84 \pm 0.05$ &\\ 
		Gl 752A & M3V & $3558$ $\pm$ $60$ & $0.10$ $\pm$ $0.08$ &$0.474 \pm 0.016$ &$0.475 \pm 0.047$ &$4.76 \pm 0.05$ &\\ 
		Gl 849 & M3.5V & $3530$ $\pm$ $60$ & $0.37$ $\pm$ $0.08$ &$0.470 \pm 0.018$ &$0.482 \pm 0.048$ &$4.78 \pm 0.06$ &\\ 
		Gl 436 & M3V & $3479$ $\pm$ $60$ & $0.01$ $\pm$ $0.08$ &$0.449 \pm 0.019$ &$0.445 \pm 0.044$ &$4.78 \pm 0.06$ &\\ 
		Gl 725A & M3V & $3441$ $\pm$ $60$ & $-0.23$ $\pm$ $0.08$ &$0.351 \pm 0.013$ &$0.334 \pm 0.033$ &$4.87 \pm 0.05$ &\\
		Gl 725B & M3.5V & $3345$ $\pm$ $60$ & $-0.30$ $\pm$ $0.08$ &$0.273 \pm 0.011$ &$0.248 \pm 0.025$ &$4.96 \pm 0.06$ &\\ 
		Gl 699 & M4V & $3228$ $\pm$ $60$ & $-0.40$ $\pm$ $0.08$ &$0.186 \pm 0.007$ &$0.155 \pm 0.015$ &$5.09 \pm 0.05$ &\\ 
		Gl 15B & M3.5V & $3218$ $\pm$ $60$ & $-0.30$ $\pm$ $0.08$ &$0.192 \pm 0.008$ &$0.159 \pm 0.016$ &$5.07 \pm 0.06$ &\\ 
		Gl 905 & M5.0V & $2930$ $\pm$ $60$ & $0.23$ $\pm$ $0.08$ &$0.189 \pm 0.008$ &$0.145 \pm 0.015$ &$5.04 \pm 0.06$ &\\
		\hline 
	\end{tabular}
\end{table*}

\begin{table*}
	\caption{{\paul Same as Table~\ref{tab:ref_stars} for 16 additional stars included in both~M15 and this study.} 
	}
	\label{tab:ref_stars_2}
	\begin{tabular}{cccccccc}
		\hline
		Star & Spectral type & $\teff$ & $\mh$ & Radius & Mass & $\logg$ \\
		\hline 
		Gl 205 & M1.5V & $3801$ $\pm$ $60$ & $0.49$ $\pm$ $0.08$ &$0.581 \pm 0.019$ &$0.633 \pm 0.063$ &$4.71 \pm 0.05$ &\\ 
		Gl~514 & M1.0V &  3727 $\pm$ 60 &  -0.09  $\pm$ 0.08 & 0.483 $\pm$ 0.016 & 0.527  $\pm$ 0.053 & 4.79 $\pm$ 0.05  \\
		Gl 382 & M2V & $3623$ $\pm$ $60$ & $0.13$ $\pm$ $0.08$ &$0.522 \pm 0.019$ &$0.525 \pm 0.053$ &$4.72 \pm 0.05$ &\\ 
		Gl 412A & M1.0V & $3619$ $\pm$ $60$ & $-0.37$ $\pm$ $0.08$ &$0.383 \pm 0.013$ &$0.390 \pm 0.039$ &$4.86 \pm 0.05$ &\\ 
		Gl 480 & M3.5V & $3463$ $\pm$ $60$ & $0.26$ $\pm$ $0.08$ &$0.466 \pm 0.025$ &$0.467 \pm 0.047$ &$4.77 \pm 0.06$ &\\ 
		Gl 251 & M3V & $3448$ $\pm$ $60$ & $-0.02$ $\pm$ $0.08$ &$0.358 \pm 0.013$ &$0.352 \pm 0.035$ &$4.88 \pm 0.05$ &\\ 
		Gl 687 & M3.0V & 3439  $\pm$ 60 & 0.050 $\pm$ 0.080 & 0.414 $\pm$ 0.015 & 0.405 $\pm$ 0.041 & 4.81 $\pm$ 0.05 \\
		Gl 581 &  M3V &  3395 $\pm$  60  & -0.150  $\pm$  0.080 & 0.311 $\pm$ 0.012 & 0.292 $\pm$ 0.029  &  4.92 $\pm$ 0.06 \\
		PM J09553-2715 &  M3V & 3346 $\pm$  60  & 0.01  $\pm$ 0.080  &  0.321  $\pm$ 0.016  &  0.299  $\pm$  0.030 & 4.90 $\pm$ 0.06 \\
		GJ 3378 & M4.0V & $3340$ $\pm$ $60$ & $-0.09$ $\pm$ $0.08$ &$0.269 \pm 0.011$ &$0.245 \pm 0.024$ &$4.97 \pm 0.06$ &\\ 
		GJ 4333 & M3.5V & $3324$ $\pm$ $60$ & $0.24$ $\pm$ $0.08$ &$0.416 \pm 0.020$ &$0.391 \pm 0.039$ &$4.79 \pm 0.06$ &\\ 
		GJ 1148 & M4.0V & $3304$ $\pm$ $61$ & $0.07$ $\pm$ $0.08$ &$0.376 \pm 0.018$ &$0.336 \pm 0.034$ &$4.81 \pm 0.06$ &\\ 
		Gl 876 & M3.5V & $3247$ $\pm$ $60$ & $0.17$ $\pm$ $0.08$ &$0.363 \pm 0.014$ &$0.328 \pm 0.033$ &$4.83 \pm 0.06$ &\\ 
		Gl 447 & M4V & 3192 $\pm$ 60 & $-0.020$ $\pm$ 0.080 & 0.197 $\pm$ 0.008 & 0.168 $\pm$ 0.017 & 5.08 $\pm$ 0.06 \\
		GJ 1289 & M4.5V & $3173$ $\pm$ $60$ & $0.05$ $\pm$ $0.08$ &$0.238 \pm 0.013$ &$0.202 \pm 0.020$ &$4.99 \pm 0.06$ &\\ 
		GJ 1151 & M4.5V & $3118$ $\pm$ $60$ & $0.03$ $\pm$ $0.08$ &$0.190 \pm 0.009$ &$0.154 \pm 0.015$ &$5.07 \pm 0.06$ &\\ 
		\hline 
	\end{tabular}
\end{table*}

\section{Observations and reduction}
\label{sec:observations}
\subsection{Selecting targets}
\label{sec:target_selection}
Most stars were monitored several tens of times  over successive seasons with the widest possible range of Barycentric Earth Radial Velocites (BERV).
In this work, we focus on \nbMdwarfs{} M dwarfs for which at least 20 SPIRou spectra were collected in order to build high-SNR stellar templates (see Sec.~\ref{sec:building_templates}, Table~{\ref{tab:observations}}). For now, we exclude highly active targets, for which stellar line profiles are likely to be impacted by magnetic fields  and chromospheric activity.
Several publications assessed the activity level from H$\alpha$ equivalent width for most targets of our sample~\citep{schoefer_2019, fouque_2018}, confirming that they are no more than weakly active. We further performed visual inspection of the spectra to ensure that the stellar lines were not {\p significantly} affected by activity, {\p e.g. with core reversals in strong lines like those seen in the spectra of more active targets (such as GJ~3622).}

Out of our \nbMdwarfs{} stars, we use 12 (the same as in C22{\p , see Table~\ref{tab:ref_stars}}) to improve our tools and calibrate our analysis procedure. We consider the parameters published by~\citet[hereafter M15]{mann_2015} as a reference for these stars, given that this study relies on methods that are largely independent from ours {\p (e.g. SED fits to low resolution spectra, equivalent widths and empirical mass-magnitude relations)}, and agree well with other literature studies. Table~\ref{tab:ref_stars_2} presents the stellar parameters {\paul for 16 additional} stars included in our sample for which M15 reported stellar properties.

\subsection{Building templates from SPIRou spectra}
\label{sec:building_templates}
All SPIRou spectra are processed through the SPIRou reduction pipeline, APERO (version 0.6.132, Cook et al., in prep). A correction of the telluric absorption and emission lines is performed by APERO, relying on telluric templates built from telluric standards ~(Artigau et al., in prep).
A blaze profile estimated from flat-field exposures is used to flatten the extracted spectra, and each order is normalized using a third degree polynomial.

Stellar templates are built by taking the median of the telluric corrected spectra in the barycentric reference frame. 
Because of the relative motion of telluric lines with respect to spectral features due to the Earth revolution around the Sun, having spectra observed at various BERV  (with typical maximum difference between observations ranging from 10 to 30~$\kms$) allows one to minimize telluric correction errors, and to obtain a template spectrum even in regions where telluric lines are deep enough to render a single observation hardly usable over the corresponding range.
All telluric-corrected spectra recorded with a SNR per 2-$\kms$ pixel in the $H$ band exceeding 50 are used to build the stellar templates. The typical SNR per pixel of these template spectra reaches up to 2000.

\section{Deriving fundamental stellar parameters from SPIRou template spectra}
\label{sec:method}

In C22, we described and tested a method for determining atmospheric parameters from SPIRou template spectra. We discussed the use of two different models, \texttt{PHOENIX-ACES}~\citep{husser_2013} and \texttt{MARCS}, the differences in the synthetic spectra computed with both models, and their impact on the results. In this work, we update the method to improve the framework and produce more reliable results. Some of these improvements include the implementation of a new continuum normalisation procedure and an empirical revision of line parameters for some of the atomic lines used (see Sec~\ref{sec:improvements}). We then further improve the method to retrieve the alpha enhancement ($\afe$) as an additional free parameter of our model (see Sec~\ref{sec:alpha}).
We concentrate our efforts on \texttt{MARCS} model atmospheres, readily available for different values of $\afe$ and computed with up-to-date line lists\footnote{ The grid of \texttt{PHOENIX-ACES} synthetic spectra was not published with multiple $\afe$ values for $\teff>3500$~K, and updating the line list is not an easy task, hence why we focused on \texttt{MARCS} models in this new study.}.

\subsection{The grid of synthetic spectra}
\label{sec:marcs_grid}

\begin{table}
	\centering
	\caption{Parameter range covered by the computed grid of \texttt{MARCS} synthetic spectra. The range and initial step size are listed along with the level to which the grid is interpolated to reach the final step size.}
	\resizebox{\columnwidth}{!}{
		\begin{tabular}{cccccc}
			\hline
			Variable & Range (and step size)  & \thead{Interp. factor \\ (and final step size)} \\
			\hline
			$\teff$~(K)  & 3000 -- 4000 (100) & 20 (5)\\
			$\logg$~(dex) & 3.5 -- 5.5 (0.5) & 50 (0.01) \\
			$\mh$~(dex) & $-1.5$ -- $+1.0$ (0.25) & 25 (0.01) \\
			$\afe$~(dex) & $-0.25$ -- $+0.5$ (0.25) & 25 (0.01) \\
			\hline
		\end{tabular}
	}
	\label{tab:marcs_range}
\end{table}

We use a grid of synthetic spectra computed from \texttt{MARCS} model atmospheres with Turbospectrum for several $\teff$, $\logg$ and $\mh$ values. This grid is the same as that used in C22, augmented with models computed for $\afe$ values ranging from -0.25 to 0.50~dex in steps of 0.25~dex (see Table~\ref{tab:marcs_range}).
{\paul Spectra were computed for all available $\logg$, although values below 4.5~dex are not expected to be used in the case of main-sequence~\citep{baraffe_2015}. 
}


\subsection{Stellar analysis procedure}
The parameter determination procedure used in this paper is similar to that described in C22. In this section, we briefly summarize the main steps of this process.

\subsubsection{Comparison of models to observation templates}
 SPIRou template spectra are compared to synthetic spectra in order to identify the best-fitting model. Prior to this comparison, the synthetic spectra are  binned on the wavelength grid of the  SPIRou template. This  binning operation is performed through a cubic interpolation and convolution with a rectangular function of width 2~$\kms$ (representing pixels). The synthetic spectra are also convolved with a Gaussian profile of full width at half maximum (FWHM) of 4.3~$\kms$ to account for  instrumental broadening (resolving power 70,000). We finally consider the effect of both rotation and macroturbulence on stellar spectra, that we approximate as a Gaussian broadening of FWHM $\vb=3~\kms$  as in C22. 
We then extract 400-bin windows around selected lines and adjust the local continuum of the synthetic spectra to match that of the observation template spectrum. This step is particularly challenging in the NIR  spectra of M dwarfs, where the large density of atomic and molecular lines renders the pseudo continuum  hard to locate. The comparison of synthetic spectra and observation templates is performed  on a total of $\sim$ 70 lines, found to be more or less adequately reproduced in synthetic spectra, and sensitive to the atmospheric parameters of interest.


\subsubsection{$\chi^2$ minimization}
Synthetic spectra for {a given range of} $\teff$, $\logg$, $\mh$ and $\afe$ are compared to the  SPIRou template for  a given star of our sample,  yielding a 4D grid of $\chi^2$ values. Given the rough step size of this initial grid  (see Sec.~\ref{sec:marcs_grid}), we interpolate the synthetic spectra to reach steps of 5~K in $\teff$ and 0.01~dex in $\logg$ and $\mh$ around the grid minimum in order to locate the grid minimum and determine the curvature at this position as accurately as possible. A new 4D $\chi^2$ landscape is computed, and a 4D second degree polynomial is fitted on the 3000 points with smallest $\chi^2$ values.

\subsubsection{Error estimation} 
To estimate error bars on the retrieved parameters, we measure the curvature of the fitted paraboloid. More specifically, we search for the ellipsoid  where the $\chi^2$ increases by 1 from the minimum, and project it on each parameter axis. The projected intervals should contain 68.3\% of normally distributed data~\citep{numerical_recipes}, which we refer to as formal error bars. In C22, 
we observed that the choice of model has a significant impact on the results, introducing systematics that are not accounted for by our formal error bars computation.  To take this effect into account, C22 introduced a second error  bar, derived from the root mean square (RMS) diference between the parameters retrieved with both sets of atmospheric models.

In the present work, we consider a single model  and thus cannot perform a similar  operation. We therefore rely on the results of C22 to increase our error bars, by quadratically adding 30~K, 0.05~dex and 0.1~dex to the computed  formal error bars on $\teff$, $\logg$ and $\mh$ respectively,  and refer to these as empirical error bars.

 Since we have no means to retrieve an empirical error bar for $\afe$, we estimate it from those derived on $\mh$. We typically compute smaller formal error bars on $\afe$ than on $\mh$, with average values of about 0.015~dex and 0.005~dex respectively. The median of the ratio between our formal error bars on $\mh$ and on $\afe$ is of 2.5. To account for some of the systematics and provide a conservative estimate of the error bars on $\afe$, we choose to quadratically add 0.04~dex to our formal error bars for this parameter. This is consistent with the dispersion of the retrieved $\afe$ values for stars having $\mh$~>~$-0.1$~dex, for which thin and thick disc populations blend together.


\subsection{Adjustment of the continuum}
\label{sec:improvements}


 In this paper, we also revised our continuum adjustment procedure. We extract 400-bin windows around all selected lines for both the SPIRou template and the synthetic spectrum. In  each window, we exclude  all points of the SPIRou template that fall above the $98^{\rm th}$ percentile,  and may correspond to poorly corrected telluric emission lines. We then  sub-divide the 400-bin windows into 40-bin windows, in which we consider all points above the $90^{\rm th}$ percentile as  tracing the continuum. We then fit a  straight line through these points to retrieve two continua, one for the template spectrum and one for the synthetic spectrum, which  are then used to bring the  continua of the template and model  spectra to the same level. {\paul This procedure sets in the local continuum of both the template and the synthetic spectrum to unity.}


\section{The impact of $\afe$ on the recovered fundamental parameters.}
\label{sec:alpha}

Several studies~\citep{passegger_2019, schweitzer_2019} assume that the abundances of elements with respect to  those of the Sun all differ by the same amount, and typically report values of $\mh$ where  [X/H]~=~$\mh$ for all elements X  with atomic numbers $\ge$ 3. This assumption  simplifies the modelling  but likely affects the estimation of the other parameters. In particular, the abundance of alpha elements (O, Ne, Mg, Si, S, Ar, Ca, and Ti)  was shown to depend on  the considered stellar population~\citep{fuhrmann_1998, adibekyan_2013}, and models  were modified to  incorporate an alpha-enhancement parameter~\citep[$\afe$,][]{allard_2011, husser_2013, gustafsson_2008}.
In the rest of the paper, $\mh$  is used to designate the overall metallicity of all elements but the alpha elements, whose abundances are  set to ${\rm [\alpha/H]} = \mh + \afe$.

The effect of $\afe$ is visible across the entire SPIRou domain where molecular lines are numerous, and where variations in the abundances of alpha elements, in particular oxygen, leads to significant changes of the model atmospheres.

\subsection{$\afe$--$\mh$ relations}

Previous publications analysing M dwarfs analysis adopted a unique $\afe$--$\mh$ relations for their analysis~\citep{rajpurohit_2017, marfil_2021}. These  assume that $\afe = -0.4 \rm{[Fe/H]}$ for $-1 \leq \rm{[Fe/H]}<0$, $\afe = 0$ for $\rm{[Fe/H]}\geq0$ and $\afe = -1$ for $\rm{[Fe/H]}<-1$. This relation was also  used for the PHOENIX BT-Settl grid of synthetic spectra~\citep{allard_2011}.

Thanks to ongoing spectroscopic large surveys, such relations can nowadays be refined more empirically. For example, this relation can be derived by looking at abundances in giants (4000<$\teff$<5000~K and $\logg$<3.5~dex) estimated from the APOGEE survey~\citep{jonsson_2020}.
 These stars  can be split into 2 groups corresponding to 2 galactic populations, with  the ones from the thick galactic disc having typically larger $\afe$ values than  those from the thin galactic disc.  This suggests that distinct $\afe$--$\mh$ relations should be considered for thin and thick disc stars. It is however  still unclear whether these relations  also apply to M dwarfs, due to the lack of accurate data for these stars. In this work, we place a fiducial boundary between the low-$\afe$  and high-$\afe$ stars to define the thin and thick disc populations respectively. This simplistic classification aims at providing an a posteriori verification that our derived $\afe$ values for the targets in our sample are consistent with the literature, rather than investigating the distribution of the stars across the galactic populations.

Several studies attempted to estimate individual abundances of elements in M dwarfs spectra, from fits of synthetic spectra~\citep[Jahandar et al. in prep]{souto_2022} or equivalent widths~\citep{ishikawa_2020, ishikawa_2022}. In particular,~\citet{souto_2022} derived the element abundances for several targets included in our study (Gl~411, Gl~15A, Gl~725A, Gl~725B, and Gl~880) and obtained $\afe$--$\mh$ trends suggesting that $\afe$ increases for metal-poor stars, consistent with the relations  derived for giant stars from APOGEE data.\\

\subsection{Classification of stellar populations from dynamics}

\begin{figure}
	\includegraphics[width=\linewidth]{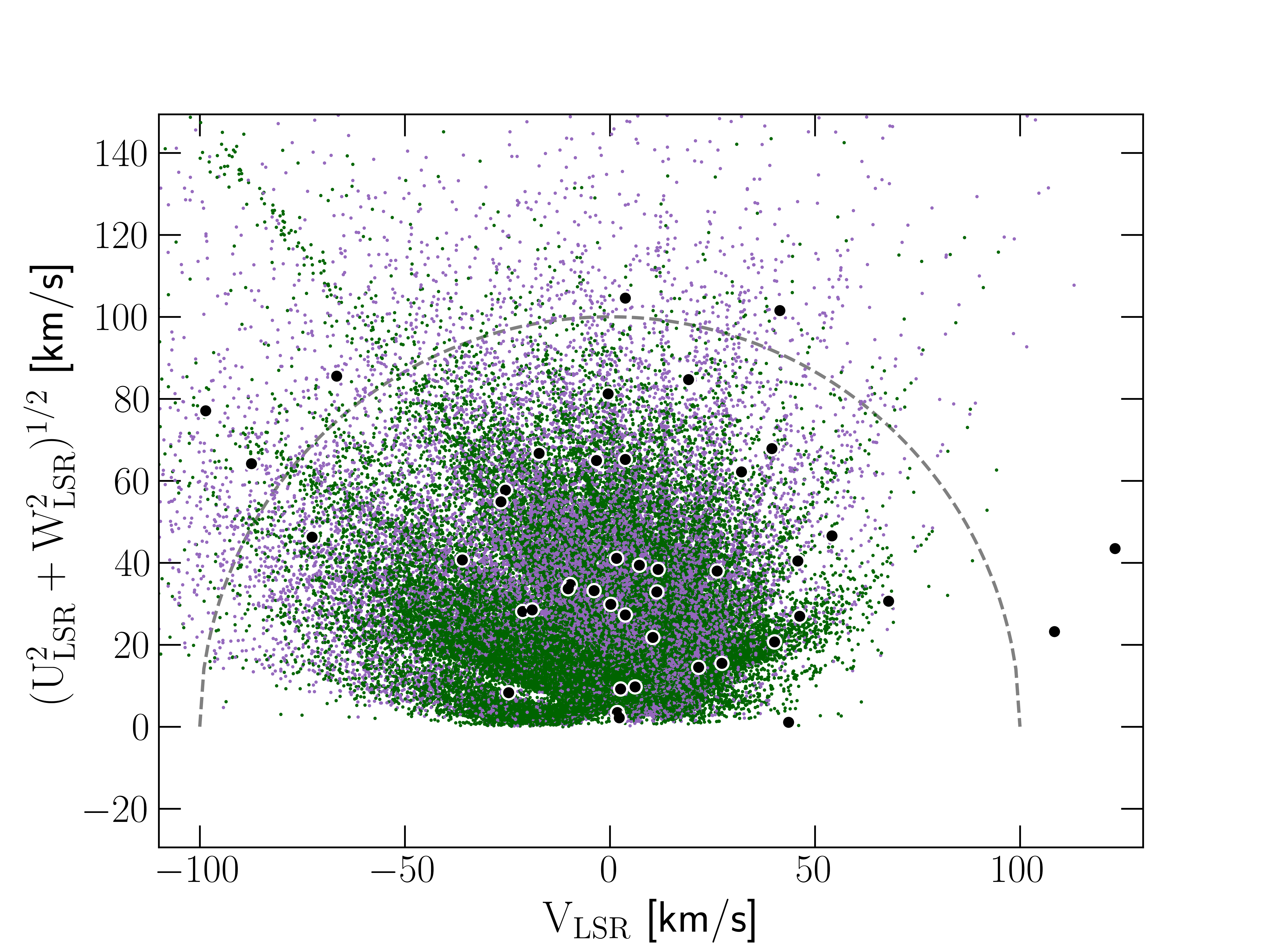}
	\caption{Toomre diagram for the  giant stars studied with APOGEE.  U, V and W are the velocities in the galactic coordinate system, corrected for solar motion (LSR). Purple and green pixels show stars from the thick and thin disc respectively, distinguished from their elemental abundances. The grey dashed line marks a fiducial boundary at 100~$\kms$. The stars studied in this work are marked with a black dot. {\paul An alternative figure with labels identifying the stars is presented in Fig.~\ref{fig:toomre_diagram_labels}.}}
	\label{fig:toomre_diagram}
\end{figure}

\begin{figure}
	\includegraphics[width=\columnwidth]{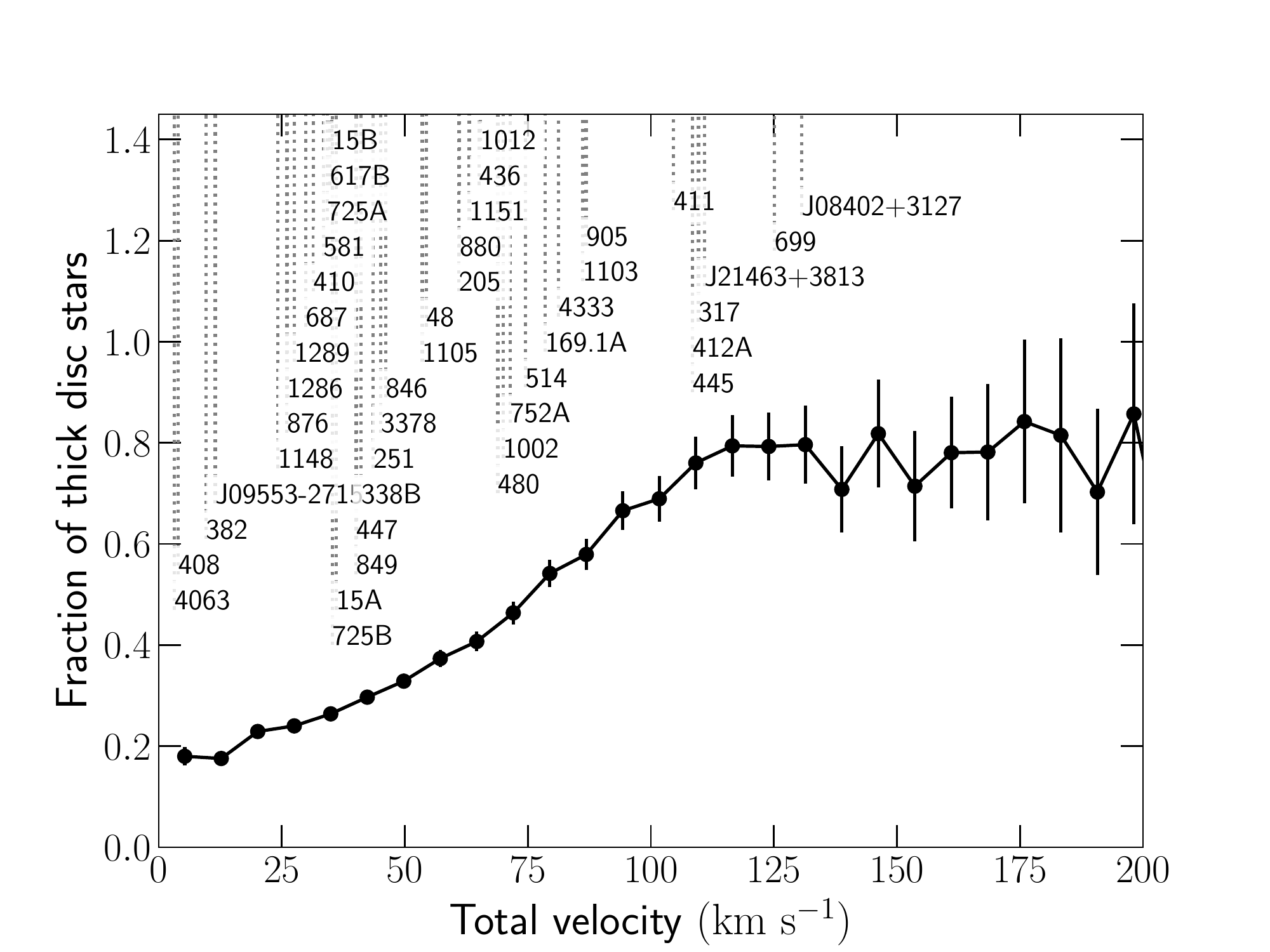}
	\caption{Thick-to-thin disk stars ratio per total velocity bin. Labels mark the velocities of the stars in our sample.  This ratio suggests that stars with total velocities >~100~$\kms$ have a probability >~70\% to belong to the thick disc.}
	\label{fig:histogram_pop}
\end{figure}

\begin{figure}
	\includegraphics[width=\linewidth]{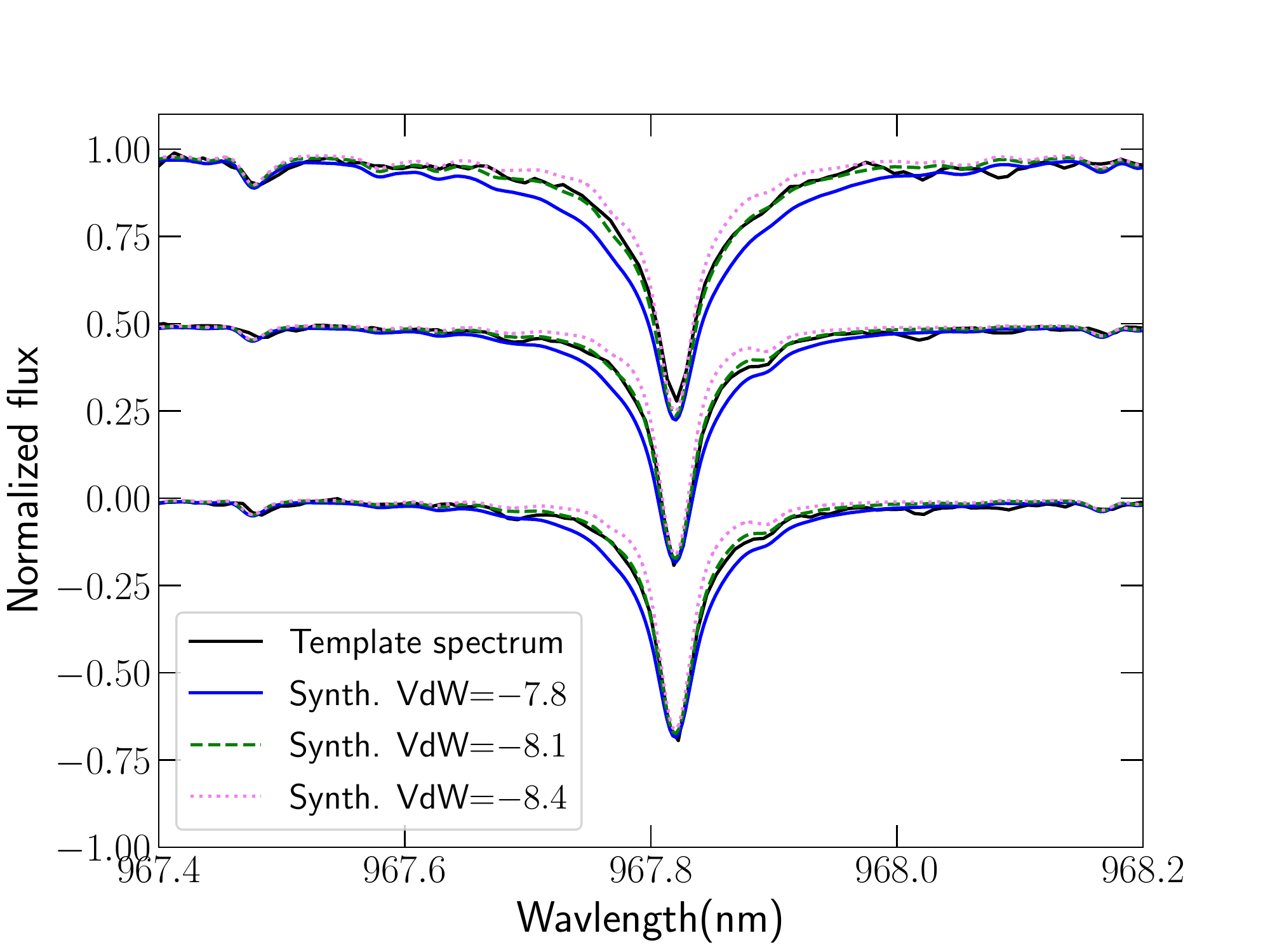}
	\caption{Example of Ti line. Black lines present the template spectra of 3 stars, from top to bottom: Gl~699, Gl~15A and Gl~411. Synthetic spectra with 3 different value of Van der Waals damping parameter are plotted for each star. The initial value found in the list was of $-7.8$, and we adopt a value of $-8.1$ for our analysis.}
	\label{fig:example_adj_line}
\end{figure}

Placing the giants studied  with APOGEE on a Toomre diagram, we find that the thick disc stars tend to have higher total velocity than thin disc stars (see Figs.~\ref{fig:toomre_diagram}~\&~\ref{fig:toomre_diagram_labels}), and that most of the stars in our sample are found  to feature a peculiar velocity below 100~$\kms$.  Besides, looking at the proportion of thin and thick disc  giants with a given velocity (see Fig.~\ref{fig:histogram_pop})  provides an estimate of the probability for a star to belong to either population based on its velocity. In particular, stars with a total velocity above 100~$\kms$  likely belong to the thick { disc with a probability~$>70\%$}.
 Assuming that M dwarfs behave as giant stars in this respect suggests that most of ours stars,  featuring velocities  $<75\,\kms$, are likely to belong to the thin disc.  Only 7 of our stars (PM~J08402+3127, PM~J21463+3813, Gl~699, Gl~411, Gl~317, Gl~445 and Gl~412A) have a total velocity $>100\,\kms$, and  are thus more likely  to belong to the thick disc.  We come back on this point further in the paper.



Because the choice of $\afe$ has a strong impact on the other three parameters, and because we cannot  arbitrarily set its value for each star, we  chose to fit $\afe$ in our analysis procedure.


%

\section{Line selection and adjustment}
\label{sec:line_selection}

The analysis must rely on well modelled spectral lines in order to provide accurate stellar parameters. Selecting such lines is particularly challenging in the NIR where molecular lines may blend with atomic features, and where models may not accurately reproduce line profiles.  SPIRou allows us to select several lines from multiple bands thanks to its large wavelength coverage. In this work, we revised the line selection performed in C22 and adjusted the properties of some lines, assuming known stellar parameters for 3 of our calibration stars: Gl~699, Gl~15A and Gl~411.

\subsection{Selecting the stellar lines of interest}

\begin{table}
	\center
	\caption{{\paul Full list of spectral lines used. Lines were identified  by depth and wavelength using the VALD database.}}
	\begin{tabular}[h]{cc}
		\hline
		Species & Wavelength (Å)\\
		\hline
		Ti I & 9678.198, 9691.527, 9708.327, 9721.626 \\ & 22969.597 \\
		Fe I & 10343.719 \\
		Ca I & 16201.500 \\
		K I & 15167.211 \\
		Mn I & 12979.459 \\
		Al I & 13126.964, 16723.524, 16755.203 \\
		Mg I & 15044.357, 15051.818 \\
		Na I & 22062.420, 22089.692 \\
		OH & 1672.3418, 1675.3831, 1675.6299 \\
		CO & 22935.233, 22935.291, 22935.585, 22935.754 \\ & 22936.343, 22936.627, 22937.511, 22937.900 \\ & 22939.094, 22939.584, 22941.089, 22941.668 \\ & 22943.494, 22944.163, 22946.311, 22947.059 \\ & 22949.544, 22953.195, 22954.059, 22957.263 \\ & 22958.159, 22961.743, 22962.671, 22966.648 \\ & 22967.576, 22971.971, 22972.884, 22977.719 \\ & 22978.596, 22983.888, 22984.707, 22990.488 \\ & 22991.222, 23112.404, 23124.542, 23150.029, 23163.381 \\
		\hline
	\end{tabular}
	\label{tab:selected_lines}
\end{table}

\begin{table}
	\caption{Line list used for the analysis. Column 1 to 5 presents the parameters found in the original list. Modifications to the oscillator strength ($\rm \Delta lggf$) and Van de Waals parameter ($\rm \Delta VdW$) are specified in columns 6 and 7, when applicable.  When the hyperfine structure (HFS) is available, we display data for all subcomponents.  Two distinct prescriptions are found in the Van der Waals column: the commonly reported Van der Waals damping parameter $\gamma_6$ is considered if the value is negative; values between 0 and 20 give the value of the fudge factor within the Unsöld approximation.
	 	}
	\label{tab:line_parameters} 
	\resizebox{.97\columnwidth}{!}{
		\begin{tabular}{rcccrrcc}
				\hline
				& Vac. wvl. (Å) &  $\chi_{\rm low}$ & $\rm lggf$ & VdW & species & $\rm \Delta lggf$ & $\rm \Delta VdW$\\
				\hline
		&  \ \    9678.198 & $0.84$ & $-0.80$ & $-7.80$ & Ti I & -- & $-0.3$ \\ 
		&   \ \    9691.527 & $0.81$ & $-1.61$ & $-7.80$ & Ti I & -- & $-0.2$ \\ 
		&  \ \    9708.327 & $0.83$ & $-1.01$ & $-7.80$ & Ti I & -- & $-0.2$ \\ 
		&  \ \   9721.626 & $1.50$ & $-1.18$ & $-7.78$ & Ti I & $-0.1$ & $-0.2$ \\ 
		&    10343.719 & $2.20$ & $-3.58$ & $-7.80$ & Fe I & -- & -- \\ 
		&    10968.389 & $5.93$ & $-2.16$ & $2.50$ & Mg I & -- & -- \\ 
		\multirow{15}{*}{HFS $\begin{dcases*} \\ \\  \\ \\ \\ \\ \\ \\ \\ \\  \\  \\  \end{dcases*}$ } & 12979.260 & $2.89$ & $-2.65$ & $2.50$ & Mn I & -- & -- \\ 
		& 12979.277 & $2.89$ & $-2.36$ & $2.50$ & Mn I & -- & -- \\ 
		& 12979.295 & $2.89$ & $-2.62$ & $2.50$ & Mn I & -- & -- \\ 
		& 12979.320 & $2.89$ & $-2.14$ & $2.50$ & Mn I & -- & -- \\ 
		& 12979.347 & $2.89$ & $-2.44$ & $2.50$ & Mn I & -- & -- \\ 
		& 12979.364 & $2.89$ & $-3.39$ & $2.50$ & Mn I & -- & -- \\ 
		& 12979.387 & $2.89$ & $-1.96$ & $2.50$ & Mn I & -- & -- \\ 
		& 12979.423 & $2.89$ & $-2.37$ & $2.50$ & Mn I & -- & -- \\ 
		& 12979.450 & $2.89$ & $-3.31$ & $2.50$ & Mn I & -- & -- \\ 
		& 12979.478 & $2.89$ & $-1.80$ & $2.50$ & Mn I & -- & -- \\ 
		& 12979.524 & $2.89$ & $-2.40$ & $2.50$ & Mn I & -- & -- \\ 
		& 12979.560 & $2.89$ & $-3.44$ & $2.50$ & Mn I & -- & -- \\ 
		& 12979.592 & $2.89$ & $-1.66$ & $2.50$ & Mn I & -- & -- \\ 
		& 12979.647 & $2.89$ & $-2.58$ & $2.50$ & Mn I & -- & -- \\ 
		& 12979.692 & $2.89$ & $-3.79$ & $2.50$ & Mn I & -- & -- \\ 
		\multirow{6}{*}{HFS $\begin{dcases*} \\ \\  \\ \\  \\  \end{dcases*}$ } & 13126.957 & $3.14$ & $-0.62$ & $2.50$ & Al I & -- & -- \\ 
		&  13126.962 & $3.14$ & $-0.52$ & $2.50$ & Al I & -- & -- \\ 
		&  13126.965 & $3.14$ & $-0.63$ & $2.50$ & Al I & -- & -- \\ 
		&  13127.024 & $3.14$ & $-0.16$ & $2.50$ & Al I & -- & -- \\ 
		&  13127.030 & $3.14$ & $-0.52$ & $2.50$ & Al I & -- & -- \\ 
		&  13127.035 & $3.14$ & $-1.06$ & $2.50$ & Al I & -- & -- \\ 
		&   15044.357 & $5.11$ & $0.12$ & $-7.20$ & Mg I & -- & -- \\ 
		&   15051.818 & $5.11$ & $-0.40$ & $-7.19$ & Mg I & -- & -- \\ 
		&   15167.211 & $2.67$ & $0.63$ & $-6.82$ & K I & -- & -- \\ 
		&   16201.500 & $4.54$ & $0.09$ & $-6.59$ & Ca I & -- & -- \\ 
		\multirow{6}{*}{HFS $\begin{dcases*} \\ \\  \\ \\  \\  \end{dcases*}$ } &  16723.478 & $4.08$ & $-0.66$ & $-7.15$ & Al I & -- & -- \\ 
		&  16723.492 & $4.08$ & $-0.55$ & $-7.15$ & Al I & -- & -- \\ 
		&  16723.510 & $4.08$ & $-1.09$ & $-7.15$ & Al I & -- & -- \\ 
		&  16723.512 & $4.08$ & $-0.65$ & $-7.15$ & Al I & -- & -- \\ 
		&  16723.530 & $4.08$ & $-0.55$ & $-7.15$ & Al I & -- & -- \\ 
		&  16723.557 & $4.08$ & $-0.19$ & $-7.15$ & Al I & -- & -- \\ 
		\multirow{12}{*}{HFS $\begin{dcases*} \\ \\  \\ \\  \\   \\ \\  \\ \\  \end{dcases*}$ }  & 16755.031 & $4.09$ & $-0.02$ & $-7.22$ & Al I & -- & -- \\ 
		&  16755.115 & $4.09$ & $-0.23$ & $-7.22$ & Al I & -- & -- \\ 
		&  16755.126 & $4.09$ & $-0.71$ & $-7.22$ & Al I & -- & -- \\ 
		&  16755.183 & $4.09$ & $-0.51$ & $-7.22$ & Al I & -- & -- \\ 
		&  16755.192 & $4.09$ & $-0.56$ & $-7.22$ & Al I & -- & -- \\ 
		&  16755.203 & $4.09$ & $-1.66$ & $-7.22$ & Al I & -- & -- \\ 
		&  16755.236 & $4.09$ & $-0.92$ & $-7.22$ & Al I & -- & -- \\ 
		&  16755.241 & $4.09$ & $-0.58$ & $-7.22$ & Al I & -- & -- \\ 
		&  16755.249 & $4.09$ & $-1.28$ & $-7.22$ & Al I & -- & -- \\ 
		&  16755.274 & $4.09$ & $-0.74$ & $-7.22$ & Al I & -- & -- \\ 
		&  16755.279 & $4.09$ & $-1.11$ & $-7.22$ & Al I & -- & -- \\ 
		&  16755.293 & $4.09$ & $-1.06$ & $-7.22$ & Al I & -- & -- \\ 
		\multirow{3}{*}{HFS $\begin{dcases*} \\ \\  \\ \\  \\  \end{dcases*}$ } &   22062.379 & $3.19$ & $-0.52$ & $2.00$ & Na I & -- & -- \\ 
		&   22062.381 & $3.19$ & $-0.52$ & $2.00$ & Na I & -- & -- \\ 
		&   22062.381 & $3.19$ & $-0.92$ & $2.00$ & Na I & -- & -- \\ 
		&   22062.442 & $3.19$ & $-0.07$ & $2.00$ & Na I & -- & -- \\ 
		&   22062.446 & $3.19$ & $-0.52$ & $2.00$ & Na I & -- & -- \\ 
		&   22062.448 & $3.19$ & $-1.22$ & $2.00$ & Na I & -- & -- \\ 
		\multirow{3}{*}{HFS $\begin{dcases*} \\ \\   \\  \end{dcases*}$ } &   22089.645 & $3.19$ & $-0.52$ & $2.00$ & Na I & -- & -- \\ 
		&   22089.655 & $3.19$ & $-1.22$ & $2.00$ & Na I & -- & -- \\ 
		&   22089.712 & $3.19$ & $-0.52$ & $2.00$ & Na I & -- & -- \\ 
		&   22089.721 & $3.19$ & $-0.52$ & $2.00$ & Na I & -- & -- \\ 
		&   22969.597 & $1.89$ & $-1.53$ & $-7.79$ & Ti I & -- & -- \\ 
				\hline
			\end{tabular} 
		}
\end{table}

Stellar lines are selected by comparing observation templates to synthetic spectra  assuming atmospheric parameters as derived from M15,  identifying those that are well reproduced by the models, and sensitive to the fundamental parameters we want to constrain. This selection is performed by comparing spectra of calibration stars to model spectra computed for expected parameters. In C22, we selected a set of 26 atomic lines and 40 molecular lines, mainly located in the CO band between 2290 and 2300 nm.  In this new study, we added several atomic and OH lines, and  rejected some atomic lines  that are found to be poorly informative, leading to a new line list containing 17 atomic lines, 9 OH lines and  CO lines from the afore mentioned {\paul (see Table~\ref{tab:selected_lines})}. The selected atomic lines are reported in Table~\ref{tab:line_parameters}, and include 7 lines from non-alpha elements (Fe, Mn, Al, K, and Na). The table also lists the parameters of the atomic lines, with the hyperfine structure when included in our line lists. This data is used to compute the emergent spectra with the Turbospectrum radiative transfer code.

To exclude some lines, we compared the $\chi^2$ values computed for the expected model (assuming the parameters of M15) and the best fit obtained ( whose parameters may  differ from the expected values).  Whenever,  for our calibration stars,  the computed $\chi^2$  is found to be much larger for the expected atmospheric parameters than  for those derived with our process, we adjusted the line parameters (see Sec.~\ref{sec:adjust_lines}) or excluded the region from  our analysis.

\subsection{Adjusting line parameters on reference stars}
\label{sec:adjust_lines}

The adjustments were performed on  3 of our best calibration stars (Gl~699, Gl~15A, Gl~411), by comparing the modelled spectra with various values  of the Van Der Waals  broadening parameter and oscillator strengths to the SPIRou stellar template spectra. For this step, the parameters published by M15 are  assumed for  our calibration stars,  and $\afe$ values were set to 0.2~dex for Gl~699 and Gl~411 and  0.08~dex for Gl~15A, assuming thick and thin disc populations based on velocity.

Significant differences are observed between models and observations, in particular for Ti lines, whose wings appear wider in the models than in observations ; this is likely to affect determinations of $\logg$ if not corrected for.  Since the wings of these lines are very sensitive to the Van Der Waals  collisional broadening parameter, as illustrated on Fig.~\ref{fig:example_adj_line}, we decreased the value of this parameter for these lines  until a good fit was achieved for all 3 reference stars, and re-computed a grid of spectra with these adjustments. All corrections applied to the line list are specified in Table~\ref{tab:line_parameters}. Some lines were attributed an Unsöld factor~\citep{unsold_1995} when no value of the Van der Waals damping parameter ($\gamma_6$) was reported in the VALD~\citep{pakhomov_2019} line lists.



\subsection{Consequence on retrieved parameters}

To assess the impact of our adjustments on the retrieved stellar parameters, we perform the analysis on our calibration stars with the new set of synthetic models computed with these adjustments, and derived for each star the 4 atmospheric parameters of interest with the corresponding error bars.
We compare these results to those obtained with the original line list (see Fig.~\ref{fig:comparison_12stars_adj}). 
{\paul The $\mh$ and $\logg$ estimates of a few stars are found to be in better agreement with  M15.} The influence of the correction remains however small on the retrieved $\teff$, $\logg$ and $\mh$ for most stars. Similarly, we look at the effect of the correction on our estimated $\afe$ (see Fig.~\ref{fig:comparison_12stars_alpha}), and retrieve values closer to those expected from empirical relations for a few stars, such as Gl~849, Gl~880 or Gl~905.
	
We also perform a comparison between the results obtained while fitting $\afe$ or if the parameter is set to 0 (see Fig.~\ref{fig:comparison_12stars}). 
We find that fitting $\afe$ allows us to significantly reduce the  scatter on the retrieved $\logg$, and to obtain $\mh$ estimates in better agreement with our reference study, with the exception of Gl~905, for which $\mh$ is found about 0.2~dex smaller than  that reported by M15, who relied on empirically calibrated relations between equivalent widths of some atomic features and metallicity. Subsequent tests  showed that a variation of $\afe$ of 0.05~dex can lead to a 0.2~dex variation on $\mh$ for this star.



\begin{figure}
	\includegraphics[width=.99\linewidth]{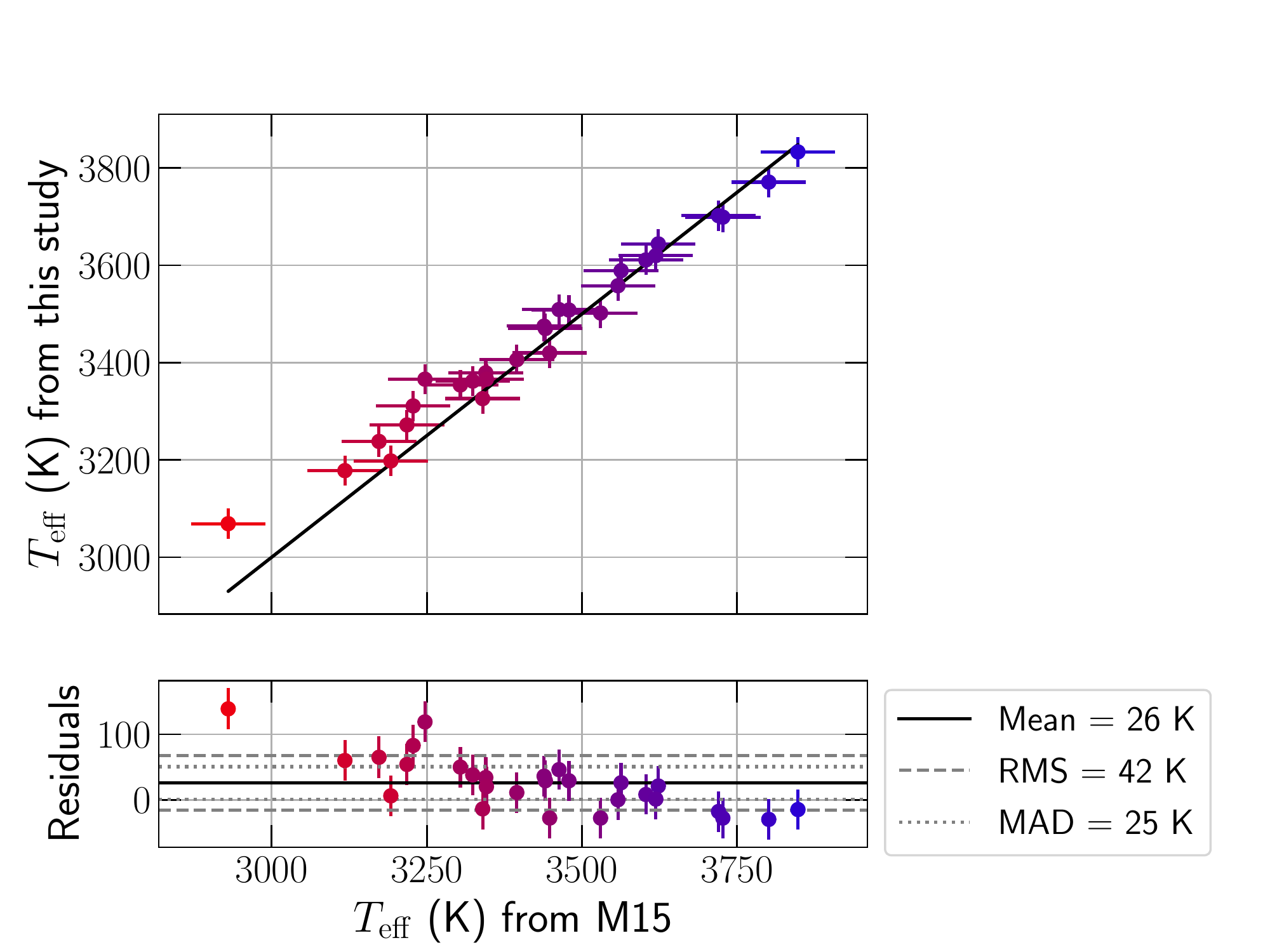}
	\includegraphics[width=.99\linewidth]{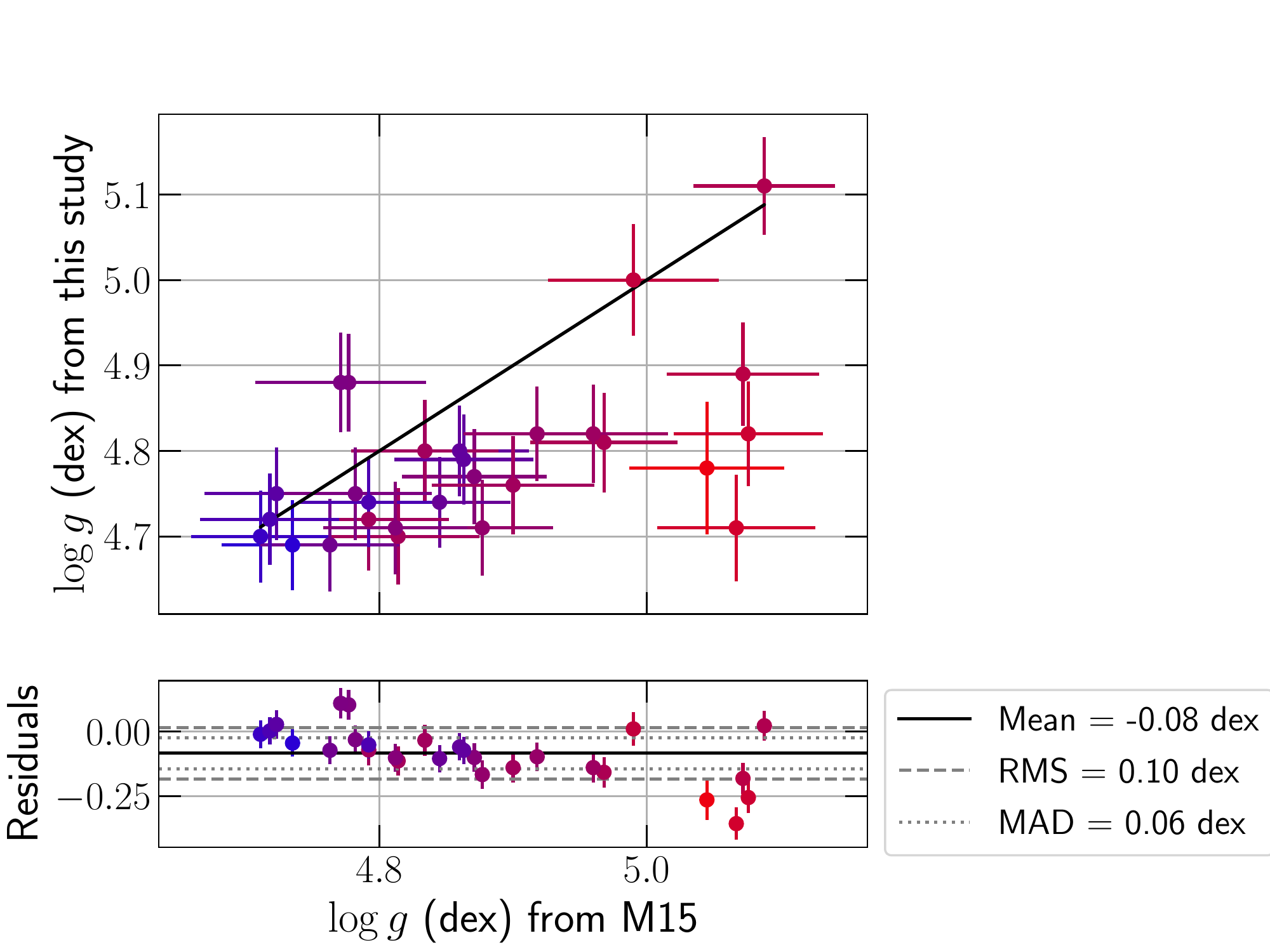}
	\includegraphics[width=.99\linewidth]{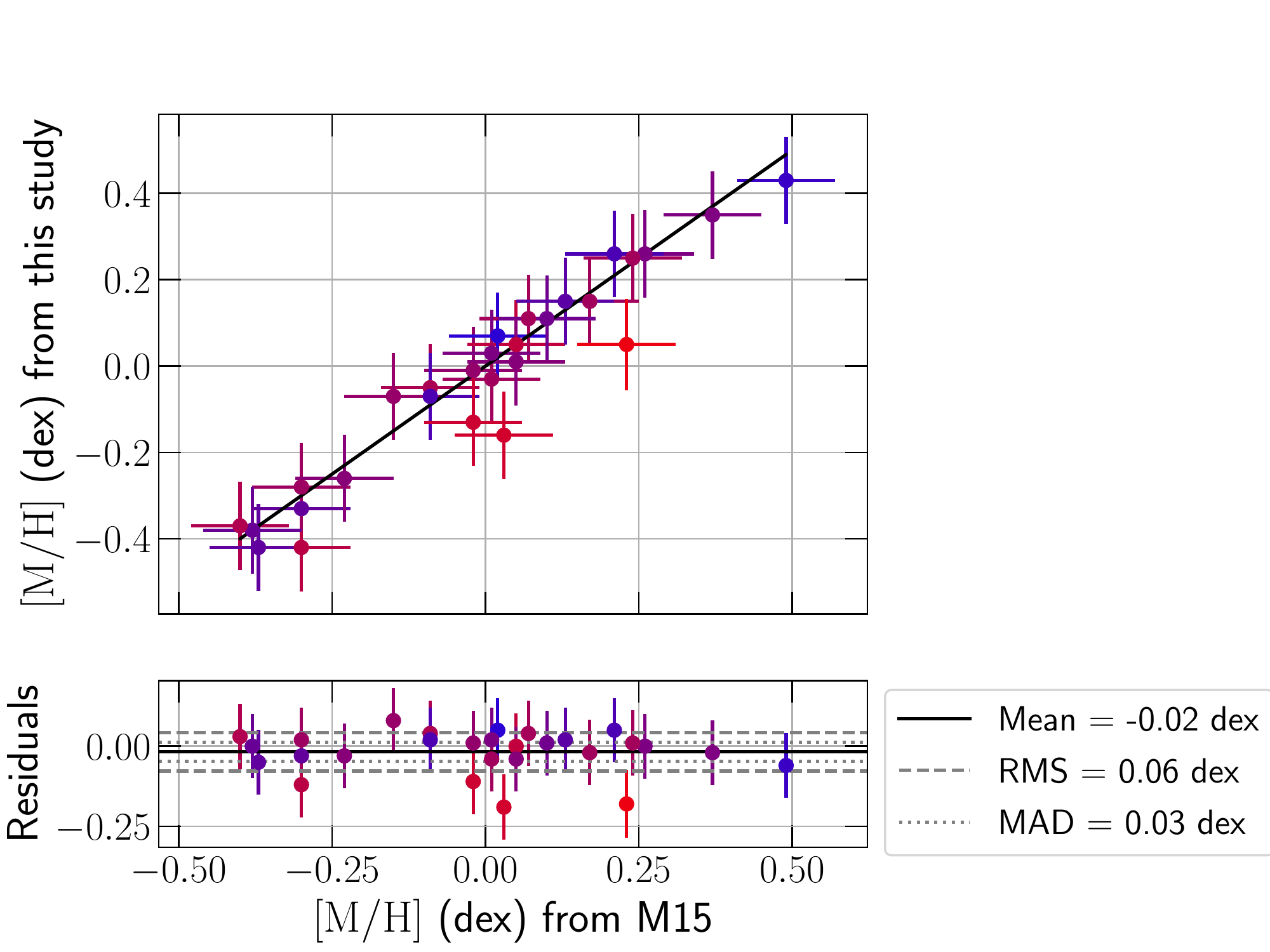}
	\caption{Comparison between retrieved parameters and value published by M15 for 23 stars  common to both samples. The temperature is colour coded from red (coolest) to blue (hottest). {\paul An alternative figure with labels identifying the stars is presented  in Fig.~\ref{fig:results_all_4d_labels}.}}
	\label{fig:results_all_4d}
\end{figure}

 Two binaries are included in our study: Gl~725 and Gl~15. For both systems, we retrieve $\mh$ for each component that are in good agreement,
 with differences of 0.02~dex for Gl~725 and 0.09~dex for Gl~15, thereby improving over our initial study where this difference reached 0.21~dex in the case of Gl~15A~(C22). For Gl~15, we also observe a small difference in the $\afe$ values of 0.06~dex, again consistent with the estimated empirical error bars.

\section{Results}
\label{sec:results}
We performed the analysis described in Sec.~\ref{sec:method} with the updated list presented in Sec.~\ref{sec:line_selection} , on our \nbMdwarfs{} selected targets (see Sec.~\ref{sec:target_selection}). Figures~\ref{fig:results_all_4d}~\&~\ref{fig:results_all_4d_labels} presents a comparison between the results and the parameters published by M15 for  the \nbMdwarfsCommon{} stars common to both samples. {\paul Figure~C1 (available as supplementary material) presents the best fit obtained on all lines for 5 stars in our sample.} Retrieved $\teff$, $\logg$, $\mh$ and $\afe$ are  listed in Table~\ref{tab:retrieved_parameters} along with  an estimate of  the stellar masses and radii.

\begin{figure}
	\includegraphics[width=\linewidth]{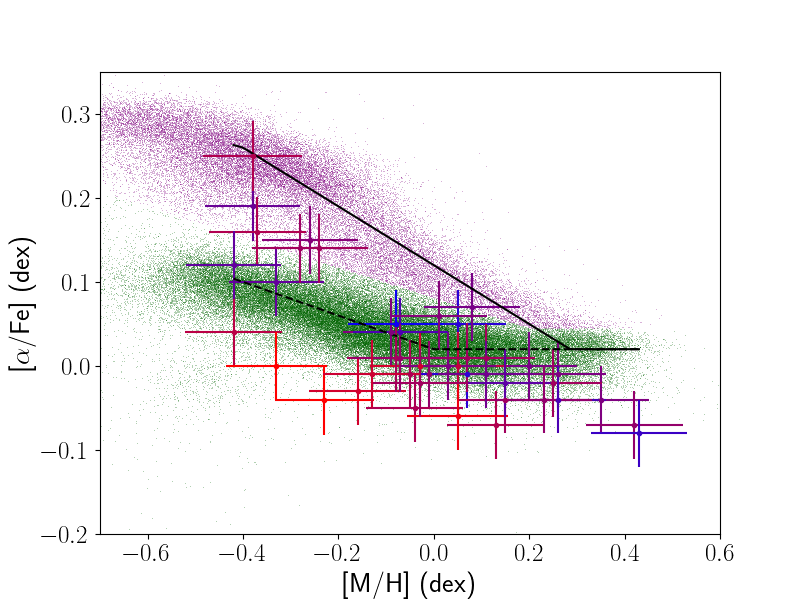}
	\caption{Retrieved $\afe$ plotted against $\mh$ for the \nbMdwarfs{} targets in our sample. {\paul Solid and dashed black lines mark empirical thick and thin disk $\mh$--$\afe$ relations, respectively. Colored pixels mark the position of giants studied by APOGEE, with purple and green colors marking those expected to be from the thick and thin disk respectively. An alternative figure with labels identifying the stars is presented in Fig.~\ref{fig:alpha_mh_all_4d_labels}.}}
	\label{fig:alpha_mh_all_4d}
\end{figure}

\subsection{Effective temperature}

For the \nbMdwarfsCommon{} stars  also studied by M15, we compare our results to the reported effective temperatures (Fig.~\ref{fig:results_all_4d}).
The overall retrieved $\teff$ are in good agreement with M15 with a RMS on the residuals of the order of 45~K, compatible with the error bars reported by M15. 
We observe a tendency to derive higher $\teff$ for cooler stars, with a deviation of up to 140~K for Gl~905. 
This trend  may reflect  discrepancies in the physics used in the \texttt{MARCS} models at the lowest side of their temperature range, or  alternatively probe systematics in M15. To assess the internal dispersion of our results, we fit a line through our retrieved results (of slope 0.85 $\pm$ 0.02). For these \nbMdwarfsCommon{} stars, the RMS about the trend in $\teff$ is of about 25~K, of the order of our estimated error bars.

Fig.~\ref{fig:results_all_4d_pass} presents a similar comparison to the parameters retrieved by~\citet{passegger_2019}, who performed fits of \texttt{PHOENIX-ACES} synthetic spectra on high-resolution CARMENES data. 
The RMS on the residuals is then of about 60~K, again, of the order of the typically published error bars.  
We point out that~\citet{passegger_2019}, as well as other references such as~\citet{marfil_2021} , also find higher $\teff$ values  than M15 for the coolest star  of our sample.


\subsection{Metallicity and alpha-enhancement}
For the \nbMdwarfsCommon{} stars studied in this work and in M15, the $\mh$ values recovered with our analysis are in good agreement, with a RMS on the residuals of about 0.1~dex, of the order of our estimated empirical error bar for this parameter. Here again, the  largest deviation is observed for the coolest stars in our sample,  for which we find lower $\mh$ than M15, but for which other studies~\citep{passegger_2019, marfil_2021} also find different values  than M15 (see Figs.~\ref{fig:results_all_4d_pass}~\&~\ref{fig:results_all_4d_marf}).

 Comparing our results to the values 
published by~\citet[Fig.~\ref{fig:results_all_4d_pass}]{passegger_2019}, we find  a much larger RMS on the residuals of about 0.16~dex. These results illustrate the difficulty to estimate  the accuracy of the parameters  derived from fits of synthetic spectra  which depends on the assumed reference on which to rely. 


Fitting $\afe$ as an additional dimension in our process  allowed us to significantly improve the  estimate of $\mh$ for  cool  metal-poor stars.  {\paul Because our line list contains several features sensitive to $\afe$ variations, we are able to obtain reliable estimates of this parameter without the need to set priors.} Figures~\ref{fig:alpha_mh_all_4d}~\&~\ref{fig:alpha_mh_all_4d_labels} present the retrieved $\afe$ as a function of the recovered $\mh$ for the \nbMdwarfs{} stars  of our sample. These results are   globally consistent with the expected trends estimated from  the APOGEE  data for giants  and suggest that most of our stars belong to the thin galactic disc, with a few exception such as Gl~699, Gl~411, PM~J21463+3813 and Gl~445 which would more likely belong to the thick galactic disc. Gl~725~A and B are found at the limit of the fiducial boundary between thick and thin disc, and are therefore difficult to classify.

\subsection{Masses and radii}
\label{subsec:masses_radii}

\begin{figure}
	\includegraphics[width=\linewidth]{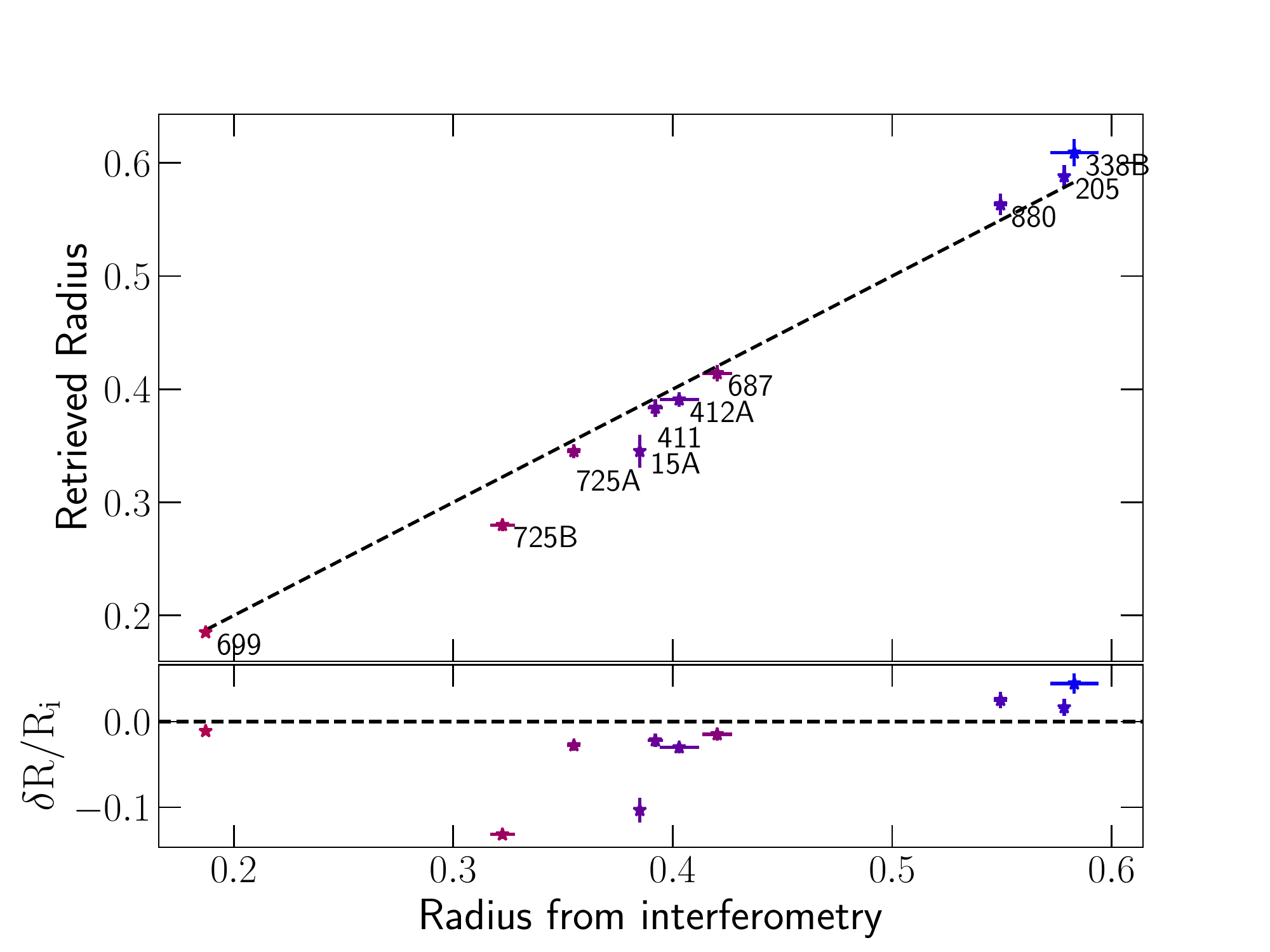}
	\caption{Comparison between radii retrieved from fits and computed from interferometric measurements~\citep{boyajian_2012} for 9 stars.  The symbol colour depicts the temperature from red (cool) to blue (hot). Larger error bars originate from uncertainties on the $\rm M_J$ measurements published by the 2MASS survey.  The bottom plot displays the relative difference between our estimated radii and those computed from interferometric measurements.}
	\label{fig:interferometry}
\end{figure}

\begin{figure*}
	\includegraphics[width=.5\linewidth]{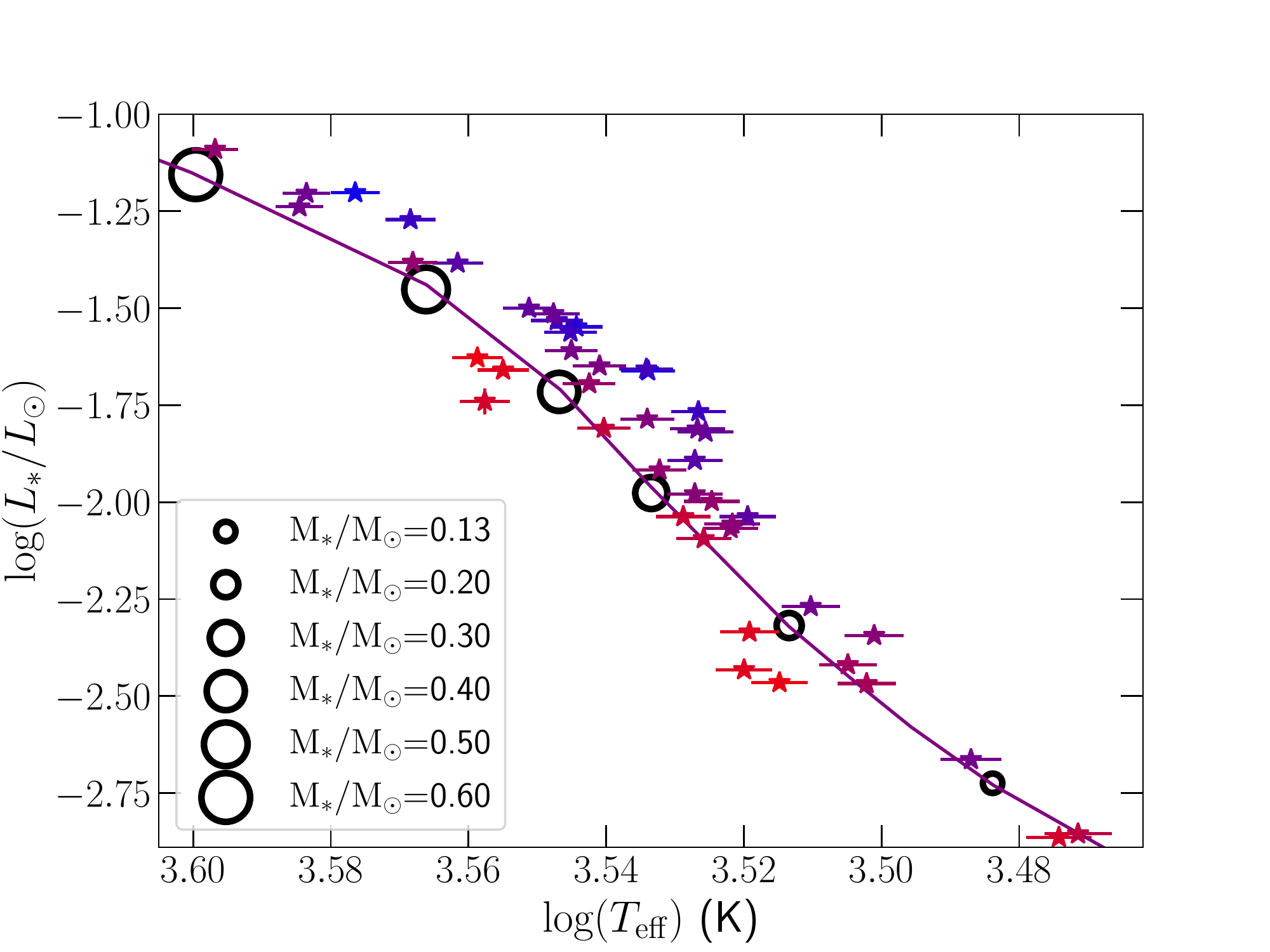}\includegraphics[width=.5\linewidth]{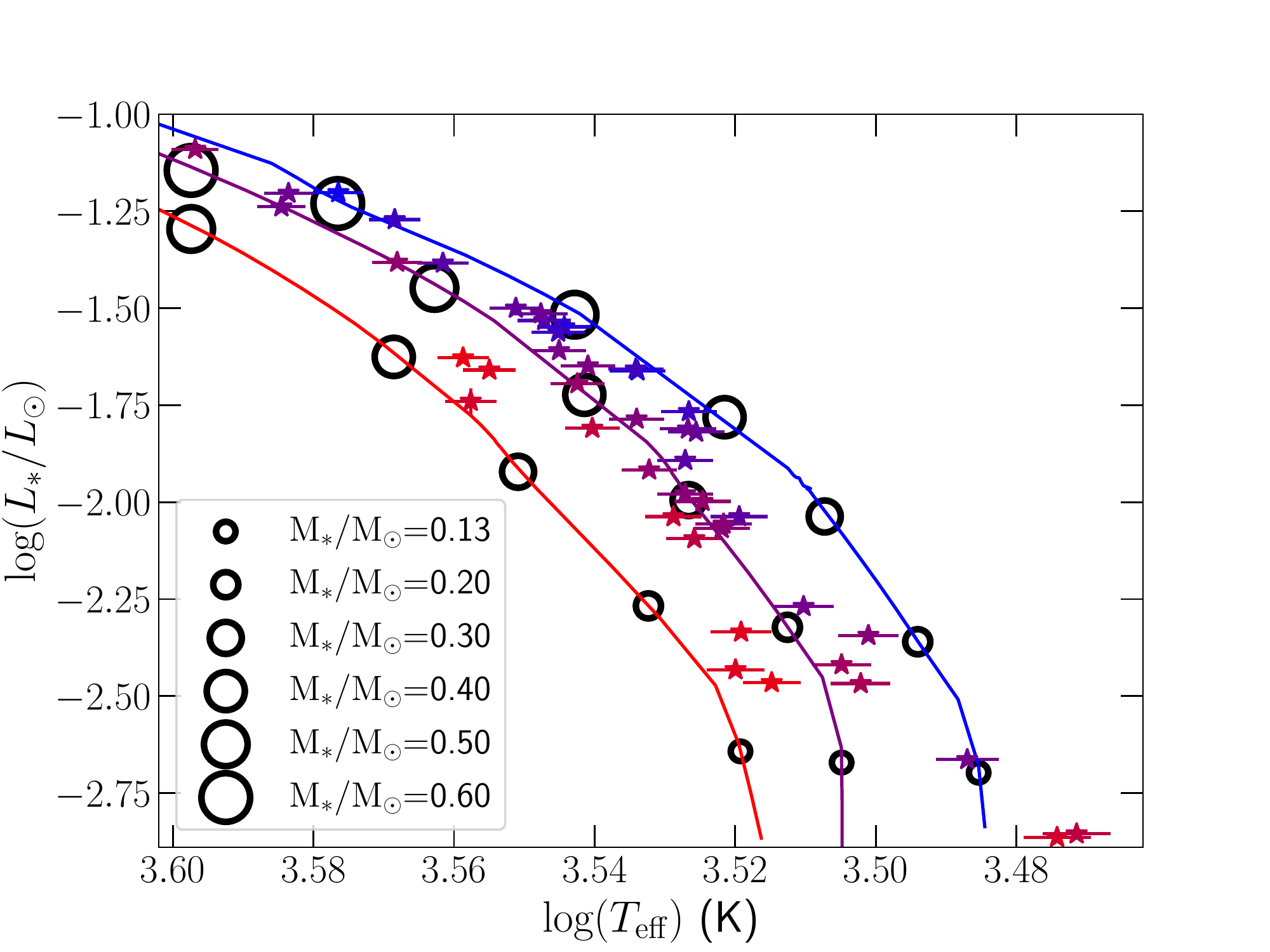}
	\caption{HR diagram showing the position of the stars in our sample. Luminosity was computed from G band magnitude retrieved through SIMBAD. The metallicity is colour coded from red to blue (low to high metallicity respectively). On the left panel, the purple solid line presents the isochrone computed by~\citet{baraffe_2015} at solar metallicity. On the right panel, the red, purple and blue solid lines present the DSEP stellar isochrones for $\mh=-0.5$~dex, $\mh=0.0$~dex and $\mh=0.5$~dex respectively. An age of 5 Gyr is assumed for all isochrones. Black circles mark the position of different stellar masses for each metallicity. {\paul An alternative figure with labels identifying the stars is presented  in Fig.~\ref{fig:hr_diagram_labels}.}}
	\label{fig:hr_diagram}
\end{figure*}

\begin{figure}
	\includegraphics[width=\columnwidth]{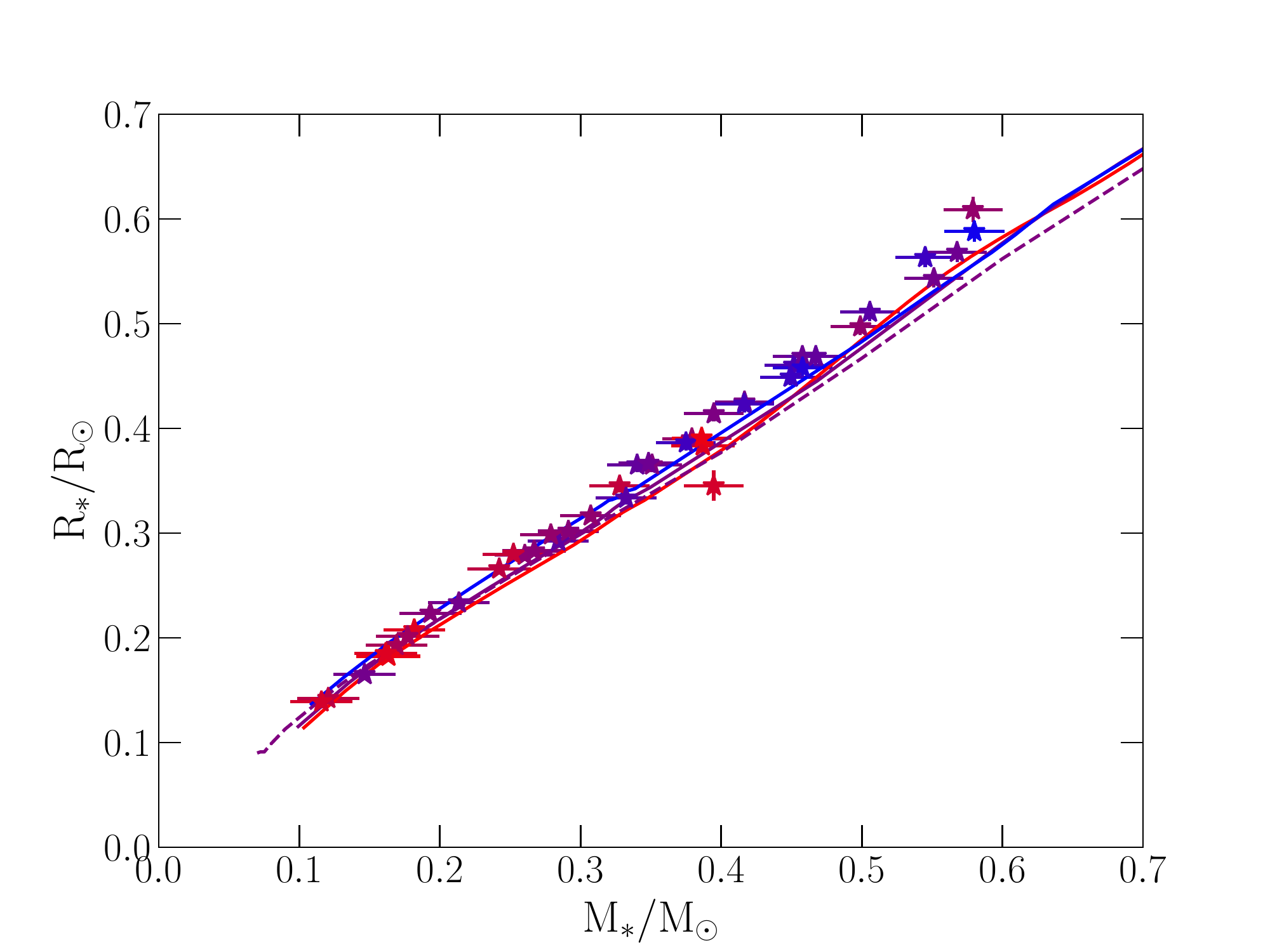}
	\caption{Mass-radius diagram showing the position of the stars in our sample. The metallicity is colour codded from red to blue (low to high metallicity respectively). The red, purple and blue  solid lines present the mass-radius relation predicted by the DSEP models for $\mh=-0.5$~dex, $\mh=0.0$~dex and $\mh=0.5$~dex respectively. The purple  dashed line presents the mass-radius relationship predicted by the models of~\citet{baraffe_2015}, at solar metallicity. An age of 5 Gyr is assumed for all models. {\paul An alternative figure with labels identifying the stars is presented  in Fig.~\ref{fig:m_r_diagram_labels}.}}
	\label{fig:m_r_diagram}
\end{figure}

 \citet{mann_2019} derived a K band magnitude ($\rm M_K$)~--~mass~--~metallicity empirical relation. We use this relation to derive the masses of the targets in our sample. 
Radii for the studied stars can be computed  from the recovered $\teff$ and  the bolometric luminosity  using Stefan-Boltzmann law.  Bolometric luminosities are directly computed from 2MASS J and Gaia (DR2) G band absolute 
magnitudes ($\rm M_{J}$ and $\rm M_{G}$ respectively) and bolometric corrections~\citep{cifuentes_2020}.
All magnitudes used in this work were extracted from SIMBAD\footnote{http://simbad.cds.unistra.fr/simbad/}. In this work we chose to derive the luminosities from bolometric corrections and absolute magnitudes rather than to rely on bolometric luminosities reported by authors such as~\citet{cifuentes_2020} or M15. This allows us to produce self-consistent results for all the stars in our sample as these studies do not typically report values for all our targets. Several tests allowed to verify that the reported values and these derived from bolometric corrections are in fair agreement for most stars (see Fig.~\ref{fig:hr_diagram_comp}). One should note that the 2MASS survey attributes a quality flag to the reported magnitudes, which may not systematically be accounted for by reported uncertainties.
We compare our retrieved radii ($R_{\rm f}$) to those computed from interferometry ($R_{\rm i}$) by~\citet[see Fig.~\ref{fig:interferometry}]{boyajian_2012}. 
We find values that are  consistent with interferometric measurements for the 9 stars studied by~\citet{boyajian_2012}, with  a dispersion on $\delta R/R_{\rm i}\approx5$~\%, with $\delta R = {R_{\rm f} - R_{\rm i}}$.
 
 We note that the radius retrieved from interferometry for Gl~725B is significantly larger than the one we  estimate with this method; coupled with  the measured magnitude, it would yield an effective temperature of $\teff=3145 \pm 10$~K ,  i.e., 200~K cooler than the values derived by  most studies~\citep[M15, C22]{fouque_2018, marfil_2021} and ours. This discrepancy was also observed and reported by M15. The apparent inconsistency in these results calls for an in-depth investigation of Gl~725B.

We  locate our stars in a Hertzsprung-Russell (HR) diagram {\paul(see Figs.~\ref{fig:hr_diagram}~\&~\ref{fig:hr_diagram_labels}).} 
 We compare our results to the isochrone computed by~\citet{baraffe_2015}. Our results tend to be in good agreement with the model, with points scattered around the isochrone, which can be attributed to metallicity. Isochrones computed with the Dartmouth stellar evolution program~\citep[DSEP,][]{dotter_2008} for different metallicities confirm the dependency on $\mh$. We also observe a strong divergence between the DSEP models and those of~\citet{baraffe_2015}, in particular for stars with masses lower than 0.3~$\rm M_{\sun}$. 

%

Our estimated radii and masses are found in good agreement with mass-radius relations expected from stellar evolution models~\citep[{\paul see Figs.~\ref{fig:m_r_diagram}~\&~\ref{fig:m_r_diagram_labels}, }][]{feiden_2012}. {\pp We further note a good agreement between our derived masses and radii and those reported by M15 (see Fig.~\ref{fig:m_r_diagram_mann}), with a relative dispersion of 4~\% on both parameters.}


\subsection{Surface gravity}

Surface gravity is known to be difficult to constrain for M dwarfs. Several studies  chose to fix this parameter from semi-empirical relations or evolutionary models~\citep{passegger_2019, rajpurohit_2017}. Following C22, we fit this parameter.  Our new estimates are in better agreement with M15 than those of C22, showing that the various improvements brought to our analysis (see Sec.~\ref{sec:method}--\ref{sec:line_selection}) helped solving the issue.

\begin{figure}
	\includegraphics[width=\linewidth]{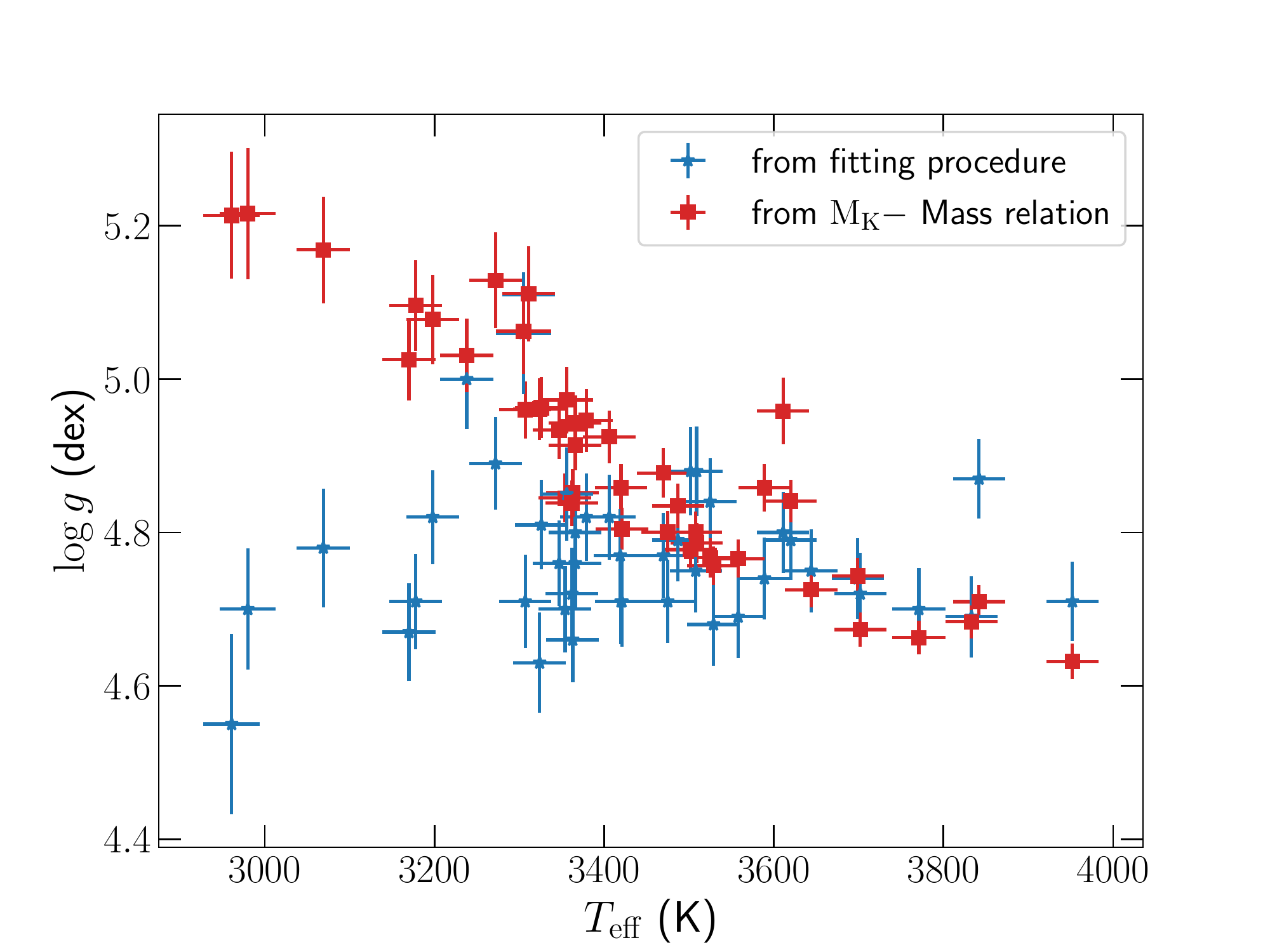}
	\caption{Comparison of the $\logg$ derived from our fitting procedure and those computed from $\rm M_{K}$--mass relation. {\paul An alternative figure with labels identifying the stars is presented  in Fig.~\ref{fig:logg_logg_labels}.}}
	\label{fig:logg_logg}
\end{figure}

 From the masses and radii derived in Sec~\ref{subsec:masses_radii} we compute new $\logg$ values, and compare these to the values obtained from the spectral fitting procedure (see {\paul Figs.~\ref{fig:logg_logg}~\&~\ref{fig:logg_logg_labels}}). We observe significant differences between the two sets of $\logg$ values, and compute a RMS on the residuals of about 0.2~dex. This dispersion is also the result of larger discrepancies at low $\teff$ and the RMS value computed  when ignoring the  6 coolest stars in our sample falls to 0.11~dex . This may suggest that,  for some yet unclear reason, we underestimate the $\logg$ of the coolest stars with our fitting procedure.



%

\begin{table*}
	\caption{Retrieved parameters of the star in our sample. Columns 2 to 5 present the stellar parameters estimated from our fitting procedure. Columns 6 through 9 respectively present the Gaia G, J and K band absolute magnitudes and parallaxes extracted from SIMBAD. Column 10 lists the log of the logarithmic bolometric luminosity (with respect to that of the Sun) derived from $M_{\rm  G}$. Columns 11 and 12 present the radius computed from $\teff$ and $M_{\rm J}$ and mass derived from $M_{\rm K}$--mass relations. Column 13 presents alternative $\logg$ estimates computed from columns 11 and 12.}
	\label{tab:retrieved_parameters}
	\center
	\resizebox{\textwidth}{!}{
	\begin{tabular}{ccccccccccccc}
		\hline
		Star & $\teff$ & $\logg$ & $\mh$ & $\afe$ & $\rm M_G$ & $\rm M_J$ & $\rm M_K$ & Parallax & $\log(L_{\rm *} / L_{\rm \odot})$ & Radius & Mass & alt. $\logg$\\
\hline 
Gl 338B & $3952$ $\pm$ $30$ & $4.71$ $\pm$ $0.05$ & $-0.08$ $\pm$ $0.10$ & $0.05$ $\pm$ $0.04$ &$8.046 \pm 0.003$ &$5.77 \pm 0.17$ &$5.13 \pm 0.02$ &$157.88 \pm 0.02$ &$-1.090 \pm 0.011$ &$0.609 \pm 0.012$ &$ 0.58 \pm 0.02$  & 4.63 $\pm$ 0.02\\ 
Gl 410 & $3842$ $\pm$ $31$ & $4.87$ $\pm$ $0.05$ & $0.05$ $\pm$ $0.10$ & $0.05$ $\pm$ $0.04$ &$8.426 \pm 0.003$ &$6.14 \pm 0.02$ &$5.30 \pm 0.02$ &$83.76 \pm 0.02$ &$-1.239 \pm 0.002$ &$0.543 \pm 0.009$ &$ 0.55 \pm 0.02$  & 4.71 $\pm$ 0.02\\ 
Gl 846 & $3833$ $\pm$ $31$ & $4.69$ $\pm$ $0.05$ & $0.07$ $\pm$ $0.10$ & $-0.01$ $\pm$ $0.04$ &$8.282 \pm 0.003$ &$6.07 \pm 0.02$ &$5.20 \pm 0.02$ &$94.56 \pm 0.05$ &$-1.204 \pm 0.002$ &$0.568 \pm 0.009$ &$ 0.57 \pm 0.02$  & 4.68 $\pm$ 0.02\\ 
Gl 205 & $3771$ $\pm$ $31$ & $4.70$ $\pm$ $0.05$ & $0.43$ $\pm$ $0.10$ & $-0.08$ $\pm$ $0.04$ &$8.327 \pm 0.003$ &$6.05 \pm 0.06$ &$5.12 \pm 0.06$ &$175.31 \pm 0.02$ &$-1.202 \pm 0.004$ &$0.588 \pm 0.010$ &$ 0.58 \pm 0.02$  & 4.66 $\pm$ 0.02\\ 
Gl 880 & $3702$ $\pm$ $31$ & $4.72$ $\pm$ $0.05$ & $0.26$ $\pm$ $0.10$ & $-0.04$ $\pm$ $0.04$ &$8.611 \pm 0.003$ &$6.18 \pm 0.02$ &$5.34 \pm 0.02$ &$145.62 \pm 0.03$ &$-1.272 \pm 0.001$ &$0.563 \pm 0.009$ &$ 0.55 \pm 0.02$  & 4.67 $\pm$ 0.02\\ 
Gl 514 & $3699$ $\pm$ $31$ & $4.74$ $\pm$ $0.05$ & $-0.07$ $\pm$ $0.10$ & $0.04$ $\pm$ $0.04$ &$8.793 \pm 0.003$ &$6.49 \pm 0.02$ &$5.62 \pm 0.03$ &$131.10 \pm 0.03$ &$-1.382 \pm 0.001$ &$0.497 \pm 0.008$ &$ 0.50 \pm 0.02$  & 4.74 $\pm$ 0.02\\ 
Gl 382 & $3644$ $\pm$ $31$ & $4.75$ $\pm$ $0.05$ & $0.15$ $\pm$ $0.10$ & $-0.02$ $\pm$ $0.04$ &$8.897 \pm 0.003$ &$6.45 \pm 0.02$ &$5.58 \pm 0.02$ &$129.75 \pm 0.03$ &$-1.384 \pm 0.001$ &$0.511 \pm 0.009$ &$ 0.51 \pm 0.02$  & 4.73 $\pm$ 0.02\\ 
Gl 412A & $3620$ $\pm$ $31$ & $4.79$ $\pm$ $0.05$ & $-0.42$ $\pm$ $0.10$ & $0.12$ $\pm$ $0.04$ &$9.460 \pm 0.003$ &$7.08 \pm 0.02$ &$6.32 \pm 0.02$ &$203.89 \pm 0.03$ &$-1.628 \pm 0.001$ &$0.391 \pm 0.007$ &$ 0.39 \pm 0.02$  & 4.84 $\pm$ 0.03\\ 
Gl 15A & $3611$ $\pm$ $31$ & $4.80$ $\pm$ $0.05$ & $-0.33$ $\pm$ $0.10$ & $0.10$ $\pm$ $0.04$ &$9.460 \pm 0.003$ &$7.49 \pm 0.26$ &$6.26 \pm 0.02$ &$280.71 \pm 0.02$ &$-1.740 \pm 0.034$ &$0.345 \pm 0.015$ &$ 0.39 \pm 0.02$  & 4.96 $\pm$ 0.04\\ 
Gl 411 & $3589$ $\pm$ $31$ & $4.74$ $\pm$ $0.05$ & $-0.38$ $\pm$ $0.10$ & $0.19$ $\pm$ $0.04$ &$9.522 \pm 0.003$ &$7.17 \pm 0.24$ &$6.31 \pm 0.05$ &$392.75 \pm 0.03$ &$-1.659 \pm 0.010$ &$0.383 \pm 0.008$ &$ 0.39 \pm 0.02$  & 4.86 $\pm$ 0.03\\ 
Gl 752A & $3558$ $\pm$ $31$ & $4.69$ $\pm$ $0.05$ & $0.11$ $\pm$ $0.10$ & $-0.01$ $\pm$ $0.04$ &$9.240 \pm 0.003$ &$6.72 \pm 0.03$ &$5.81 \pm 0.02$ &$169.06 \pm 0.02$ &$-1.500 \pm 0.001$ &$0.469 \pm 0.008$ &$ 0.47 \pm 0.02$  & 4.77 $\pm$ 0.02\\ 
Gl 48 & $3529$ $\pm$ $31$ & $4.68$ $\pm$ $0.05$ & $0.08$ $\pm$ $0.10$ & $0.07$ $\pm$ $0.04$ &$9.364 \pm 0.003$ &$6.72 \pm 0.03$ &$5.87 \pm 0.02$ &$121.46 \pm 0.02$ &$-1.514 \pm 0.002$ &$0.469 \pm 0.008$ &$ 0.46 \pm 0.02$  & 4.76 $\pm$ 0.03\\ 
Gl 617B & $3525$ $\pm$ $31$ & $4.84$ $\pm$ $0.06$ & $0.20$ $\pm$ $0.10$ & $0.00$ $\pm$ $0.04$ &$9.459 \pm 0.003$ &$6.75 \pm 0.02$ &$5.91 \pm 0.02$ &$92.90 \pm 0.02$ &$-1.532 \pm 0.002$ &$0.460 \pm 0.008$ &$ 0.45 \pm 0.02$  & 4.77 $\pm$ 0.03\\ 
Gl 480 & $3509$ $\pm$ $31$ & $4.88$ $\pm$ $0.06$ & $0.26$ $\pm$ $0.10$ & $-0.01$ $\pm$ $0.04$ &$9.565 \pm 0.003$ &$6.81 \pm 0.02$ &$5.92 \pm 0.04$ &$70.11 \pm 0.03$ &$-1.562 \pm 0.002$ &$0.449 \pm 0.008$ &$ 0.45 \pm 0.02$  & 4.79 $\pm$ 0.03\\ 
Gl 436 & $3508$ $\pm$ $31$ & $4.75$ $\pm$ $0.05$ & $0.03$ $\pm$ $0.10$ & $0.00$ $\pm$ $0.04$ &$9.631 \pm 0.003$ &$6.95 \pm 0.02$ &$6.12 \pm 0.02$ &$102.30 \pm 0.03$ &$-1.609 \pm 0.002$ &$0.425 \pm 0.008$ &$ 0.42 \pm 0.02$  & 4.80 $\pm$ 0.03\\ 
Gl 849 & $3502$ $\pm$ $31$ & $4.88$ $\pm$ $0.06$ & $0.35$ $\pm$ $0.10$ & $-0.04$ $\pm$ $0.04$ &$9.511 \pm 0.003$ &$6.78 \pm 0.02$ &$5.87 \pm 0.02$ &$113.44 \pm 0.03$ &$-1.548 \pm 0.002$ &$0.458 \pm 0.008$ &$ 0.46 \pm 0.02$  & 4.78 $\pm$ 0.03\\ 
Gl 408 & $3487$ $\pm$ $31$ & $4.79$ $\pm$ $0.05$ & $-0.09$ $\pm$ $0.10$ & $0.04$ $\pm$ $0.04$ &$9.831 \pm 0.003$ &$7.17 \pm 0.02$ &$6.36 \pm 0.01$ &$148.20 \pm 0.03$ &$-1.695 \pm 0.002$ &$0.390 \pm 0.007$ &$ 0.38 \pm 0.02$  & 4.83 $\pm$ 0.03\\ 
Gl 687 & $3475$ $\pm$ $31$ & $4.71$ $\pm$ $0.05$ & $0.01$ $\pm$ $0.10$ & $0.06$ $\pm$ $0.04$ &$9.739 \pm 0.003$ &$7.05 \pm 0.02$ &$6.26 \pm 0.02$ &$219.79 \pm 0.02$ &$-1.649 \pm 0.002$ &$0.414 \pm 0.007$ &$ 0.39 \pm 0.02$  & 4.80 $\pm$ 0.03\\ 
Gl 725A & $3470$ $\pm$ $31$ & $4.77$ $\pm$ $0.06$ & $-0.26$ $\pm$ $0.10$ & $0.15$ $\pm$ $0.04$ &$10.120 \pm 0.003$ &$7.45 \pm 0.02$ &$6.70 \pm 0.02$ &$283.84 \pm 0.02$ &$-1.809 \pm 0.002$ &$0.345 \pm 0.006$ &$ 0.33 \pm 0.02$  & 4.88 $\pm$ 0.03\\ 
Gl 317 & $3421$ $\pm$ $31$ & $4.71$ $\pm$ $0.06$ & $0.23$ $\pm$ $0.10$ & $-0.04$ $\pm$ $0.04$ &$9.859 \pm 0.003$ &$7.03 \pm 0.03$ &$6.12 \pm 0.02$ &$65.88 \pm 0.04$ &$-1.657 \pm 0.004$ &$0.423 \pm 0.008$ &$ 0.42 \pm 0.02$  & 4.80 $\pm$ 0.03\\ 
Gl 251 & $3420$ $\pm$ $31$ & $4.71$ $\pm$ $0.06$ & $-0.01$ $\pm$ $0.10$ & $-0.01$ $\pm$ $0.04$ &$10.129 \pm 0.003$ &$7.37 \pm 0.02$ &$6.54 \pm 0.02$ &$179.06 \pm 0.03$ &$-1.786 \pm 0.003$ &$0.365 \pm 0.007$ &$ 0.35 \pm 0.02$  & 4.86 $\pm$ 0.03\\ 
GJ 4063 & $3419$ $\pm$ $31$ & $4.77$ $\pm$ $0.06$ & $0.42$ $\pm$ $0.10$ & $-0.07$ $\pm$ $0.04$ &$9.982 \pm 0.003$ &$7.00 \pm 0.02$ &--- $\pm$ --- &$91.80 \pm 0.02$ &$-1.662 \pm 0.004$ &$0.422 \pm 0.008$ &--- $\pm$ ---  &  --- $\pm$ ---\\ 
Gl 581 & $3406$ $\pm$ $31$ & $4.82$ $\pm$ $0.06$ & $-0.07$ $\pm$ $0.10$ & $0.01$ $\pm$ $0.04$ &$10.425 \pm 0.003$ &$7.71 \pm 0.03$ &$6.84 \pm 0.02$ &$158.72 \pm 0.03$ &$-1.917 \pm 0.002$ &$0.317 \pm 0.006$ &$ 0.31 \pm 0.02$  & 4.92 $\pm$ 0.03\\ 
Gl 725B & $3379$ $\pm$ $31$ & $4.82$ $\pm$ $0.06$ & $-0.28$ $\pm$ $0.10$ & $0.14$ $\pm$ $0.04$ &$10.790 \pm 0.003$ &$7.99 \pm 0.02$ &$7.27 \pm 0.02$ &$283.84 \pm 0.03$ &$-2.038 \pm 0.003$ &$0.280 \pm 0.005$ &$ 0.25 \pm 0.02$  & 4.95 $\pm$ 0.04\\ 
Gl 876 & $3366$ $\pm$ $31$ & $4.80$ $\pm$ $0.06$ & $0.15$ $\pm$ $0.10$ & $-0.04$ $\pm$ $0.04$ &$10.528 \pm 0.003$ &$7.59 \pm 0.02$ &$6.66 \pm 0.02$ &$214.04 \pm 0.04$ &$-1.892 \pm 0.004$ &$0.333 \pm 0.006$ &$ 0.33 \pm 0.02$  & 4.91 $\pm$ 0.03\\ 
PM J09553-2715 & $3366$ $\pm$ $31$ & $4.76$ $\pm$ $0.06$ & $-0.03$ $\pm$ $0.10$ & $-0.02$ $\pm$ $0.04$ &$10.629 \pm 0.003$ &$7.84 \pm 0.02$ &$6.96 \pm 0.02$ &$91.74 \pm 0.04$ &$-1.979 \pm 0.003$ &$0.302 \pm 0.006$ &$ 0.29 \pm 0.02$  & 4.94 $\pm$ 0.04\\ 
GJ 1012 & $3363$ $\pm$ $31$ & $4.66$ $\pm$ $0.06$ & $0.07$ $\pm$ $0.10$ & $0.01$ $\pm$ $0.04$ &$10.268 \pm 0.003$ &$7.40 \pm 0.02$ &$6.56 \pm 0.02$ &$74.71 \pm 0.04$ &$-1.811 \pm 0.003$ &$0.367 \pm 0.007$ &$ 0.35 \pm 0.02$  & 4.85 $\pm$ 0.03\\ 
GJ 4333 & $3362$ $\pm$ $31$ & $4.72$ $\pm$ $0.06$ & $0.25$ $\pm$ $0.10$ & $-0.02$ $\pm$ $0.04$ &$10.233 \pm 0.003$ &$7.27 \pm 0.03$ &$6.38 \pm 0.02$ &$94.37 \pm 0.03$ &$-1.767 \pm 0.006$ &$0.386 \pm 0.008$ &$ 0.37 \pm 0.02$  & 4.84 $\pm$ 0.03\\ 
Gl 445 & $3356$ $\pm$ $31$ & $4.85$ $\pm$ $0.06$ & $-0.24$ $\pm$ $0.10$ & $0.14$ $\pm$ $0.04$ &$10.949 \pm 0.003$ &$8.12 \pm 0.02$ &$7.35 \pm 0.03$ &$190.33 \pm 0.02$ &$-2.094 \pm 0.003$ &$0.266 \pm 0.005$ &$ 0.24 \pm 0.02$  & 4.97 $\pm$ 0.04\\ 
GJ 1148 & $3354$ $\pm$ $31$ & $4.70$ $\pm$ $0.06$ & $0.11$ $\pm$ $0.10$ & $0.01$ $\pm$ $0.04$ &$10.370 \pm 0.003$ &$7.40 \pm 0.02$ &$6.61 \pm 0.02$ &$90.69 \pm 0.03$ &$-1.820 \pm 0.004$ &$0.365 \pm 0.007$ &$ 0.34 \pm 0.02$  & 4.84 $\pm$ 0.03\\ 
PM J08402+3127 & $3347$ $\pm$ $31$ & $4.76$ $\pm$ $0.06$ & $-0.08$ $\pm$ $0.10$ & $0.01$ $\pm$ $0.04$ &$10.739 \pm 0.003$ &$7.87 \pm 0.02$ &$7.05 \pm 0.02$ &$89.07 \pm 0.03$ &$-1.998 \pm 0.003$ &$0.299 \pm 0.006$ &$ 0.28 \pm 0.02$  & 4.93 $\pm$ 0.04\\ 
GJ 3378 & $3326$ $\pm$ $31$ & $4.81$ $\pm$ $0.06$ & $-0.05$ $\pm$ $0.10$ & $-0.01$ $\pm$ $0.04$ &$10.975 \pm 0.003$ &$8.02 \pm 0.02$ &$7.20 \pm 0.02$ &$129.30 \pm 0.03$ &$-2.068 \pm 0.004$ &$0.279 \pm 0.005$ &$ 0.26 \pm 0.02$  & 4.96 $\pm$ 0.04\\ 
GJ 1105 & $3324$ $\pm$ $31$ & $4.63$ $\pm$ $0.07$ & $-0.04$ $\pm$ $0.10$ & $-0.05$ $\pm$ $0.04$ &$10.931 \pm 0.003$ &$8.00 \pm 0.02$ &$7.14 \pm 0.03$ &$112.99 \pm 0.03$ &$-2.056 \pm 0.004$ &$0.283 \pm 0.005$ &$ 0.27 \pm 0.02$  & 4.96 $\pm$ 0.04\\ 
Gl 699 & $3311$ $\pm$ $31$ & $5.11$ $\pm$ $0.06$ & $-0.37$ $\pm$ $0.10$ & $0.16$ $\pm$ $0.04$ &$11.884 \pm 0.003$ &$8.93 \pm 0.02$ &$8.21 \pm 0.02$ &$546.98 \pm 0.04$ &$-2.432 \pm 0.004$ &$0.185 \pm 0.004$ &$ 0.16 \pm 0.02$  & 5.11 $\pm$ 0.06\\ 
GJ 169.1A & $3307$ $\pm$ $31$ & $4.71$ $\pm$ $0.06$ & $0.13$ $\pm$ $0.10$ & $-0.07$ $\pm$ $0.04$ &$10.994 \pm 0.003$ &$7.91 \pm 0.02$ &$7.01 \pm 0.02$ &$181.24 \pm 0.05$ &$-2.037 \pm 0.005$ &$0.292 \pm 0.006$ &$ 0.28 \pm 0.02$  & 4.96 $\pm$ 0.04\\ 
PM J21463+3813 & $3305$ $\pm$ $33$ & $5.06$ $\pm$ $0.08$ & $-0.38$ $\pm$ $0.10$ & $0.25$ $\pm$ $0.04$ &$11.591 \pm 0.003$ &$8.71 \pm 0.02$ &$7.96 \pm 0.02$ &$141.89 \pm 0.02$ &$-2.335 \pm 0.003$ &$0.208 \pm 0.004$ &$ 0.18 \pm 0.02$  & 5.06 $\pm$ 0.06\\ 
Gl 15B & $3272$ $\pm$ $31$ & $4.89$ $\pm$ $0.06$ & $-0.42$ $\pm$ $0.10$ & $0.04$ $\pm$ $0.04$ &$11.928 \pm 0.003$ &$9.03 \pm 0.02$ &$8.19 \pm 0.02$ &$280.69 \pm 0.03$ &$-2.465 \pm 0.004$ &$0.182 \pm 0.004$ &$ 0.16 \pm 0.02$  & 5.13 $\pm$ 0.06\\ 
GJ 1289 & $3238$ $\pm$ $32$ & $5.00$ $\pm$ $0.07$ & $0.05$ $\pm$ $0.10$ & $-0.00$ $\pm$ $0.04$ &$11.556 \pm 0.003$ &$8.50 \pm 0.03$ &$7.61 \pm 0.02$ &$119.58 \pm 0.06$ &$-2.269 \pm 0.007$ &$0.233 \pm 0.005$ &$ 0.21 \pm 0.02$  & 5.03 $\pm$ 0.05\\ 
Gl 447 & $3198$ $\pm$ $31$ & $4.82$ $\pm$ $0.06$ & $-0.13$ $\pm$ $0.10$ & $-0.01$ $\pm$ $0.04$ &$11.960 \pm 0.003$ &$8.86 \pm 0.02$ &$8.01 \pm 0.02$ &$296.31 \pm 0.03$ &$-2.419 \pm 0.006$ &$0.201 \pm 0.004$ &$ 0.18 \pm 0.02$  & 5.08 $\pm$ 0.06\\ 
GJ 1151 & $3178$ $\pm$ $31$ & $4.71$ $\pm$ $0.06$ & $-0.16$ $\pm$ $0.10$ & $-0.03$ $\pm$ $0.04$ &$12.158 \pm 0.003$ &$8.96 \pm 0.03$ &$8.11 \pm 0.02$ &$124.34 \pm 0.05$ &$-2.468 \pm 0.009$ &$0.193 \pm 0.004$ &$ 0.17 \pm 0.02$  & 5.10 $\pm$ 0.06\\ 
GJ 1103 & $3170$ $\pm$ $31$ & $4.67$ $\pm$ $0.06$ & $-0.03$ $\pm$ $0.10$ & $-0.00$ $\pm$ $0.04$ &$11.818 \pm 0.003$ &$8.66 \pm 0.02$ &$7.83 \pm 0.02$ &$107.85 \pm 0.04$ &$-2.344 \pm 0.007$ &$0.224 \pm 0.005$ &$ 0.19 \pm 0.02$  & 5.03 $\pm$ 0.05\\ 
Gl 905 & $3069$ $\pm$ $31$ & $4.78$ $\pm$ $0.08$ & $0.05$ $\pm$ $0.11$ & $-0.06$ $\pm$ $0.04$ &$12.881 \pm 0.003$ &$9.39 \pm 0.03$ &$8.43 \pm 0.02$ &$316.48 \pm 0.04$ &$-2.664 \pm 0.012$ &$0.165 \pm 0.004$ &$ 0.15 \pm 0.02$  & 5.17 $\pm$ 0.07\\ 
GJ 1002 & $2980$ $\pm$ $33$ & $4.70$ $\pm$ $0.08$ & $-0.33$ $\pm$ $0.11$ & $-0.00$ $\pm$ $0.04$ &$13.347 \pm 0.003$ &$9.90 \pm 0.02$ &$9.01 \pm 0.02$ &$206.35 \pm 0.05$ &$-2.864 \pm 0.009$ &$0.139 \pm 0.003$ &$ 0.12 \pm 0.02$  & 5.22 $\pm$ 0.09\\ 
GJ 1286 & $2961$ $\pm$ $33$ & $4.55$ $\pm$ $0.12$ & $-0.23$ $\pm$ $0.10$ & $-0.04$ $\pm$ $0.04$ &$13.344 \pm 0.003$ &$9.87 \pm 0.02$ &$8.90 \pm 0.02$ &$139.34 \pm 0.04$ &$-2.855 \pm 0.010$ &$0.142 \pm 0.004$ &$ 0.12 \pm 0.02$  & 5.21 $\pm$ 0.08\\ 
\hline 
	\end{tabular}}
\end{table*}

\section{Discussion and conclusions}
\label{sec:conclusions}

In this work, we improved and extended a method designed to retrieve the  atmospheric parameters of M dwarfs from high-resolution  spectroscopic observations using state-of-the-art synthetic spectra computed with Turbospectrum from \texttt{MARCS} model atmospheres. Our analysis consists in  comparing these models to high-SNR template spectra built from tens to hundreds of observations  collected with SPIRou. We extend the work  initiated in C22 and applied our new tool to our SLS sample of \nbMdwarfs{} M dwarfs.

Recent publications~\citep{marfil_2021, rajpurohit_2017} included empirical $\afe$--$\mh$ relations in their analysis, or relied on models that did so, in order to constrain $\teff$ or $\mh$. 
 In this work, the fitting procedure, initially developed to constrain $\teff$, $\logg$ and $\mh$ was   extended to also include a fit of $\afe$, motivated by the large impact this parameter has on the derivation of the other stellar parameters. We retrieve $\afe$ values that are consistent with empirical trends observed when studying giants~\citep{adibekyan_2013}.
%
We  find that the coolest low-metallicity stars in our sample  are the most sensitive to $\afe$. This is likely due  to the presence of strong O-bearing molecular bands (e.g. CO) in the NIR spectra at low $\teff$, strongly impacted by variations in the abundances of alpha elements, in particular oxygen.

In  the present paper, we revised the line list used in C22, and updated the continuum adjustment procedure to improve the fit quality. This updated list contains 17 atomic lines, 9 OH lines and about 40 molecular lines found in the CO band redward of 2293~nm, which represents a very small subset of the lines that are included in the models and  those  present in the observed template spectra  (which in most cases do not match well). 
 Previous studies have attempted to refine the parameters of some atomic lines for their analysis~\citep{petit_2021}. Here, we tried to improve the fits of synthetic spectra to SPIRou templates by adjusting the values of Van Der Waals  broadening parameters and oscillator strengths for  a few of the selected lines. We assumed the parameters published by M15 for 3 calibration stars (Gl~699, Gl~15A and Gl~411) to perform this step. These corrections, and in particular  those applied to the Van Der Waals parameter of Ti lines, helped to bring our $\logg$ estimates closer to those of M15 for some targets. One should note that these corrections may not be the sole result of uncertainties in the line parameters, and may  also reflect inaccuracies of the atmospheric models.  


With the implemented improvements and updated line list, we recover parameters in  good agreement with M15 for \nbMdwarfsCommon{} stars included in both studies. We retrieve $\teff$ with a typical dispersion of about 45~K, lower that the uncertainties reported by M15, although larger than our estimated error bars of about 30~K.  This difference is also the result of a trend observed in the retrieved $\teff$ values, as we tend to derive larger $\teff$ for cool stars than M15. The dispersion about this trend is of the order of 25~K, of the order of our empirical error bars. We also obtain $\mh$ values with a dispersion of 0.06~dex, consistent with our error bars estimated to about 0.1~dex. 
Finally, $\logg$ is in better agreement with M15 compared to the values reported in C22, although we tend to recover smaller estimates than M15 for the coolest stars in our sample.

 For our \nbMdwarfs{} targets, we extracted Gaia G, J and K band magnitudes from SIMBAD, along with parallaxes, when available. We computed the radii for our sample from $\teff$, absolute J band magnitude ($M_{\rm J}$) and bolometric corrections~\citep{cifuentes_2020}. 
 Interferometric data published by~\citet{boyajian_2012} for 9 of these stars reported angular diameters  that are consistent with our retrieved radii,  with a relative dispersion of about 5~\%. Additionally, we derive the masses of the stars in our sample from $\rm M_K$--mass relations~\citep{mann_2019}.  Our derived masses and radii tend to be in good agreement with mass-radius relationships predicted by evolutionary models. We note a slight tendency to estimate larger radii that those predicted by the DSEP models and those of~\citet{baraffe_2015}. This tendency was reported in the literature~\citep{feiden_2013, jackson_2018} and different hypothesises were proposed, attributing the phenomenon to metallicity, modelling assumptions or radius inflation induced by the presence of magnetic fields.
    From our masses and radii estimates, we compute new $\logg$ values, and compare them to those derived from the fitting procedure. We find significant discrepancies between the two sets of $\logg$ values, {\p especially at the lowest temperatures}. This difference suggests that we tend to underestimate $\logg$ for the coolest stars in our sample with our fitting procedure. Fixing $\logg$ to higher values for the coolest stars in our sample results in an increase in $\teff$ of 20-50~K, an increase in $\mh$ of up to 0.2~dex, and slight increases in $\afe$ by less than 0.04~dex.
    This may reflect MARCS models being less accurate at temperatures close to 3000~K, i.e., close to the lower limit of our model temperature grid.
 

 We also retrieved $\afe$ values for the \nbMdwarfs{} stars in our sample,  but lack references for most of these targets.  Given that $\teff$, $\logg$ and $\mh$ are very sensitive to small variations in $\afe$,  the latter should be carefully considered when fitting models to spectra of M dwarfs. To assess the quality of the constraint on this parameter, we place our stars in a $\afe$--$\mh$ plane, and find that the recovered $\afe$ are in good agreement with values expected from empirical relations.
  We find that a few stars, in particular Gl~699, Gl~445, PM~J21463+3813 and Gl~411, have relatively large retrieved $\afe$ values and are likely to belong to the thick galactic disc, while most of our stars are likely to belong to thin disc, with lower $\afe$ values. These results are somewhat consistent with the computed velocities, larger than 100~$\kms$ for these 4 stars. 
  Although Gl~317 and PM~J09553-2715 also feature high velocities, their supersolar metallicities make it difficult to reliably conclude about the disc population these stars belong to.
  	Gl~412A also has a velocity above 100~$\kms$, but we derive an $\afe$ value smaller than that expected for the thick disc. These results are compatible with previous classification of these stars~\citep{cortes_contreras_2016, schoefer_2019}, in which PM~J09553-2715, PM~J21463+3813, Gl~699, Gl~445, and Gl~411 were identified as belonging to the thick disc, and Gl~412A labelled as within the transition between thin and thick discs. Most other stars studied by~\citet{cortes_contreras_2016} and included in our work were classified as belonging to the thin or young disc, with a few exceptions such as Gl~880, Gl~905 and GJ~1151, placed either in the thick of transition between thick and thin discs.
  One should note that the boundary between thin and thick  disc from $\afe$ remains fuzzy even for giants making it tricky to clearly split the stars of our sample into two distinct populations.

 In subsequent works, we will perform a similar analysis with other models, such as PHOENIX, which will require to compute new grids of synthetic spectra  for different $\afe$ values, and with up-to-date line lists.  As our models evolve, we will revise the modifications performed on the line lists and identify additional stellar features to use for our purposes. This  will allow us to further investigate the differences between models, and to identify the modelling assumptions that are best suited to the computation of synthetic spectra of M dwarfs and cool stars.
Additionally, we will try to perform the same kind of analysis on more active targets that were excluded from our sample, and on the pre-main sequence stars also observed with SPIRou in the framework of the SLS. The spectra of such stars may be impacted by activity, with effects from the chromosphere~\citep{hintz_2019} or Zeemann broadening~\citep{deen_2013} and radius inflation due to stronger magnetic fields~\citep{feiden_2013}. This may require the addition of extra steps to the modelling process. Spots  are indeed likely to be present at the surface of active targets, which may require implementing a two-temperature model to reproduce their spectra~\citep{santiago_2017}.

\section*{Acknowledgements}

We acknowledge funding from the European Research Council under the H2020 \& innovation program (grant \#740651 NewWorlds).

This work is based on observations obtained at the Canada-France-Hawaii Telescope (CFHT) which is operated by the National Research Council (NRC) of Canada, the Institut National des Sciences de l’Univers of the Centre National de la Recherche Scientifique (CNRS) of France, and the University of Hawaii. The observations at the CFHT were performed with care and respect from the summit of Maunakea which is a significant cultural and historic site.

This research made use of the SIMBAD database~\citep{wenger_2000}, operated at CDS, Strasbourg, France.

TM acknowledges financial support from the Spanish Ministry of Science and Innovation (MICINN) through the Spanish State Research Agency, under the Severo Ochoa Program 2020-2023 (CEX2019-000920-S) as well as support from the ACIISI, Consejería de Economía, Conocimiento y Empleo del Gobiernode Canarias and the European Regional Development Fund (ERDF) under grant with reference  PROID2021010128.
777

{\p We acknowledge funding from the French National Research Agency (ANR) under contract number ANR18CE310019 (SPlaSH). XD and AC acknowledge support in the framework of the Investissements d'Avenir program (ANR-15-IDEX-02), through the funding of the “Origin of Life" project of the Grenoble-Alpes University.}
\section*{Data Availability}

 The data used in this work was acquired in the context of the SLS, and will be publicly available at the Canadian Astronomy Data Center one year after completion of the program.



\bibliographystyle{mnras}
\bibliography{article02} 




\appendix

\begin{figure}
	\includegraphics[width=\linewidth]{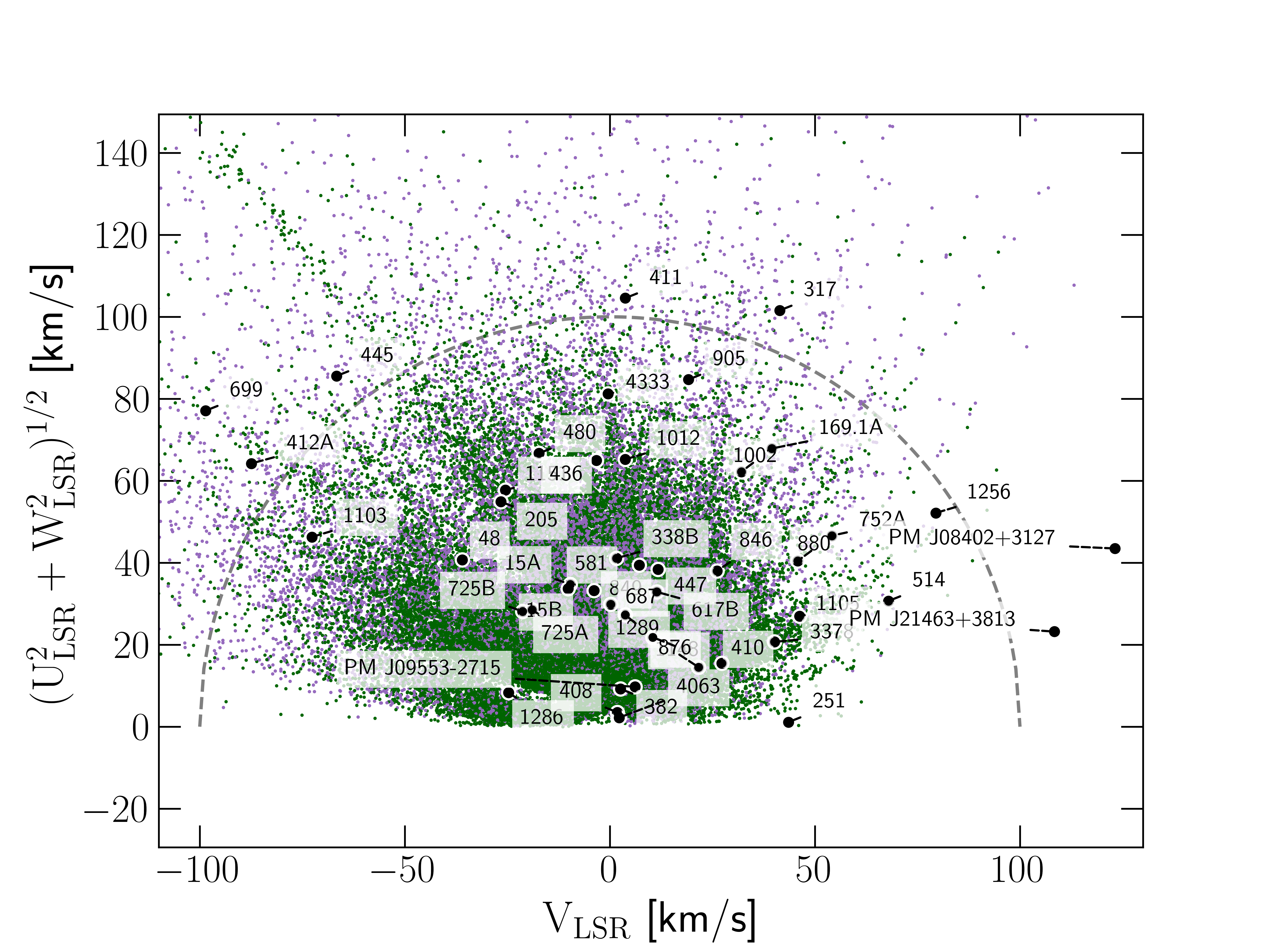}
	\caption{{\paul  Same as Fig.~\ref{fig:toomre_diagram} with labels identifying the stars.}}
	\label{fig:toomre_diagram_labels}
\end{figure}
\begin{figure}
	\includegraphics[width=.99\linewidth]{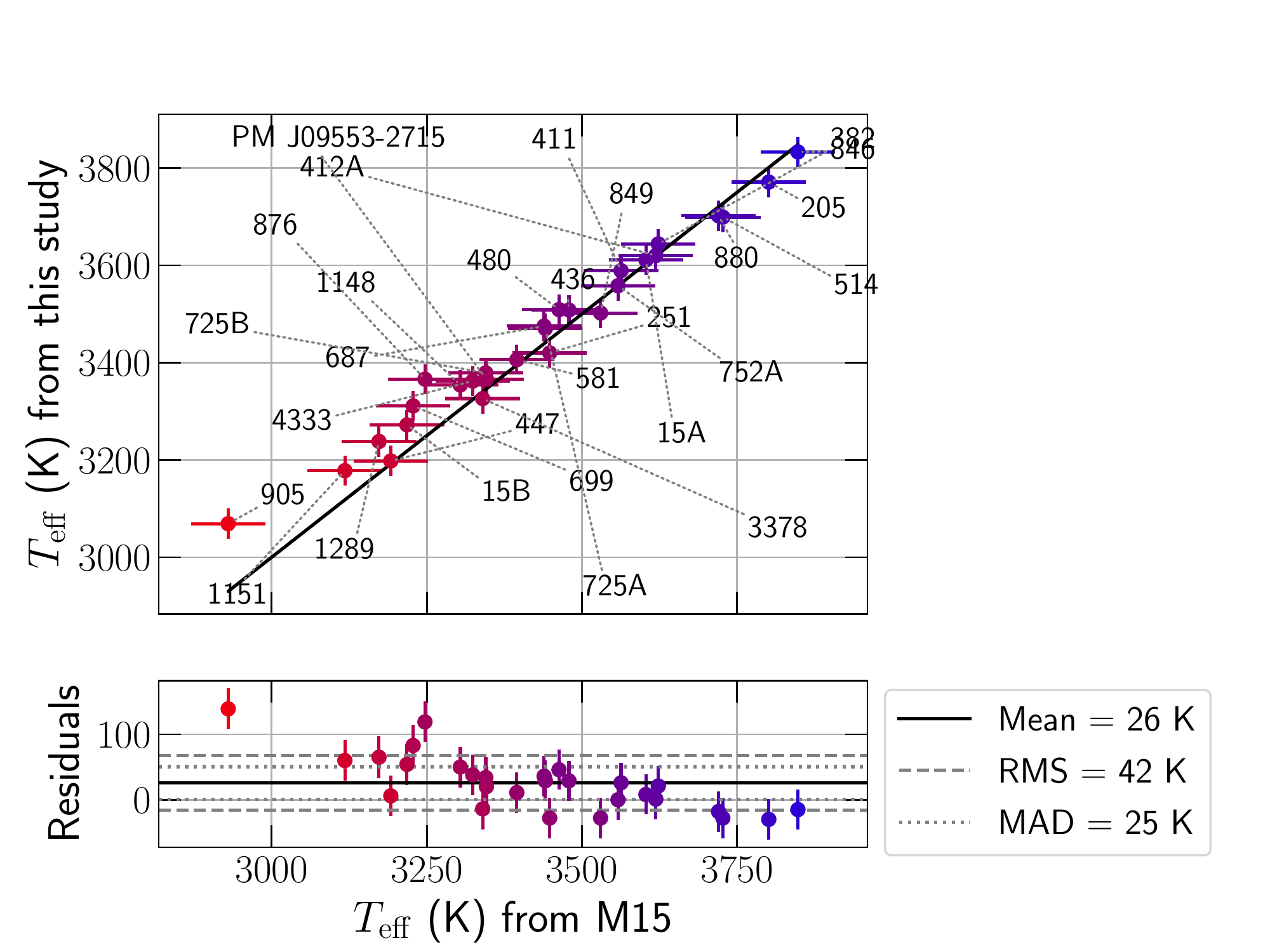}
	\includegraphics[width=.99\linewidth]{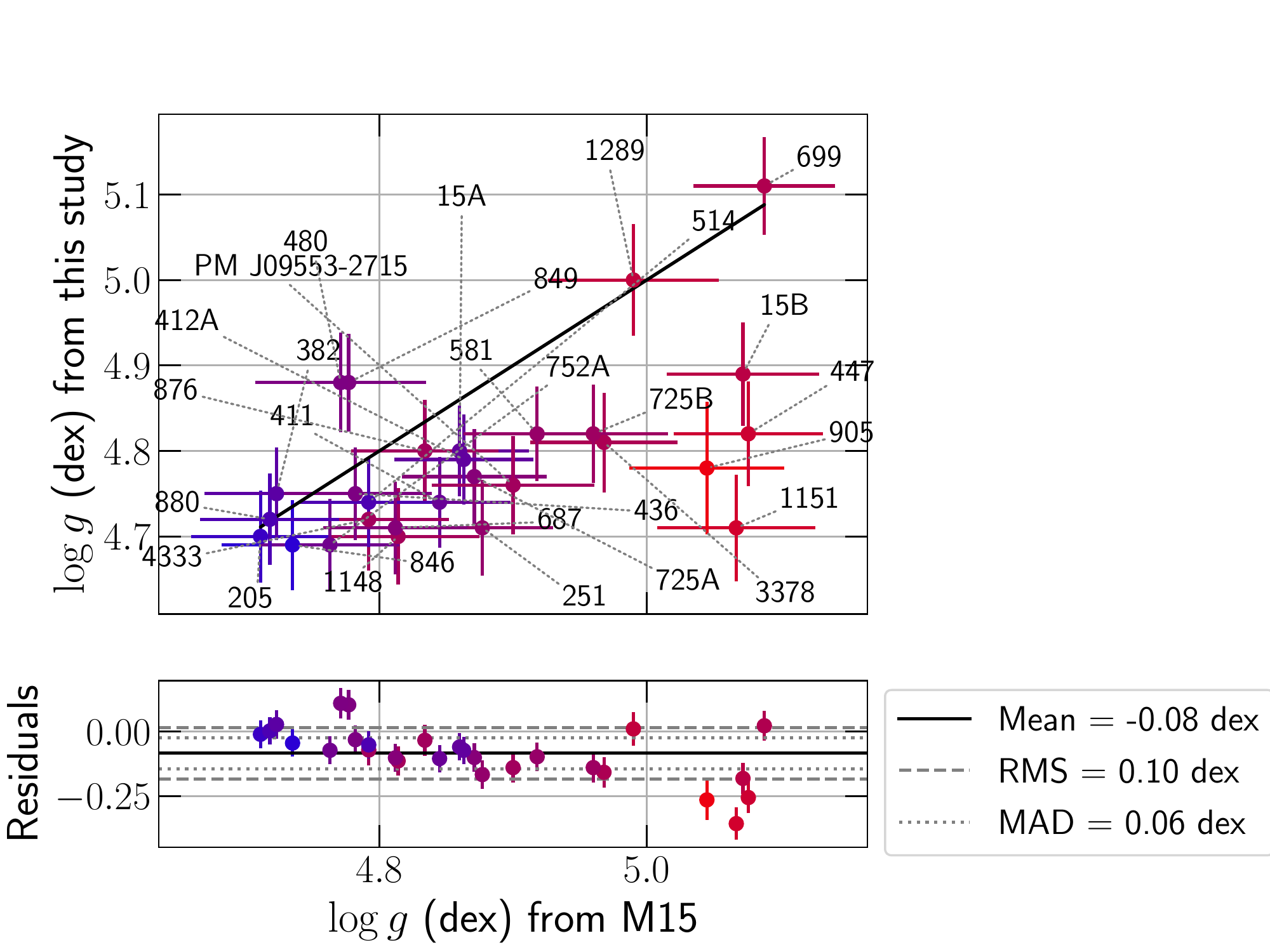}
\includegraphics[width=.99\linewidth]{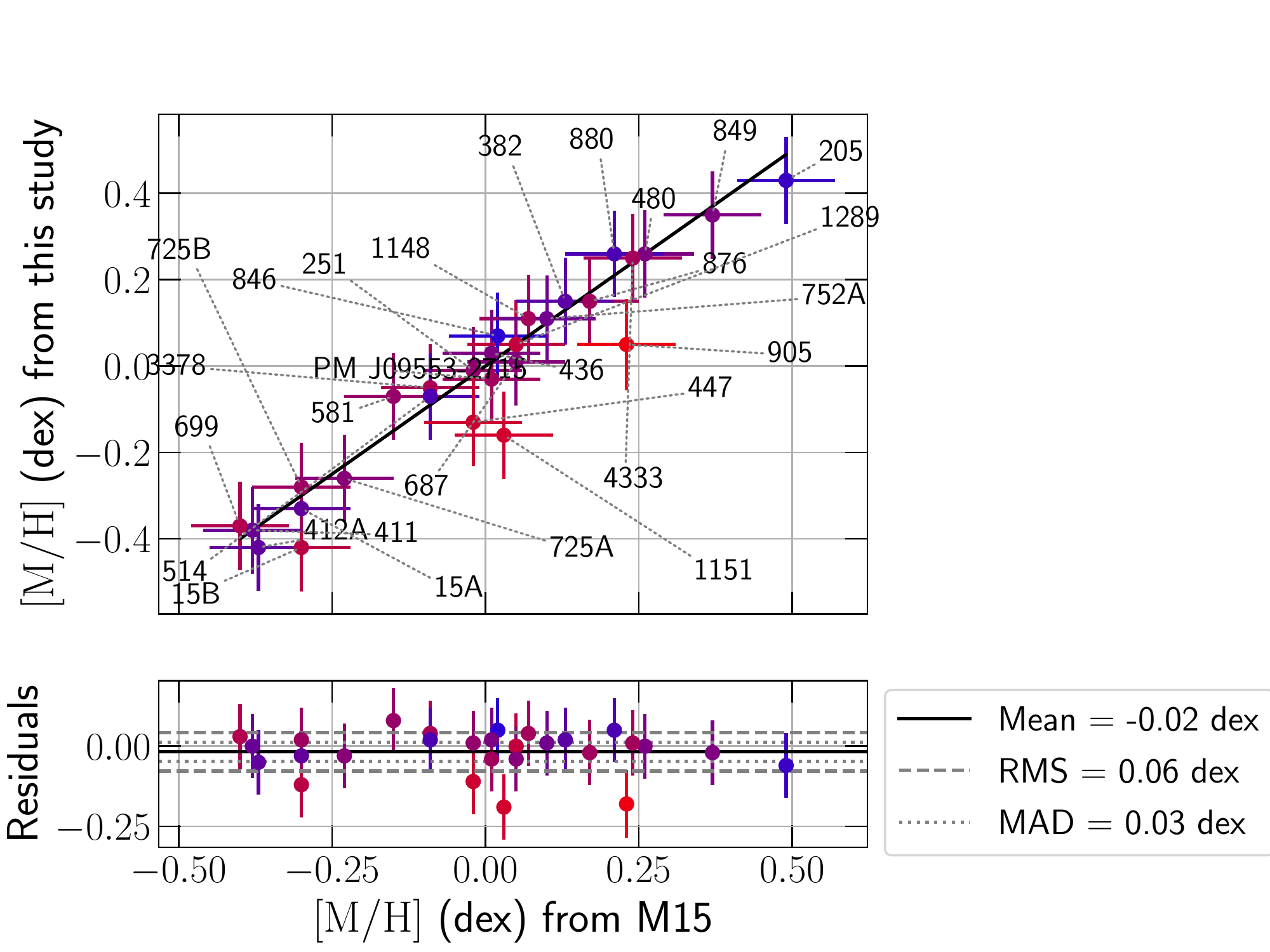}
	\caption{{\paul Same as Fig.~\ref{fig:results_all_4d} with labels identifying the stars.}}
	\label{fig:results_all_4d_labels}
\end{figure}
\begin{figure}
	\includegraphics[width=0.99\linewidth]{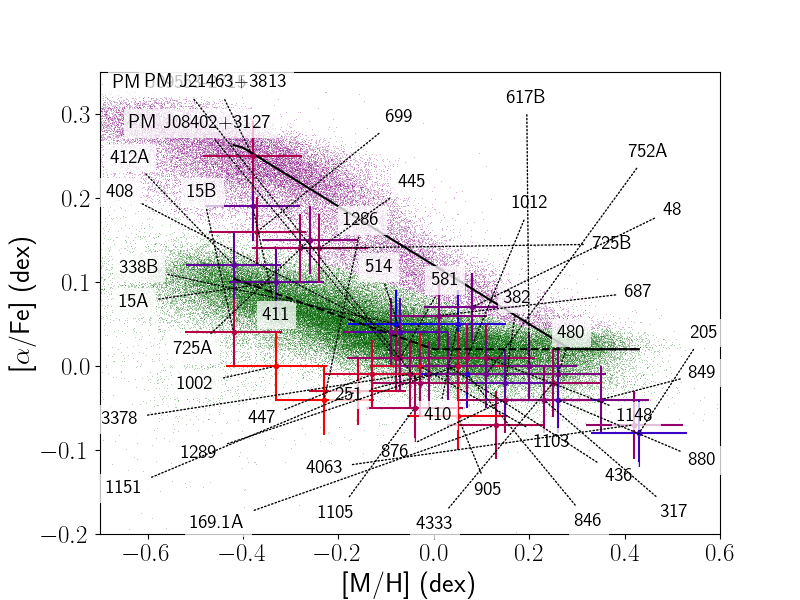}
	\caption{{\paul Same as Fig.~\ref{fig:alpha_mh_all_4d} with labels identifying the stars.}}
	\label{fig:alpha_mh_all_4d_labels}
\end{figure}
\begin{figure*}
	\includegraphics[width=.5\linewidth]{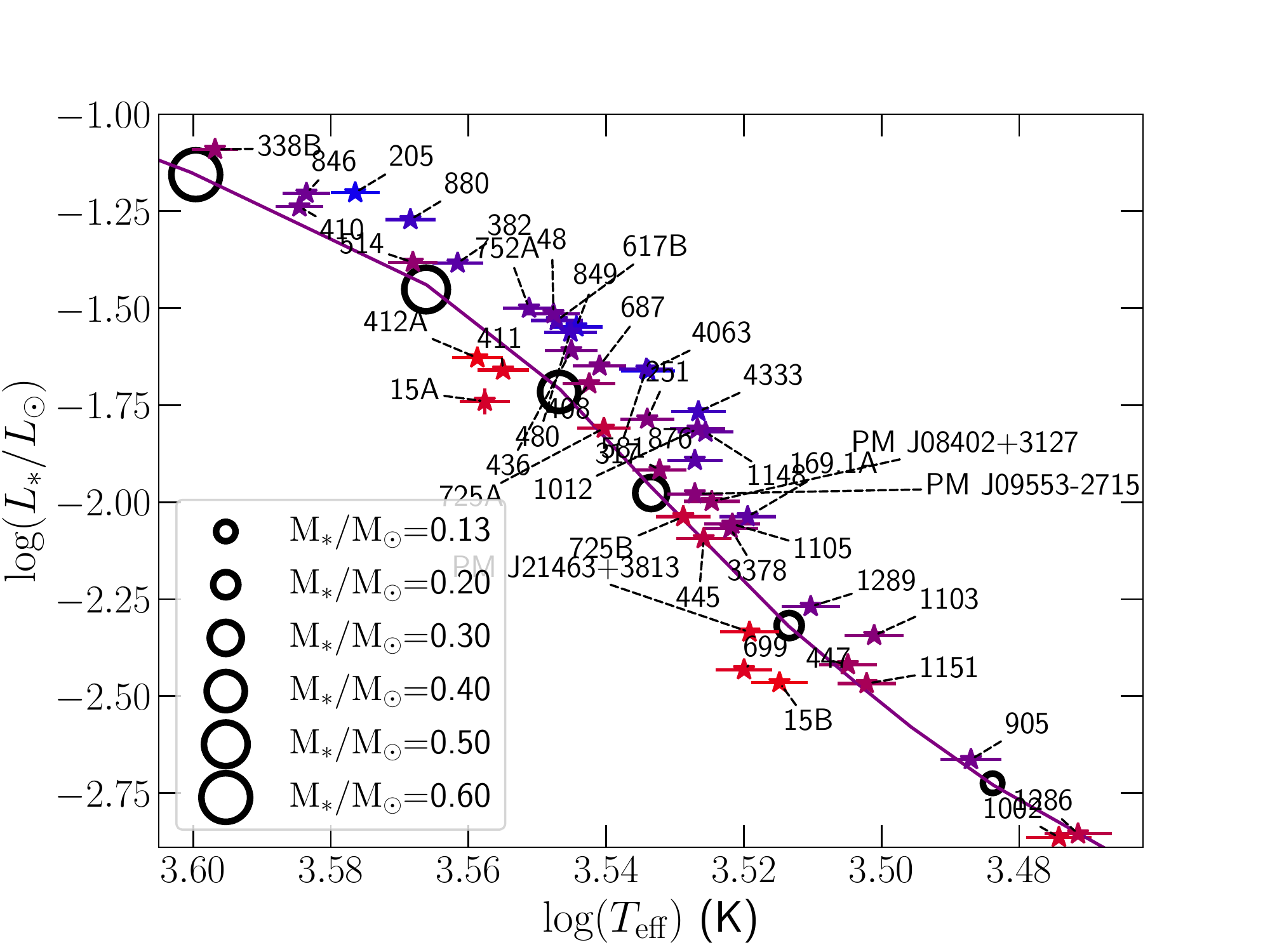}\includegraphics[width=.5\linewidth]{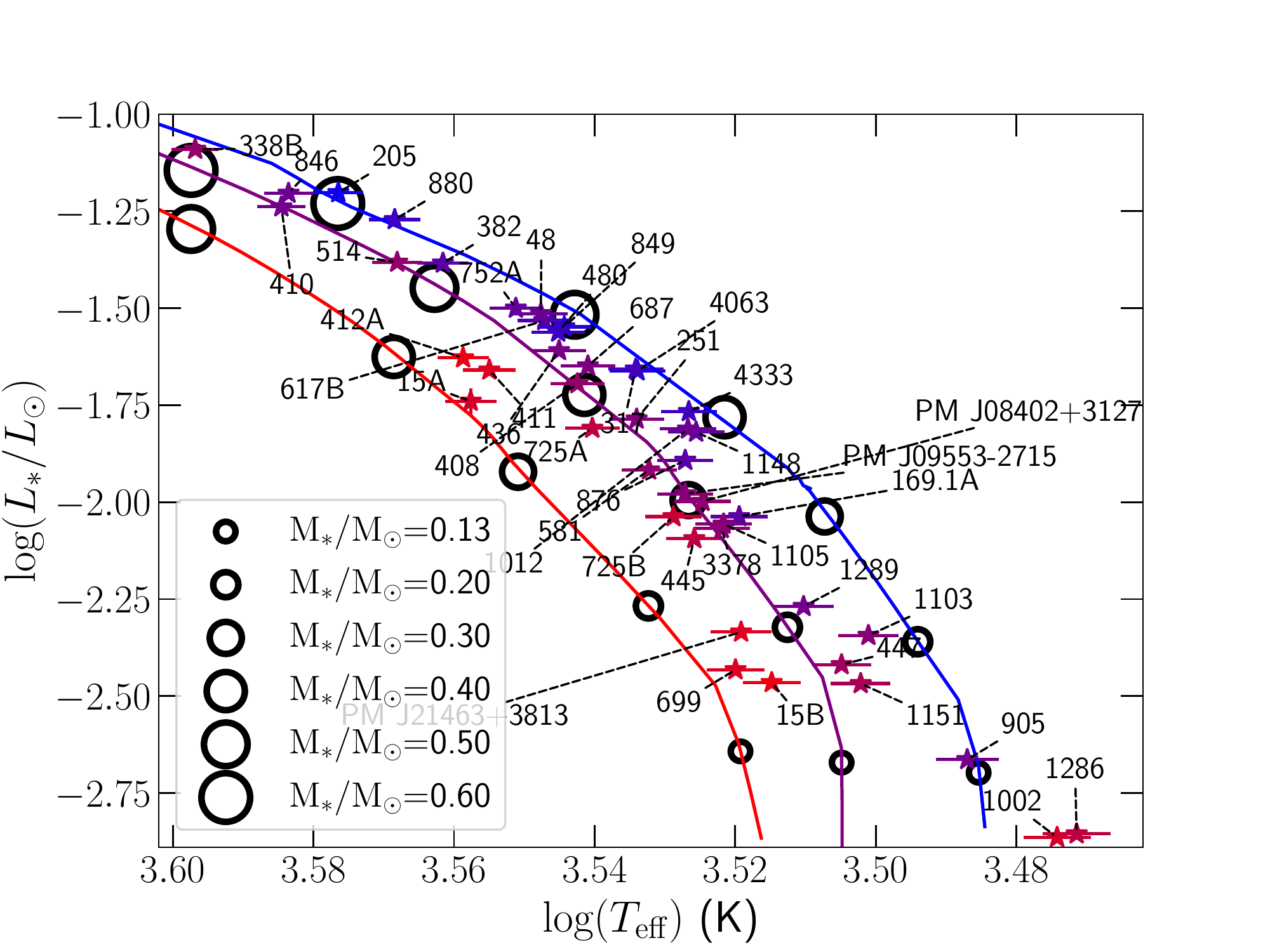}
	\caption{{\paul Same as Fig.~\ref{fig:hr_diagram} with labels identifying the stars.} }
	\label{fig:hr_diagram_labels}
\end{figure*}
\begin{figure}
	\includegraphics[width=\columnwidth]{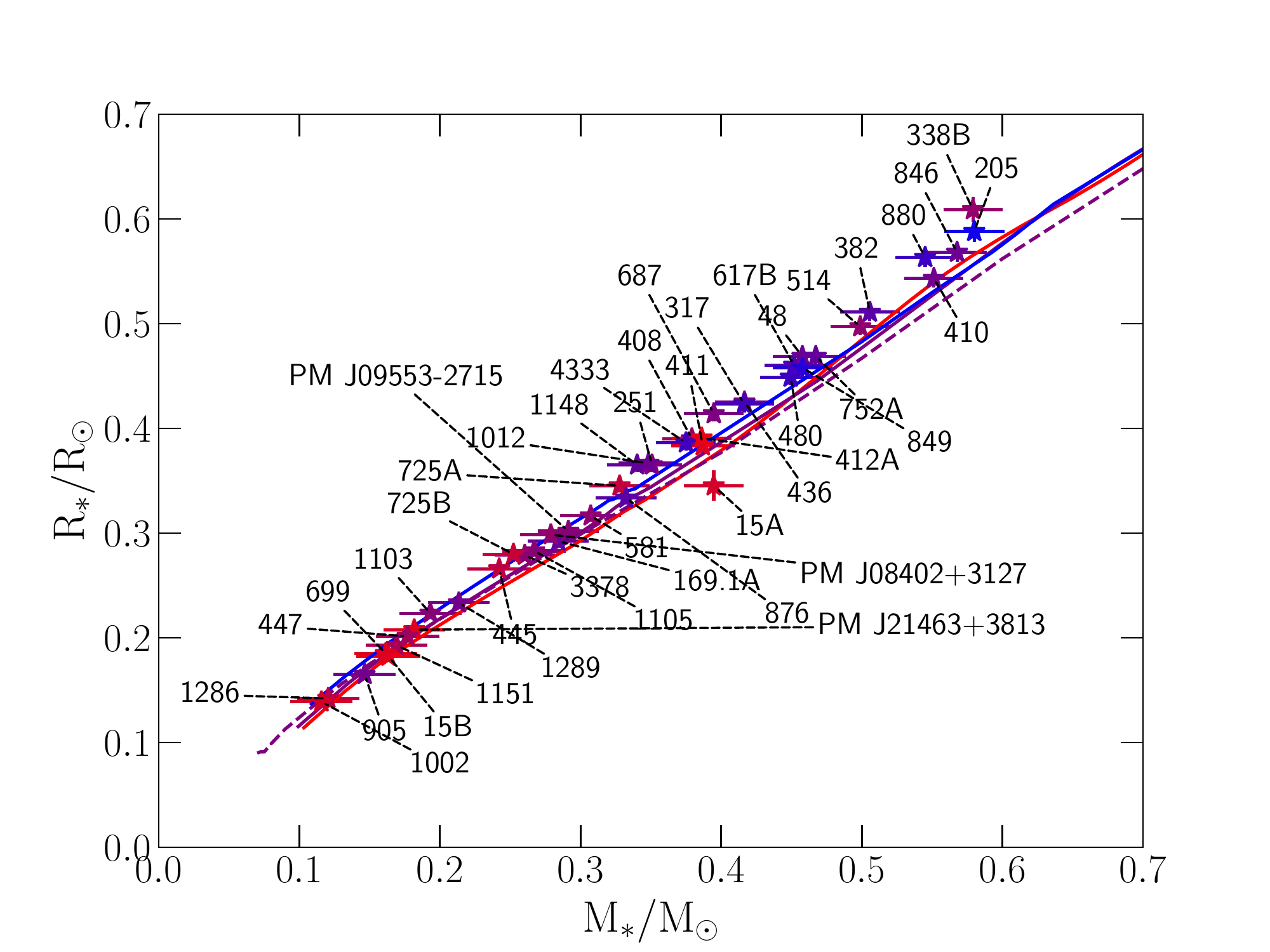}
	\caption{{\paul Same as Fig.~\ref{fig:m_r_diagram} with labels identifying the stars.} }
	\label{fig:m_r_diagram_labels}
\end{figure}
\begin{figure}
	\includegraphics[width=\linewidth]{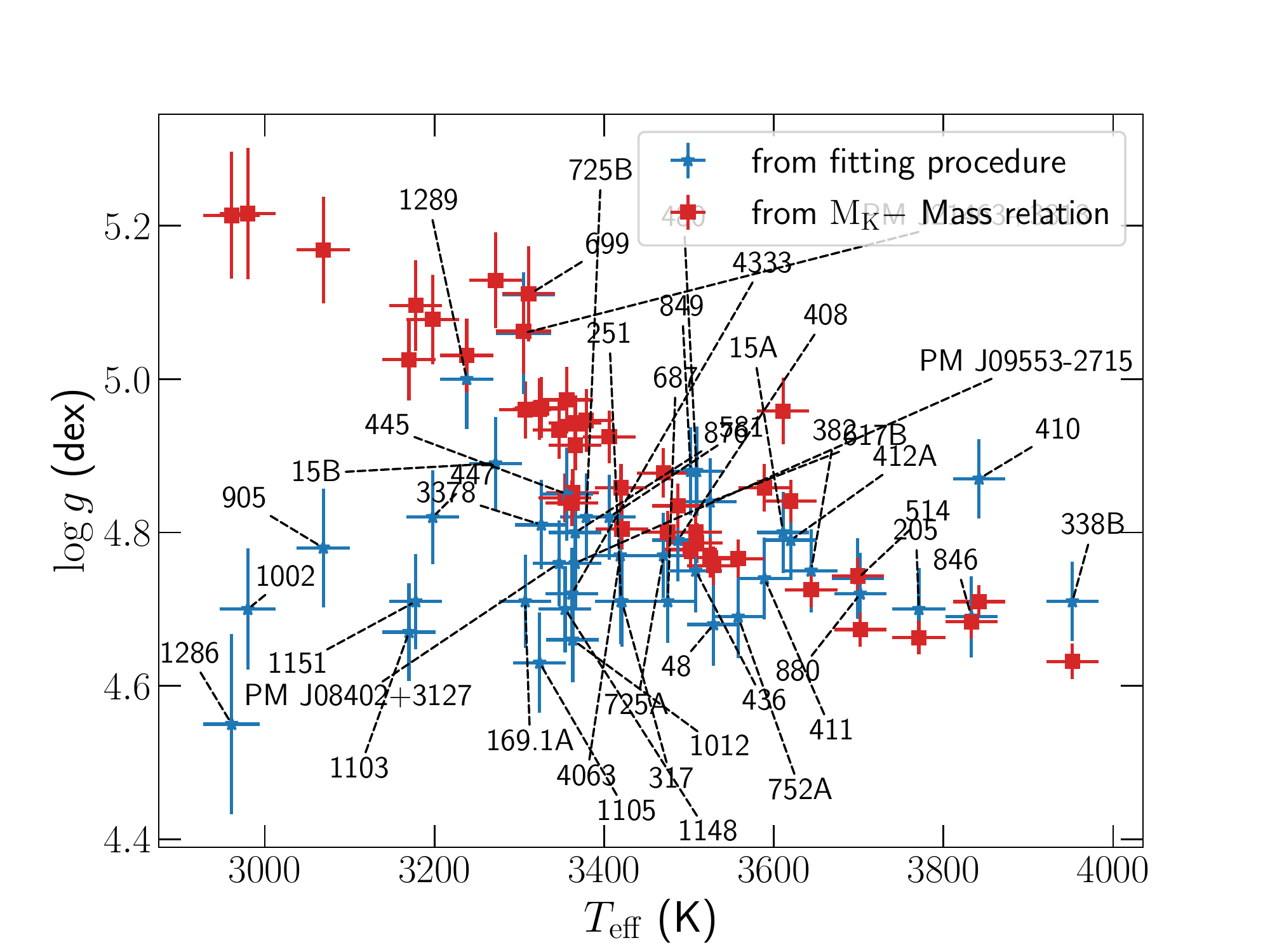}
	\caption{{\paul Same as Fig.~\ref{fig:logg_logg} with labels identifying the stars.} }
	\label{fig:logg_logg_labels}
\end{figure}

{\paul
\section{Figures with labels}

Figures~\ref{fig:toomre_diagram_labels} to ~\ref{fig:logg_logg_labels} present alternative plots to Fig.~\ref{fig:toomre_diagram}, \ref{fig:results_all_4d}, \ref{fig:alpha_mh_all_4d}, \ref{fig:hr_diagram}, \ref{fig:m_r_diagram} and  \ref{fig:logg_logg} with labels identifying the stars.
}

\section{Results on calibration stars}

\begin{figure*}
	\includegraphics[width=.5\linewidth]{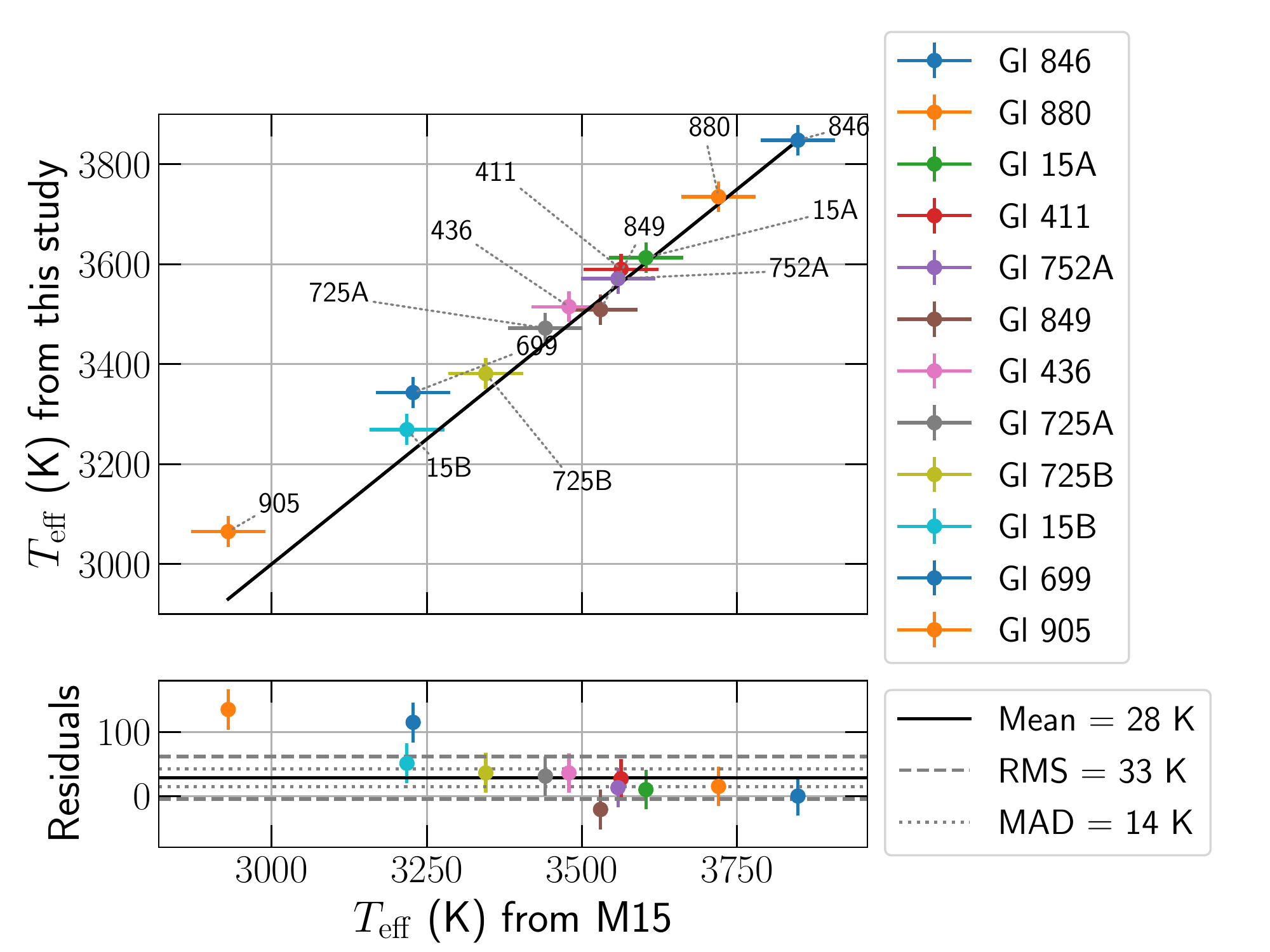}\includegraphics[width=.5\linewidth]{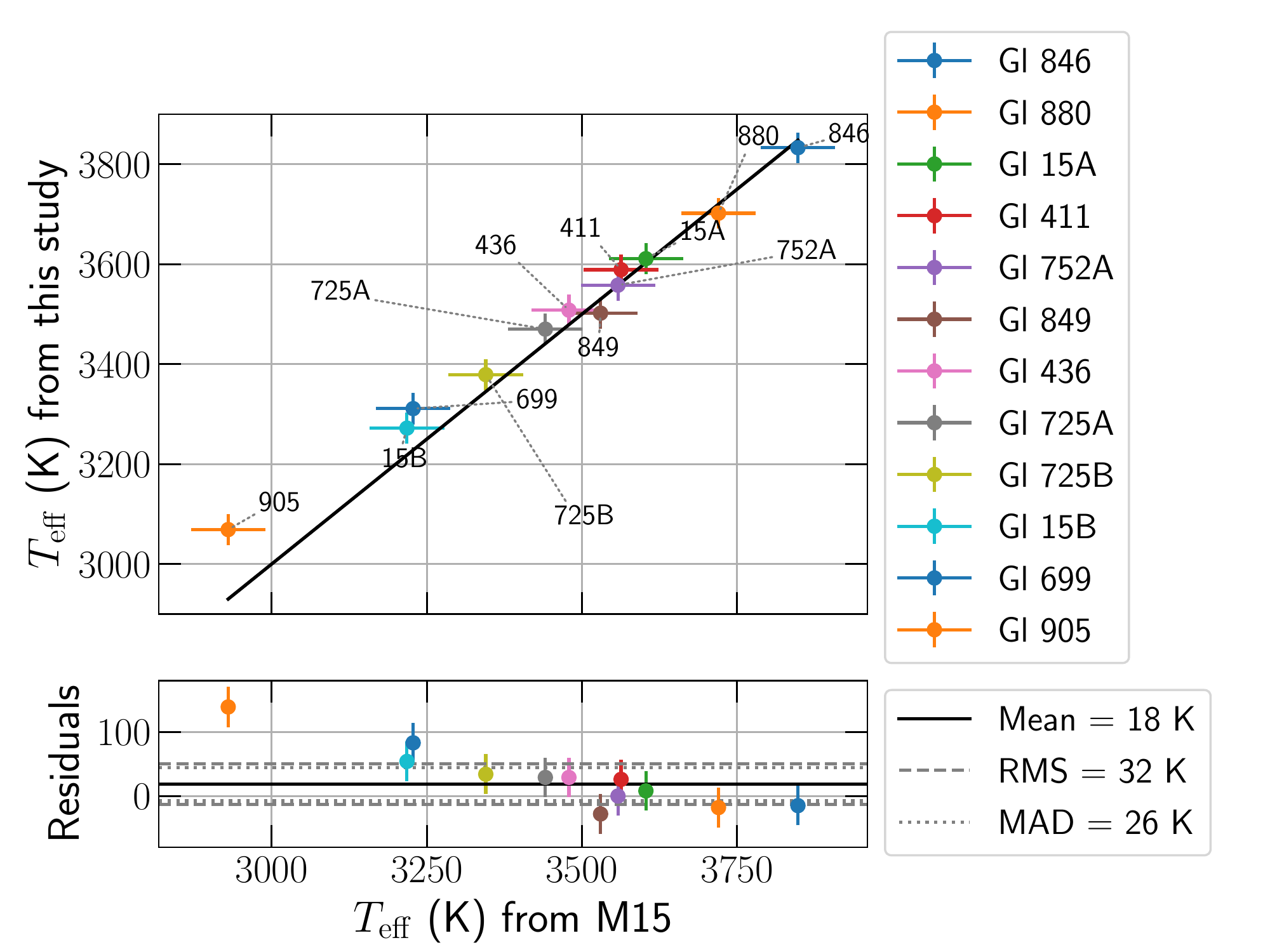}	\includegraphics[width=.5\linewidth]{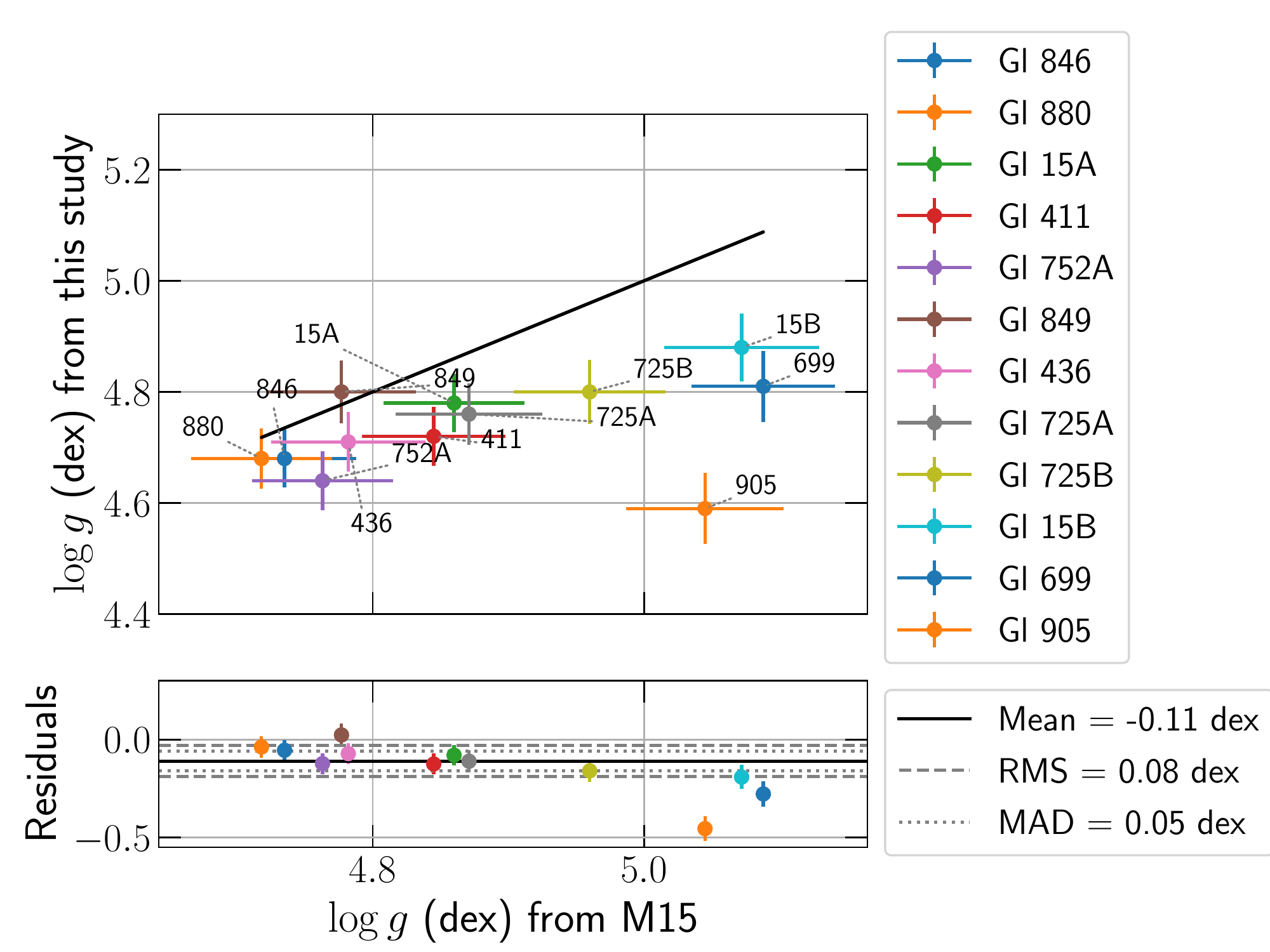}\includegraphics[width=.5\linewidth]{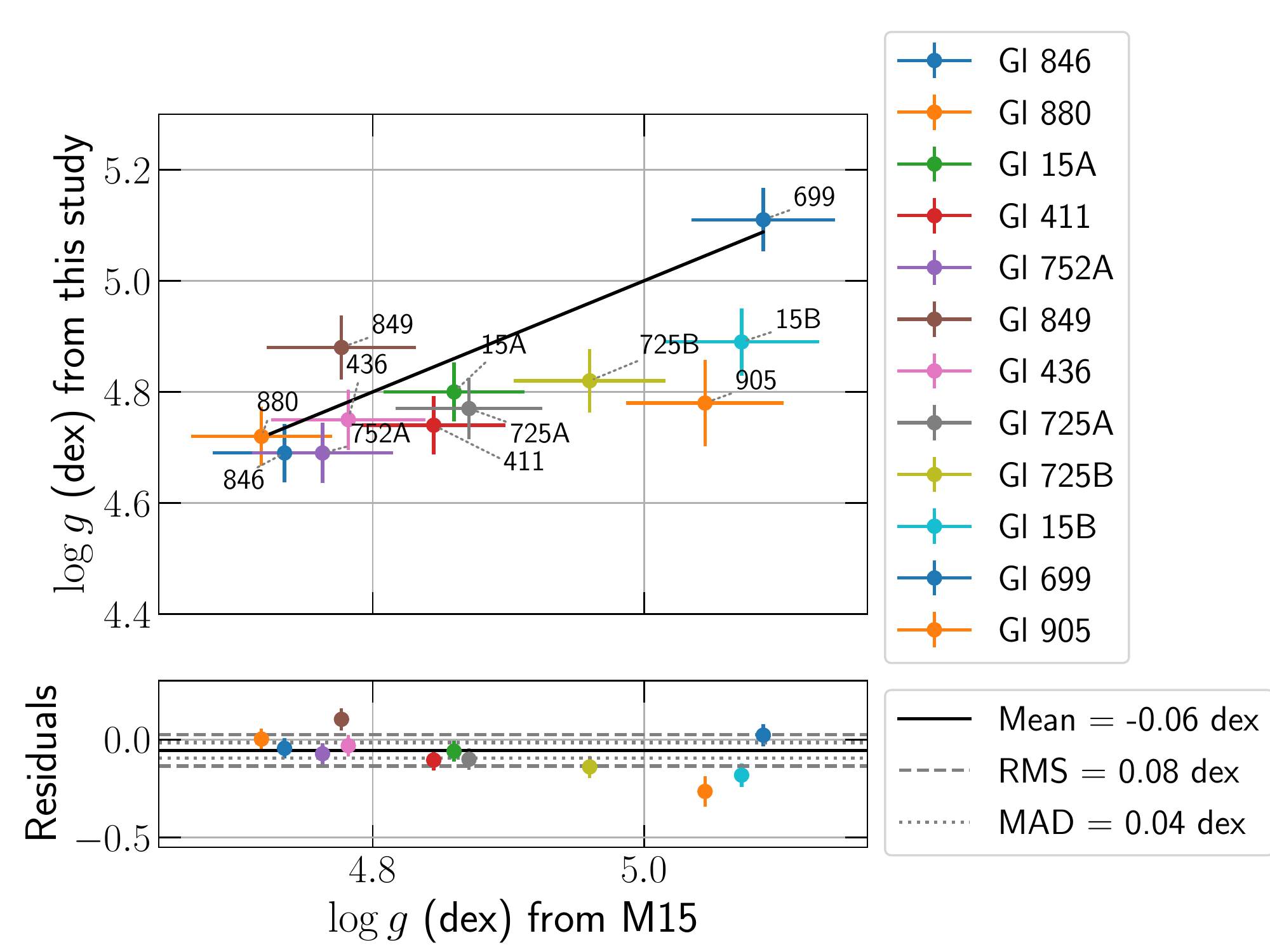}
	\includegraphics[width=.5\linewidth]{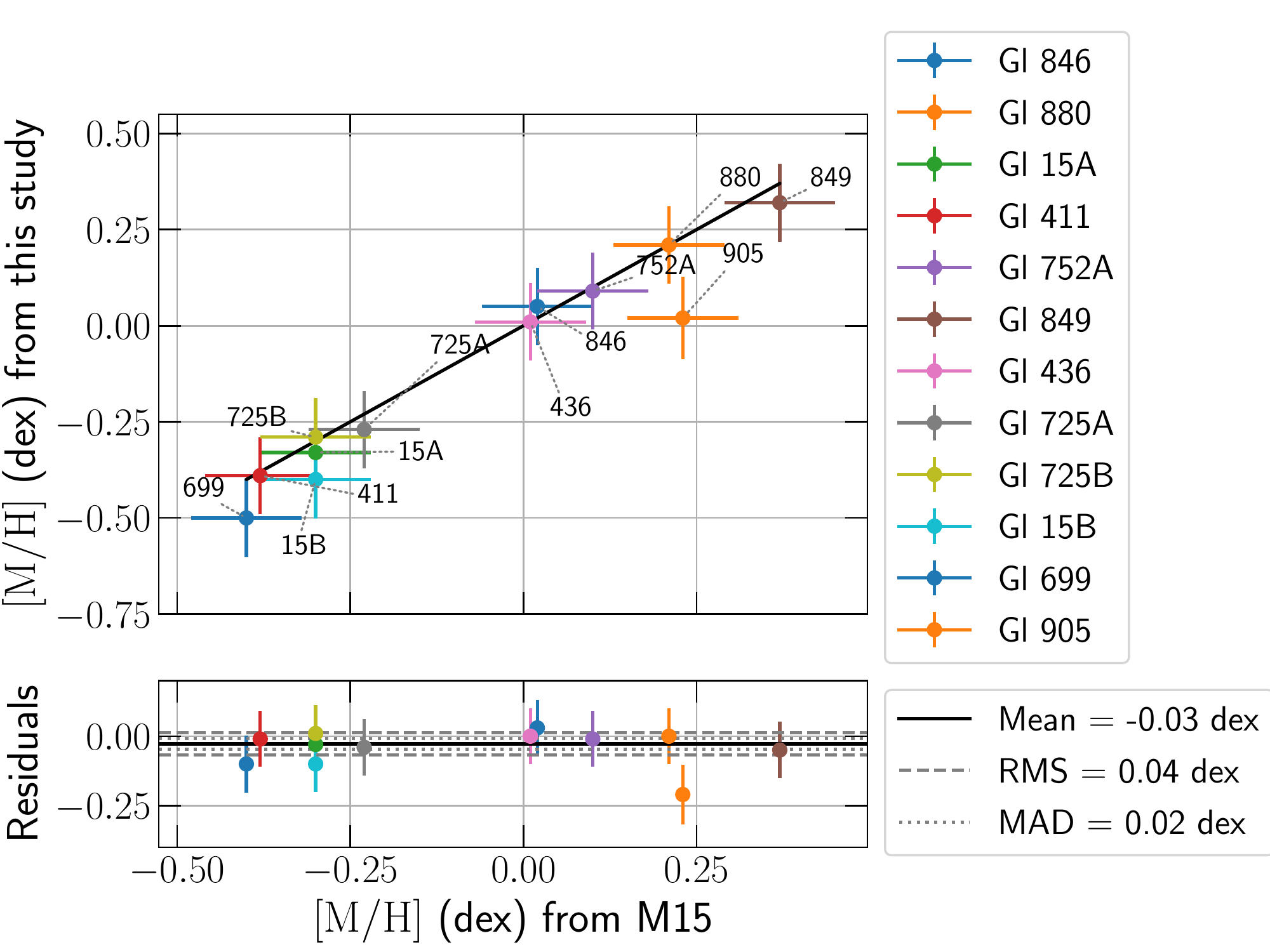}\includegraphics[width=.5\linewidth]{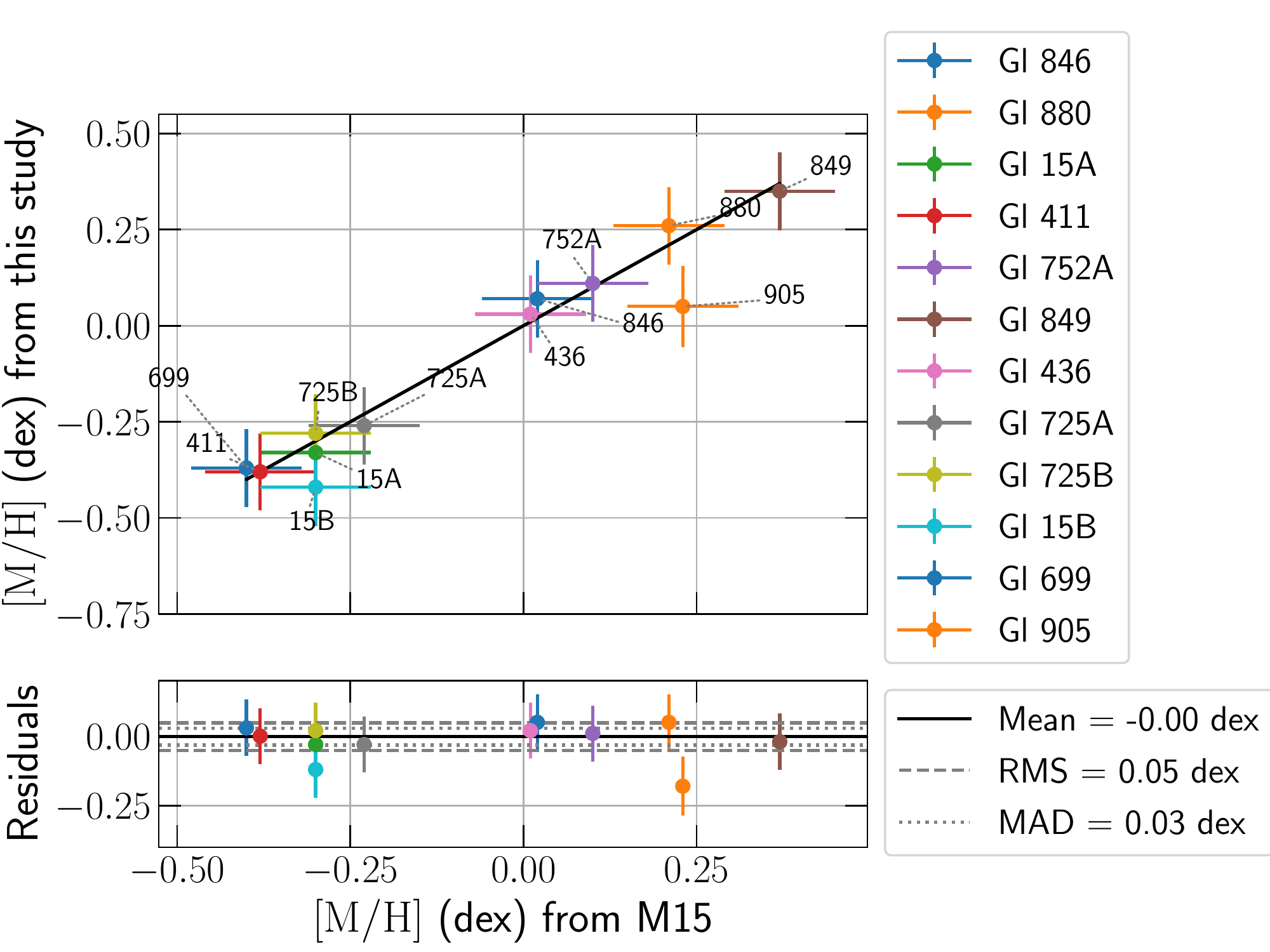}
	\caption{Comparisons between the retrieved $\teff$, $\logg$ and $\mh$ for our 12 calibration stars. Left and right panels present the results obtained before and after the corrections applied to the line list parameters listed in Sec.~\ref{sec:line_selection}.}
	\label{fig:comparison_12stars_adj}
\end{figure*}

\begin{figure*}
	\includegraphics[width=\columnwidth]{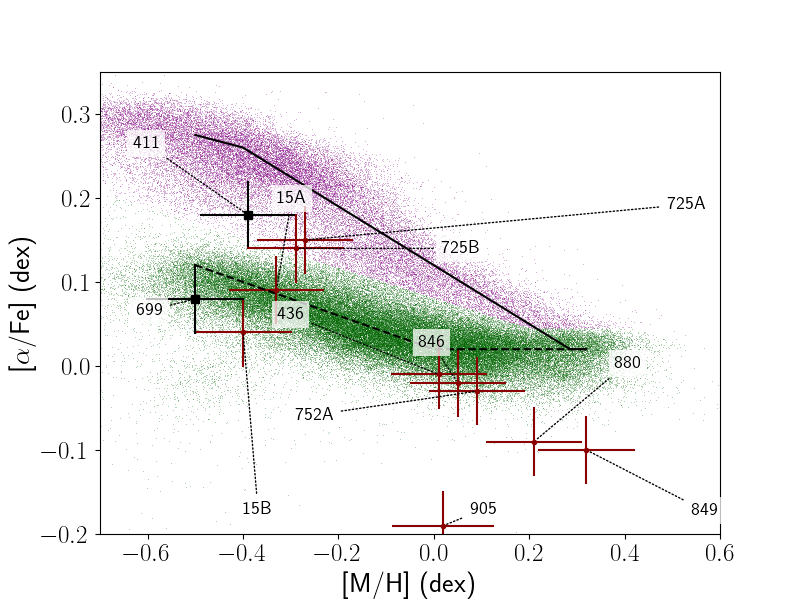}\includegraphics[width=\columnwidth]{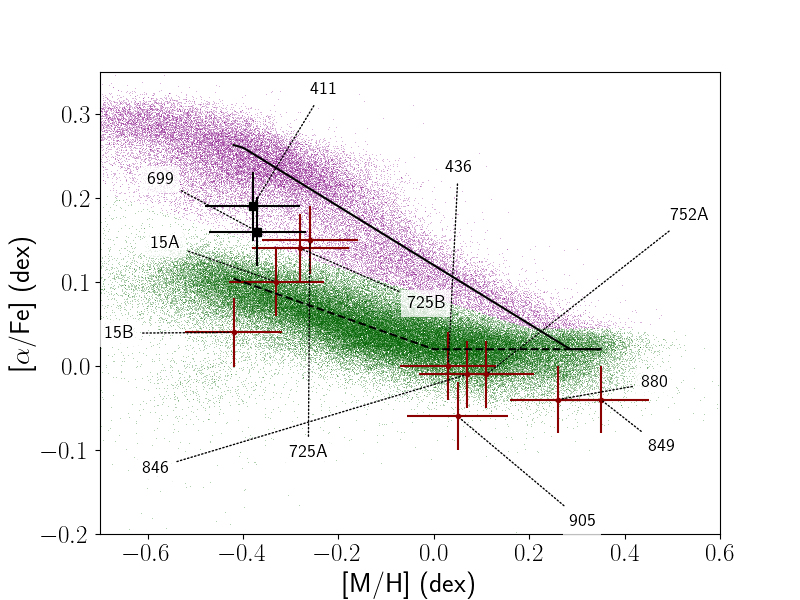}
	\caption{Retrieved $\afe$ and $\mh$ values for our 12 calibration stars. Purple and green pixels  depict APOGEE data for giants of the thick and thin disk respectively. The stars in our sample expected to belong to the thick disk from  their velocity are marked with a black square symbol. Solid and dashed black lines mark empirical thick and thin disk $\mh$--$\afe$ relations, respectively. Left and right panels present the results obtained before and after correction on some line parameters (see Sec.~\ref{sec:line_selection}), respectively.}
	\label{fig:comparison_12stars_alpha}
\end{figure*}

\begin{figure*}
	\includegraphics[width=.5\linewidth]{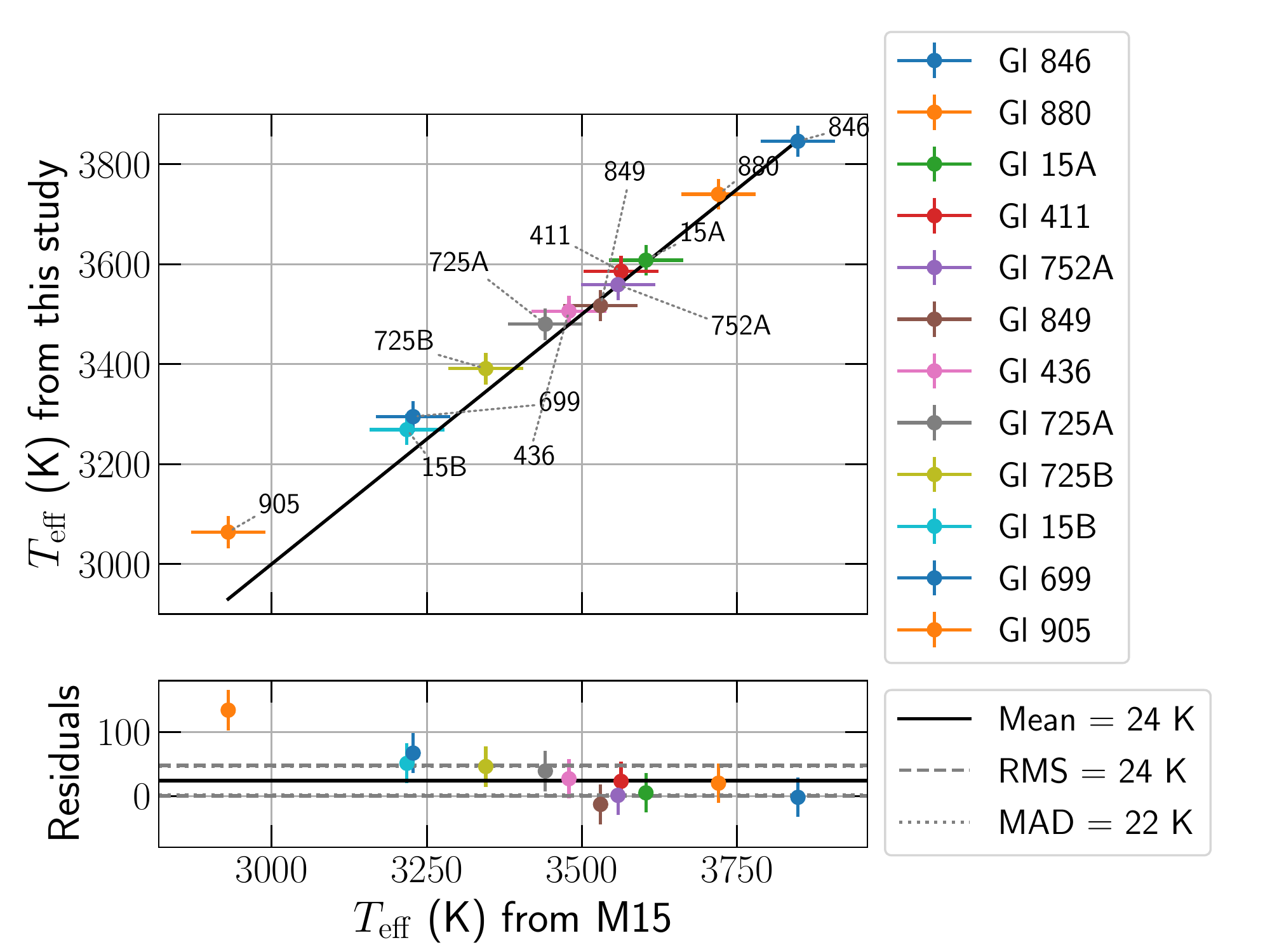}\includegraphics[width=.5\linewidth]{figures/4d_12_mann_vs_obtained_teff_vmac_3}	\includegraphics[width=.5\linewidth]{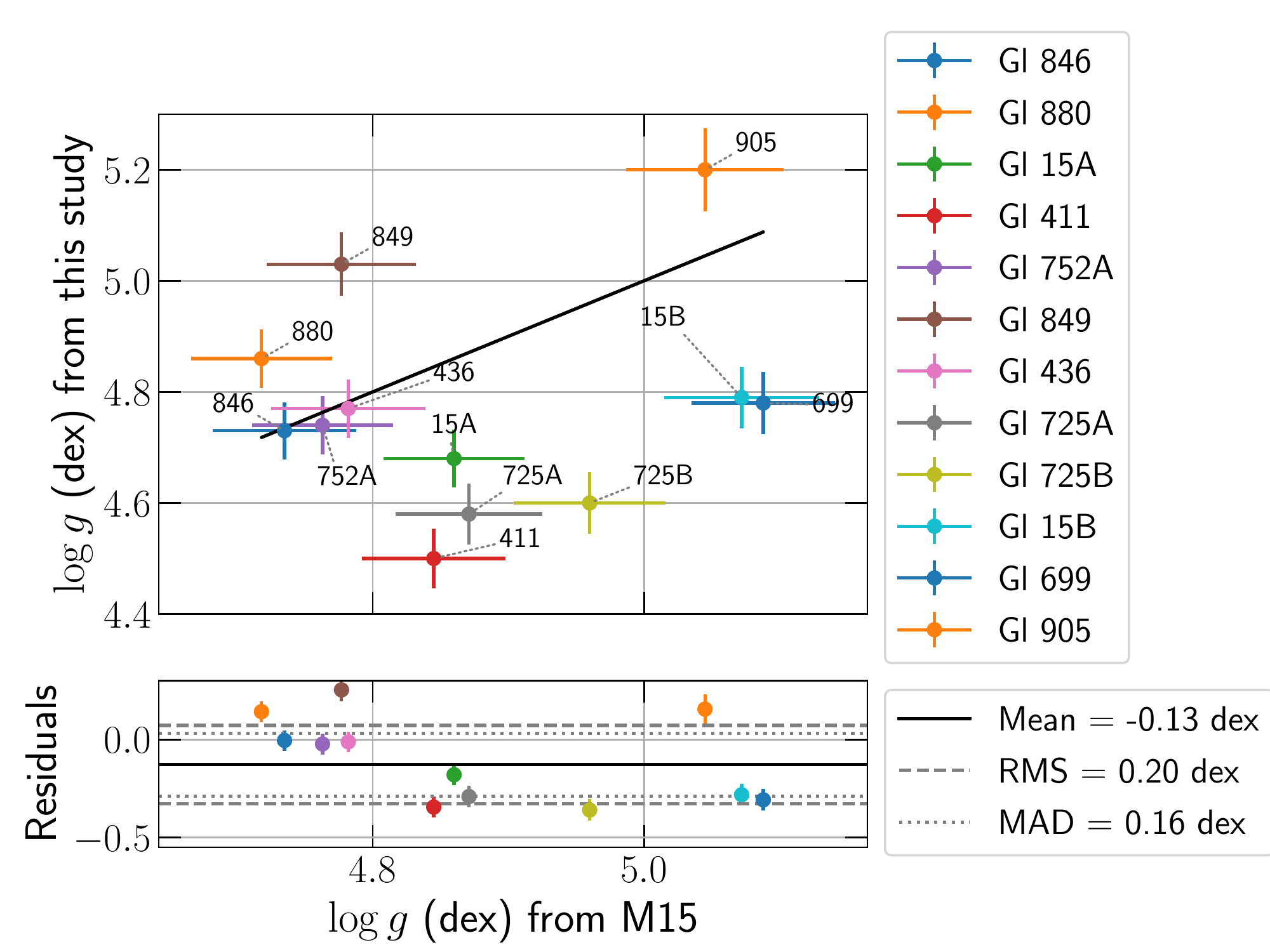}\includegraphics[width=.5\linewidth]{figures/4d_12_mann_vs_obtained_logg_vmac_3}
	\includegraphics[width=.5\linewidth]{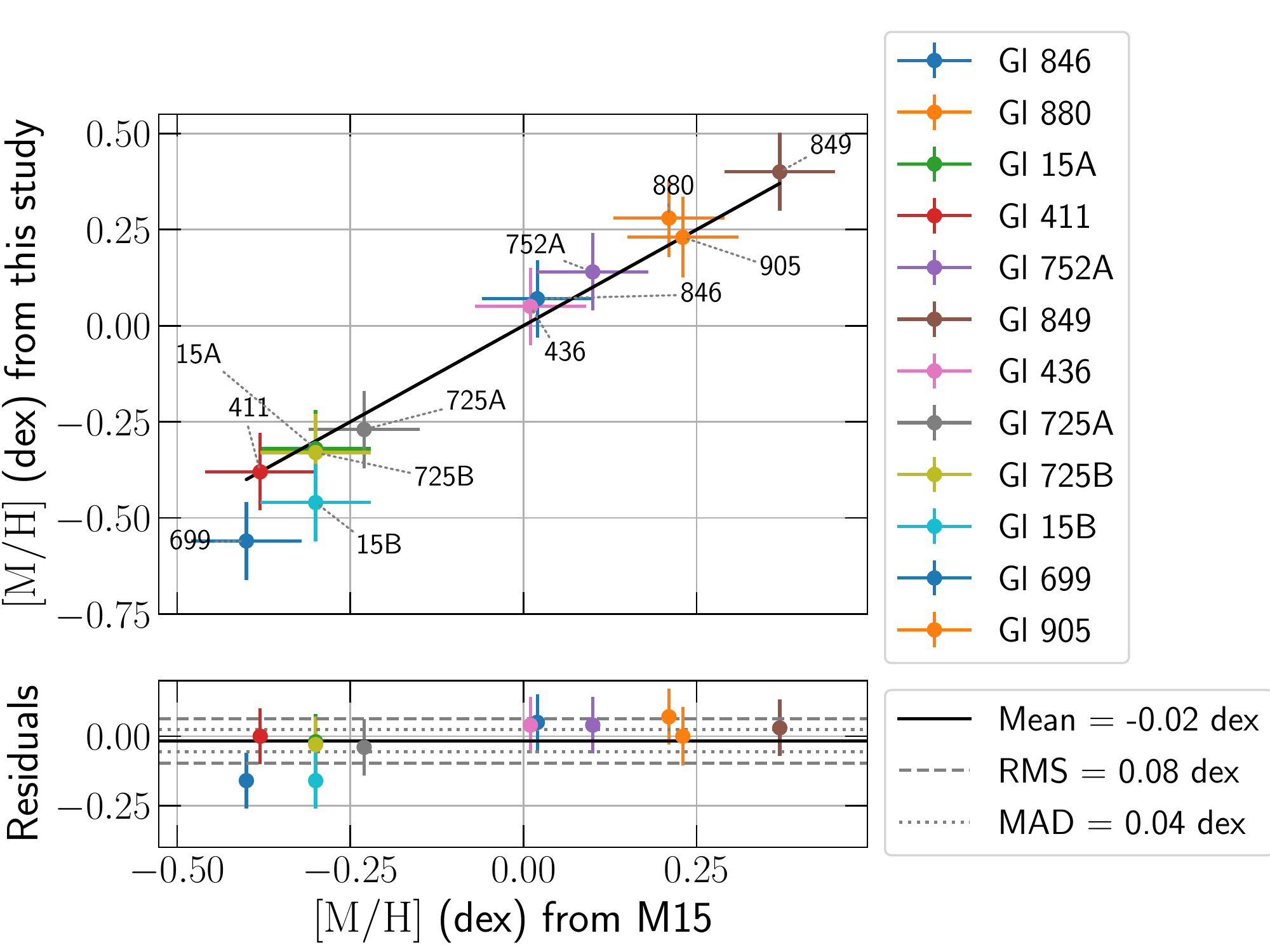}\includegraphics[width=.5\linewidth]{figures/4d_12_mann_vs_obtained_mh_vmac_3}
	\caption{  Same as Fig.~\ref{fig:comparison_12stars_adj} but comparing the results obtained with $\afe=0$~dex (left panels) and while fitting $\afe$ (right panels). These results are obtained with corrections of the line list  described in Sec.~\ref{sec:line_selection}.}
	\label{fig:comparison_12stars}
\end{figure*}

 Figures~\ref{fig:comparison_12stars_adj}~\&~\ref{fig:comparison_12stars_alpha} present a comparison of the results obtained with and without corrections applied to the line list parameters (see Sec.~\ref{sec:line_selection}). Figure~\ref{fig:comparison_12stars} illustrates the effect of fitting on $\afe$ on the retrieved parameters of our calibration stars.

{\paul
\section{Best fits on all spectral lines}

Figure~C1 available as supplementary material presents the best fits obtained for 5 stars in our sample.

}

\section{Literature parameters comparison}
\begin{figure*}
	\includegraphics[width=0.5\linewidth]{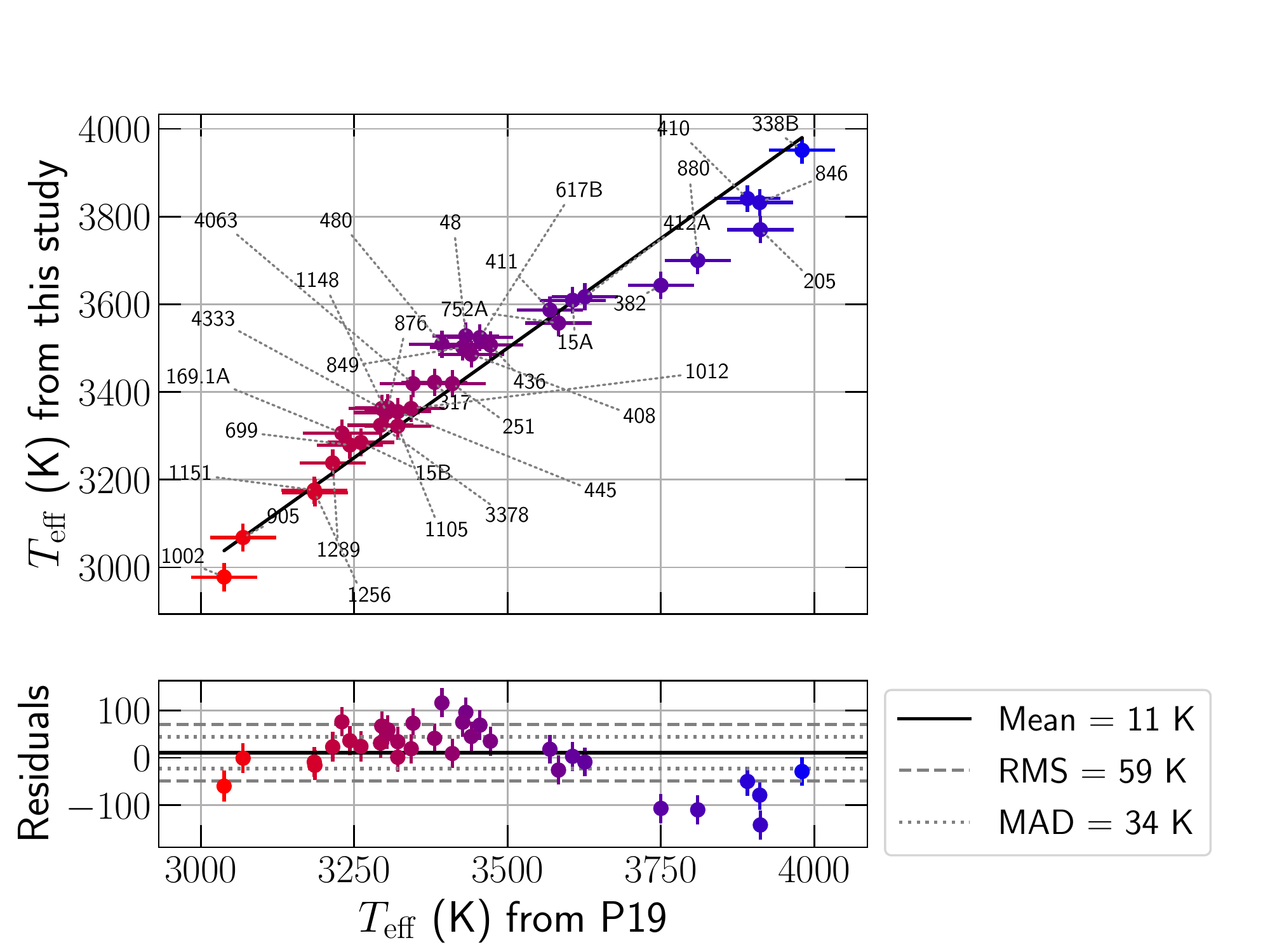}\includegraphics[width=0.5\linewidth]{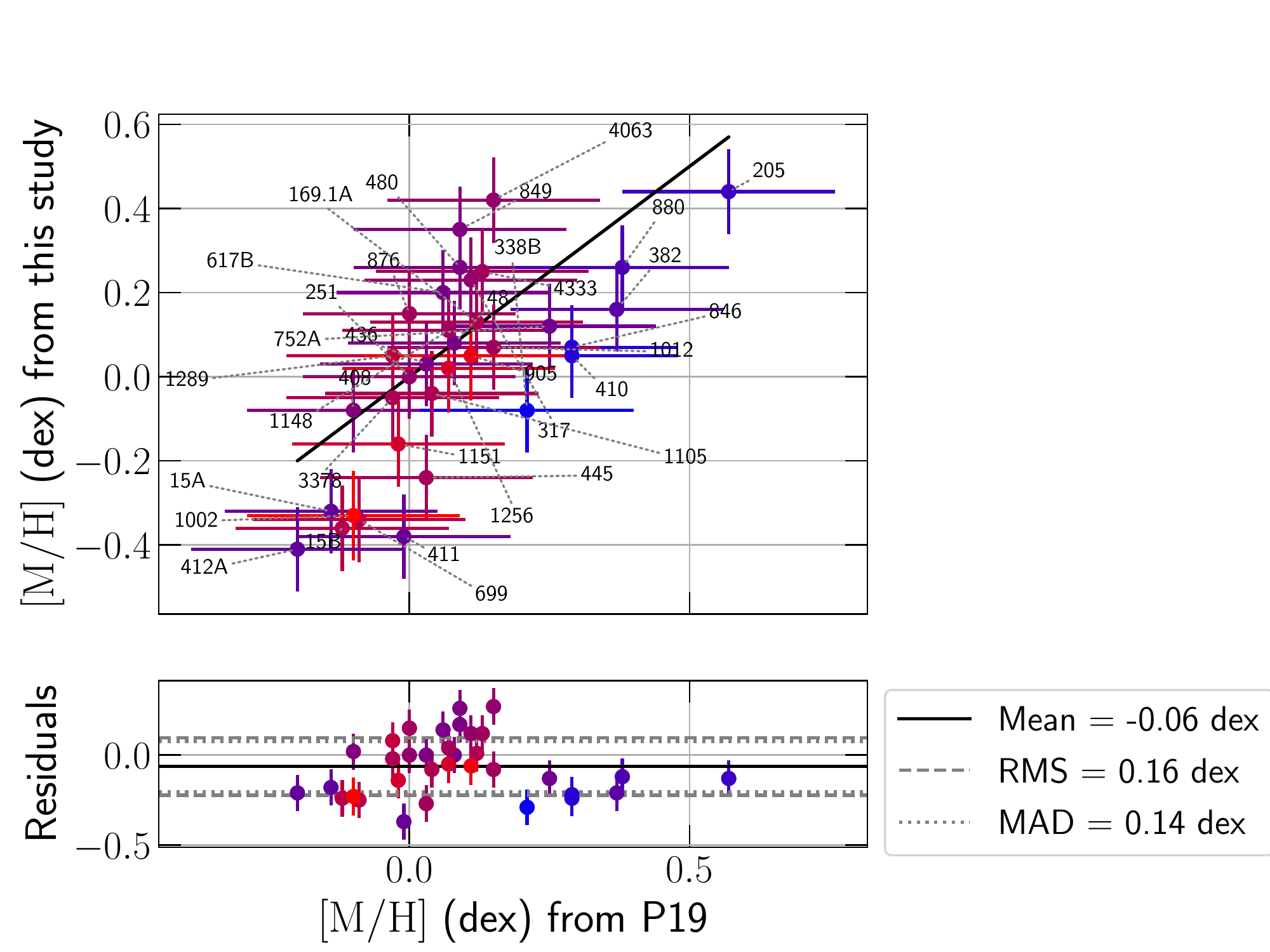}
	\caption{Comparison between retrieved parameters of 32 stars and the values published by~\citet{passegger_2019}.}
	\label{fig:results_all_4d_pass}
\end{figure*}

\begin{figure*}
	\includegraphics[width=0.5\linewidth]{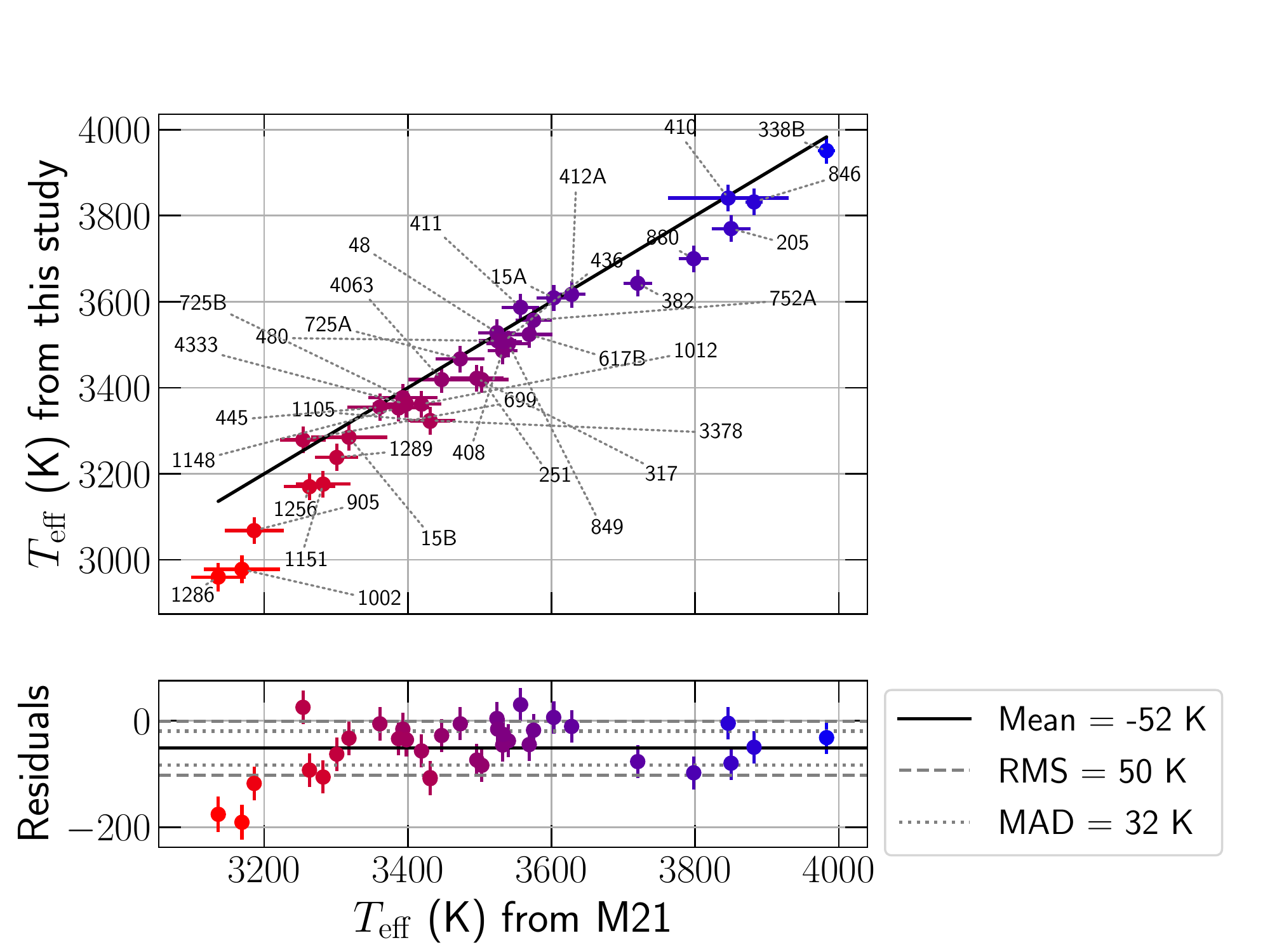}\includegraphics[width=0.5\linewidth]{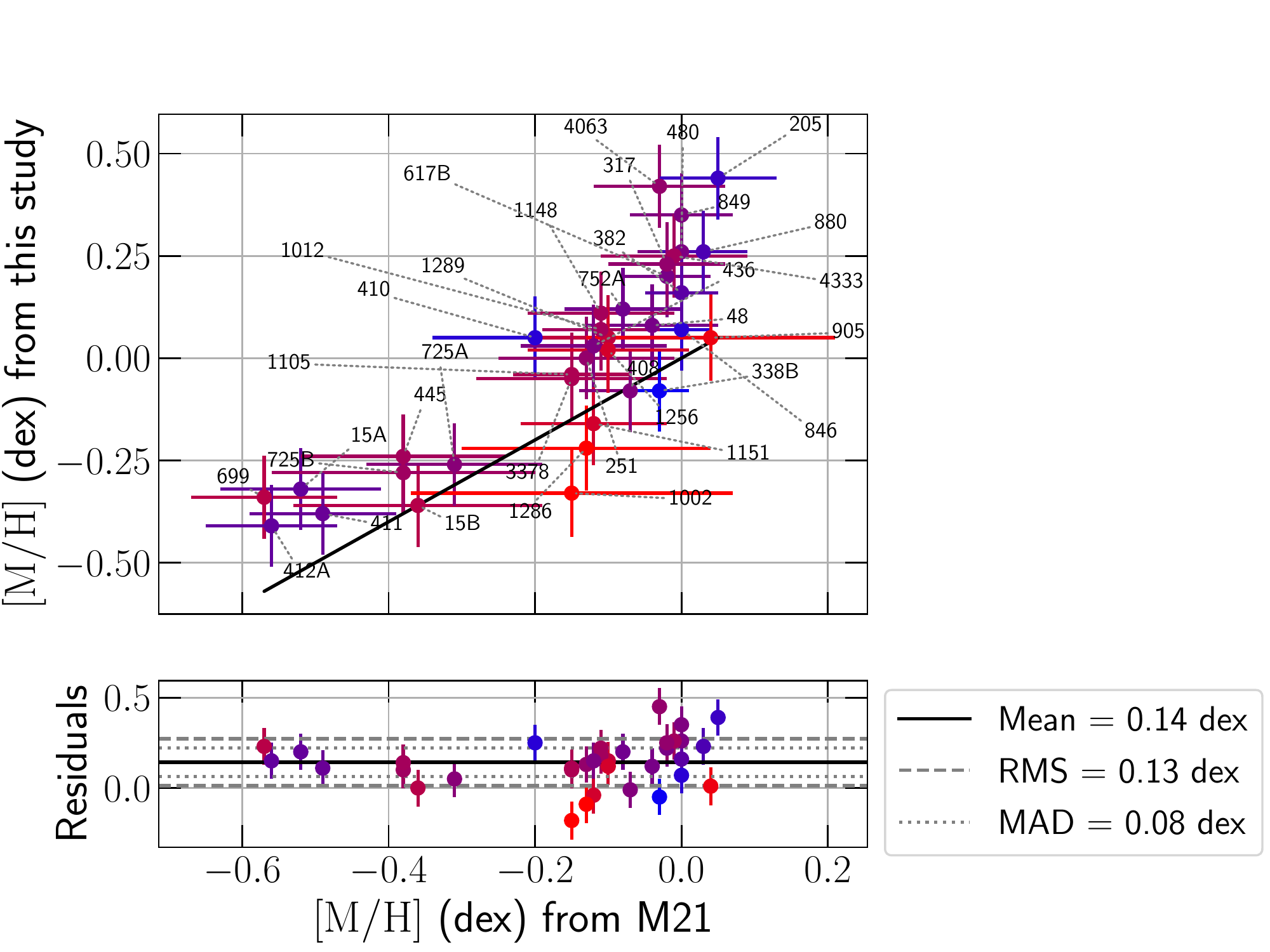}
	\caption{Same as Fig.~\ref{fig:results_all_4d_pass}  for 35 stars and values publishedter
		 by~\citet{marfil_2021}.}
	\label{fig:results_all_4d_marf}
\end{figure*}

We present comparisons of parameters recovered by several studies. Figures~\ref{fig:results_all_4d_pass}~\&~\ref{fig:results_all_4d_marf} present the results for 32 and 35 stars studied by~\citet{passegger_2019} and~\citet{marfil_2021} respectively.

\section{Estimation of luminosity}

\begin{figure}
	\includegraphics[width=1\linewidth]{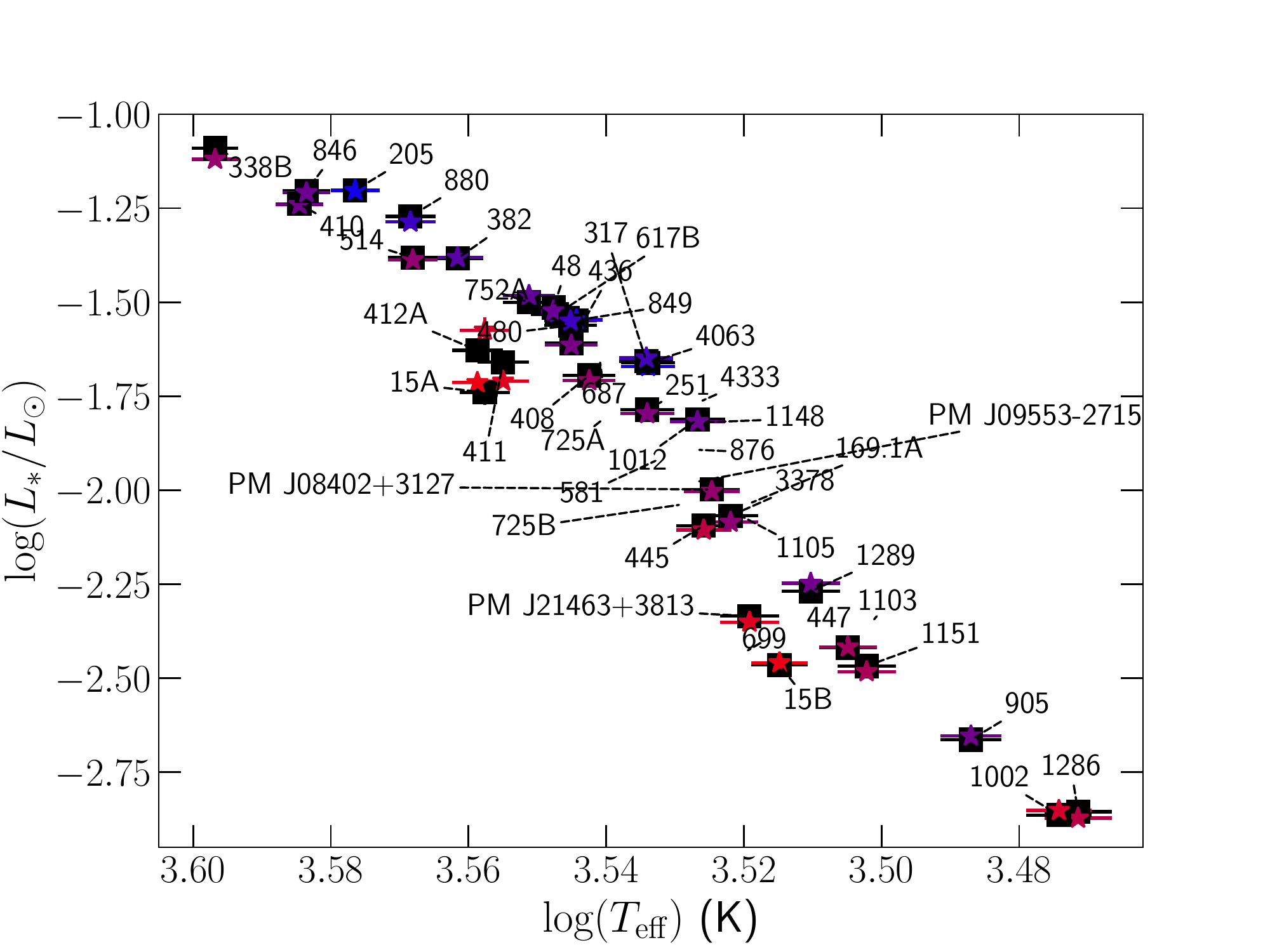}
	\caption{Comparison between the luminosities computed from bolometric corrections using the relation proposed by~\citet{cifuentes_2020} (black squares), and those reported by~\citet{cifuentes_2020} (coloured symbols) for 33 stars included in our sample. One should note that the $\teff$ values used by the authors to estimate the luminosities may differ from those estimated in this work. The symbol colours display the metallicity from low (red) to high (blue).}
	\label{fig:hr_diagram_comp}
\end{figure}

Figure~\ref{fig:hr_diagram_comp} presents a comparison between the luminosities estimated from G and J band magnitudes using bolometric corrections~\citep{cifuentes_2020} and these reported by~\citet{cifuentes_2020}.

\section{Mass-Radius relation}

\begin{figure}
	\includegraphics[width=\columnwidth]{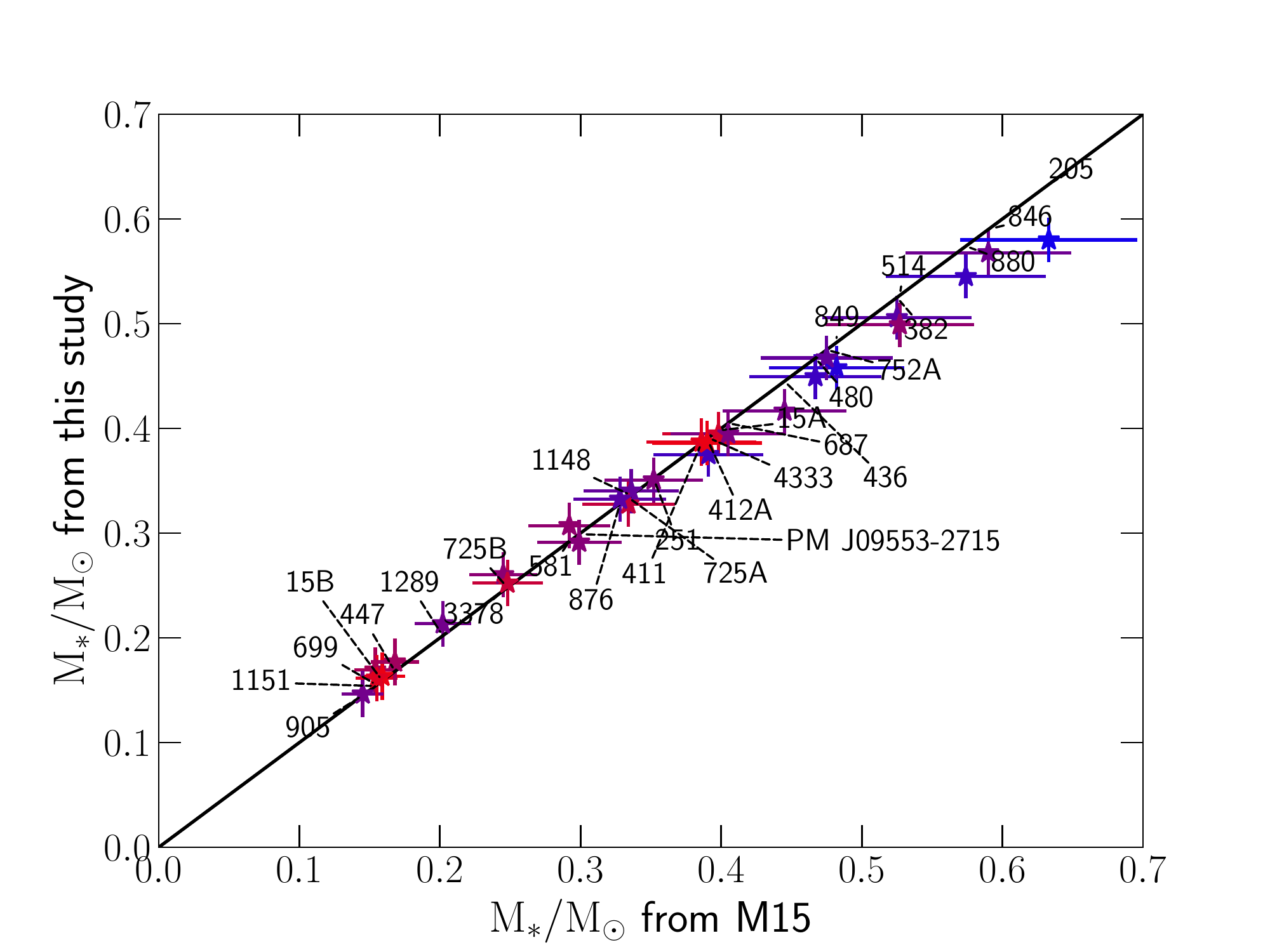}
	\includegraphics[width=\columnwidth]{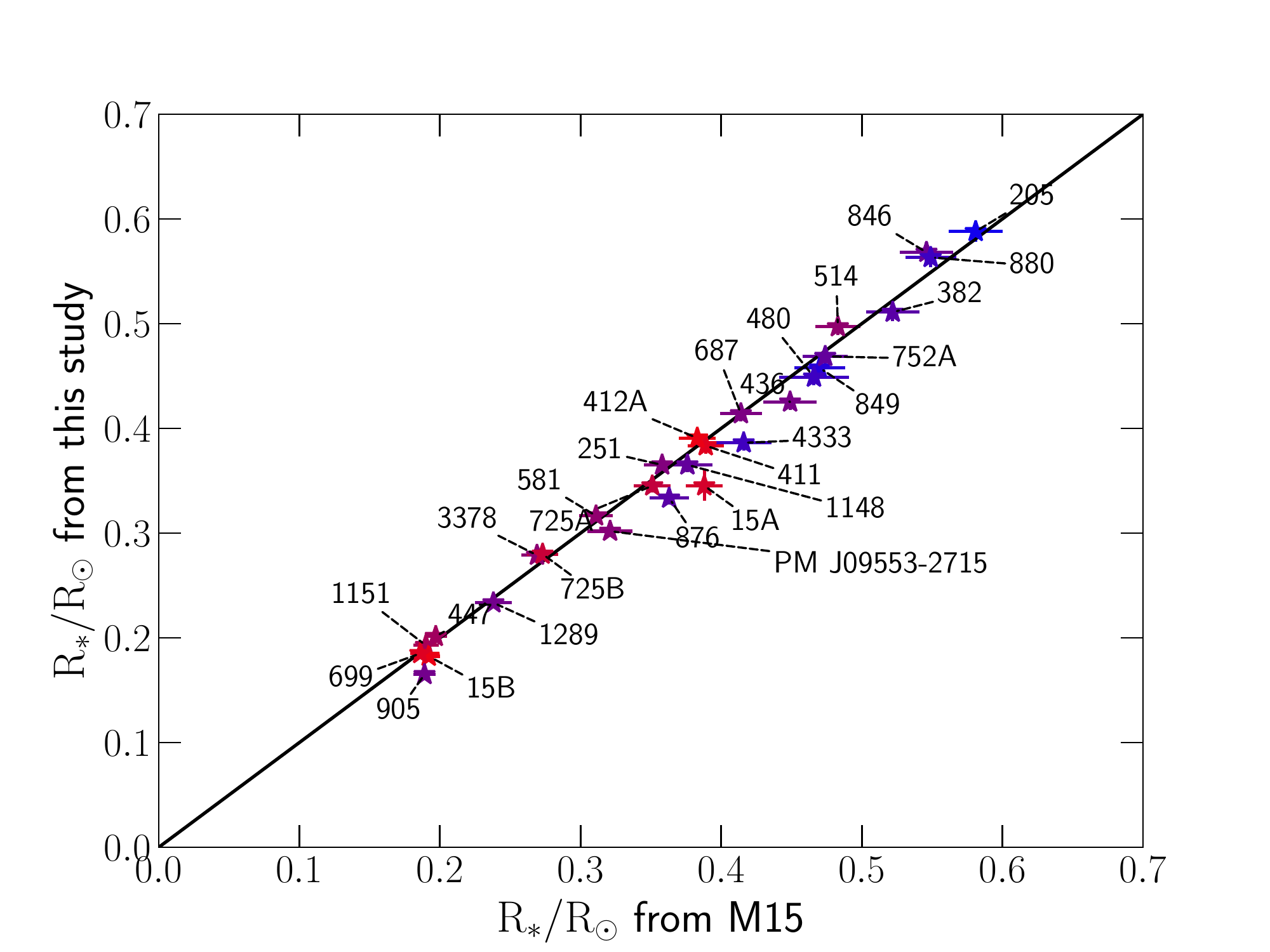}
	\caption{{\pp Comparison between our derived masses and radii and these reported by M15 for the \nbMdwarfsCommon{} stars included in both studies (top and bottom panels respectively). The black solid lines represent the equality.}}
	\label{fig:m_r_diagram_mann}
\end{figure}

{\p Figure~\ref{fig:m_r_diagram_mann} presents a comparison between the masses and radii derived in this study and these reported by~M15.}

\bsp	
\label{lastpage}
\end{document}


\maketitle

\section*{APPENDIX C: BEST FITS ON ALL SPECTRAL LINES}
Fig.~C1 presents the best fit obtained for 5 stars in our sample, and for all spectral lines used.

\begin{figure}[h]
	\includegraphics[scale=0.5]{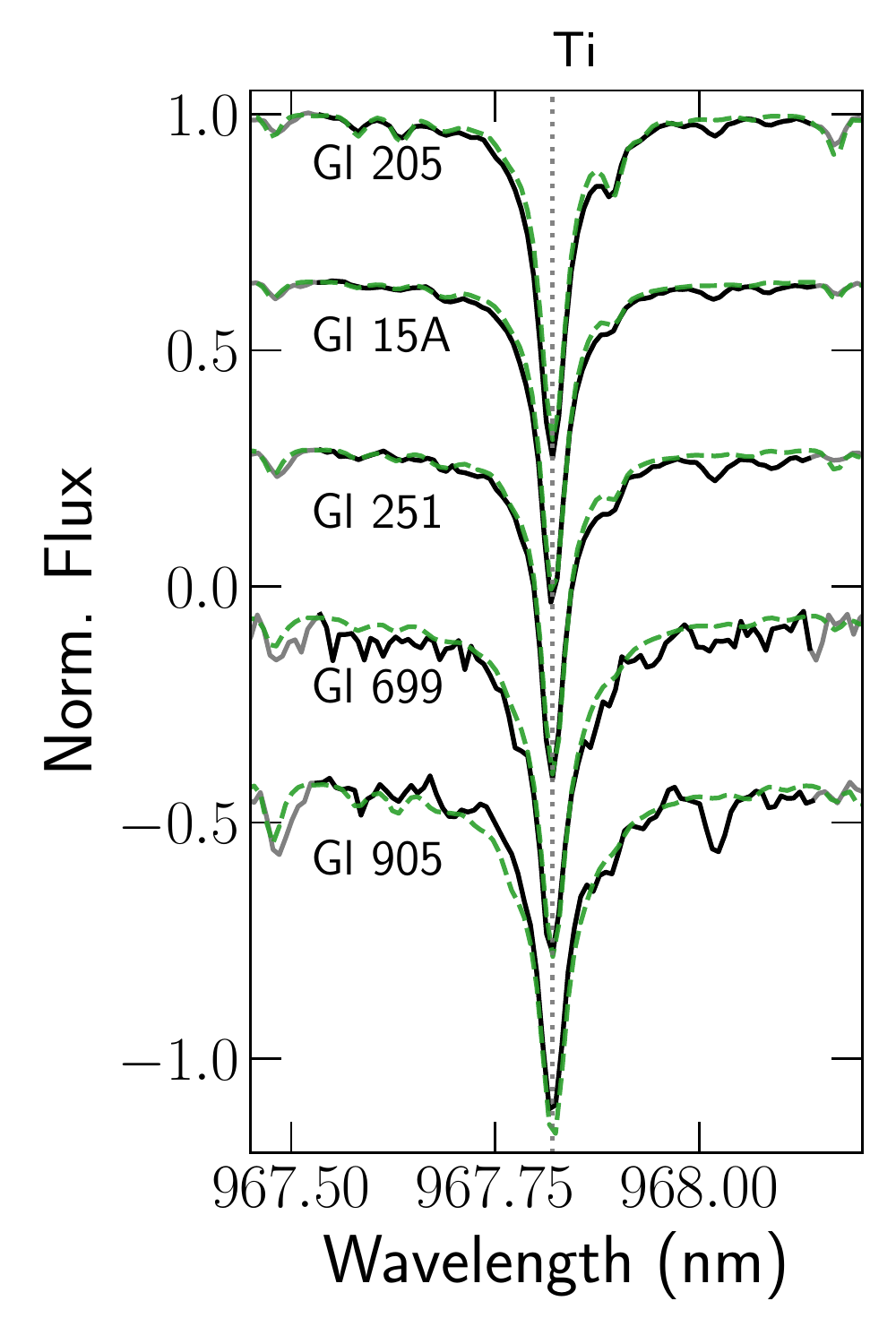}\includegraphics[trim={1cm 0  0cm 0}, clip, scale=0.5]{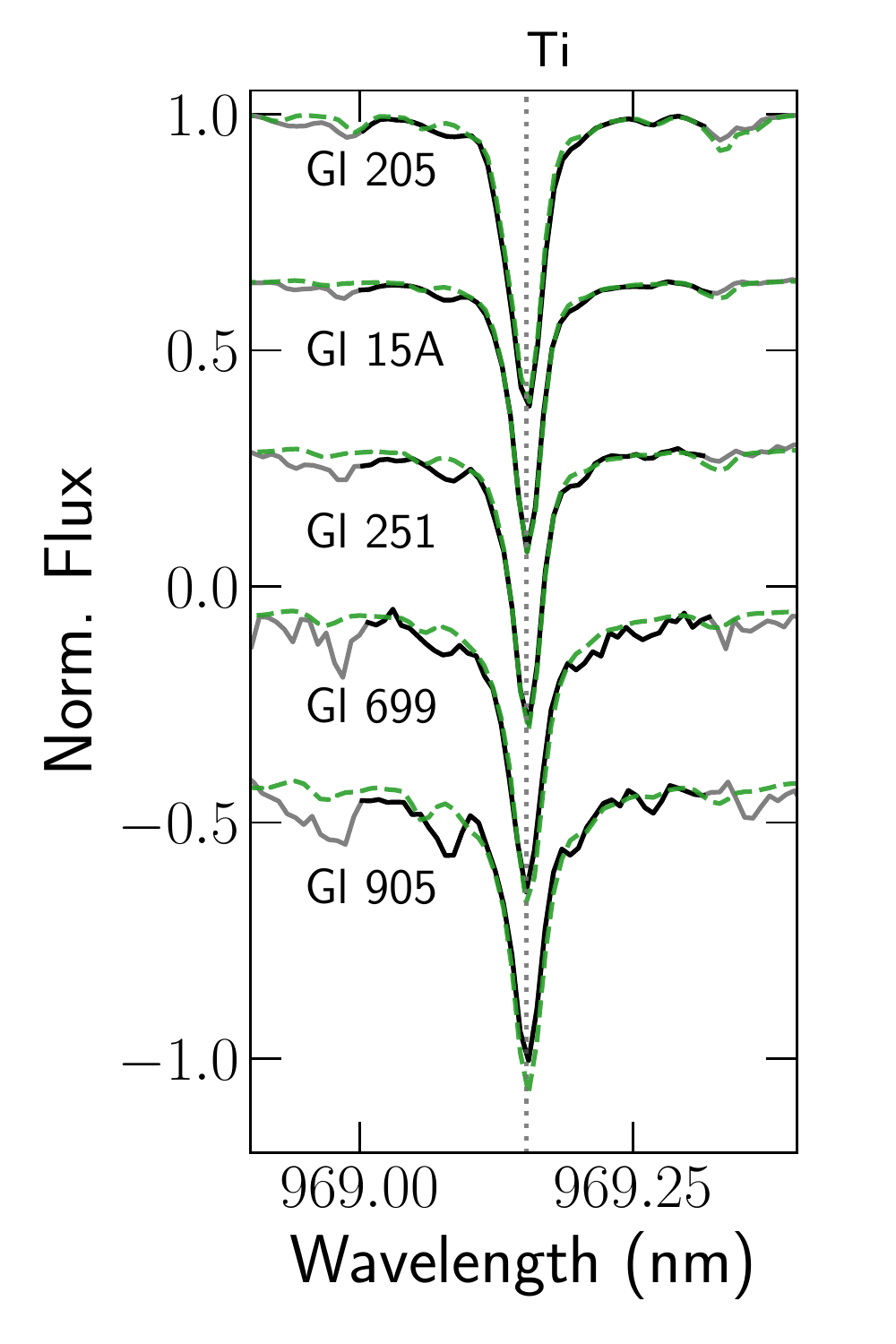}\includegraphics[trim={1cm 0  0cm 0}, clip,scale=0.5]{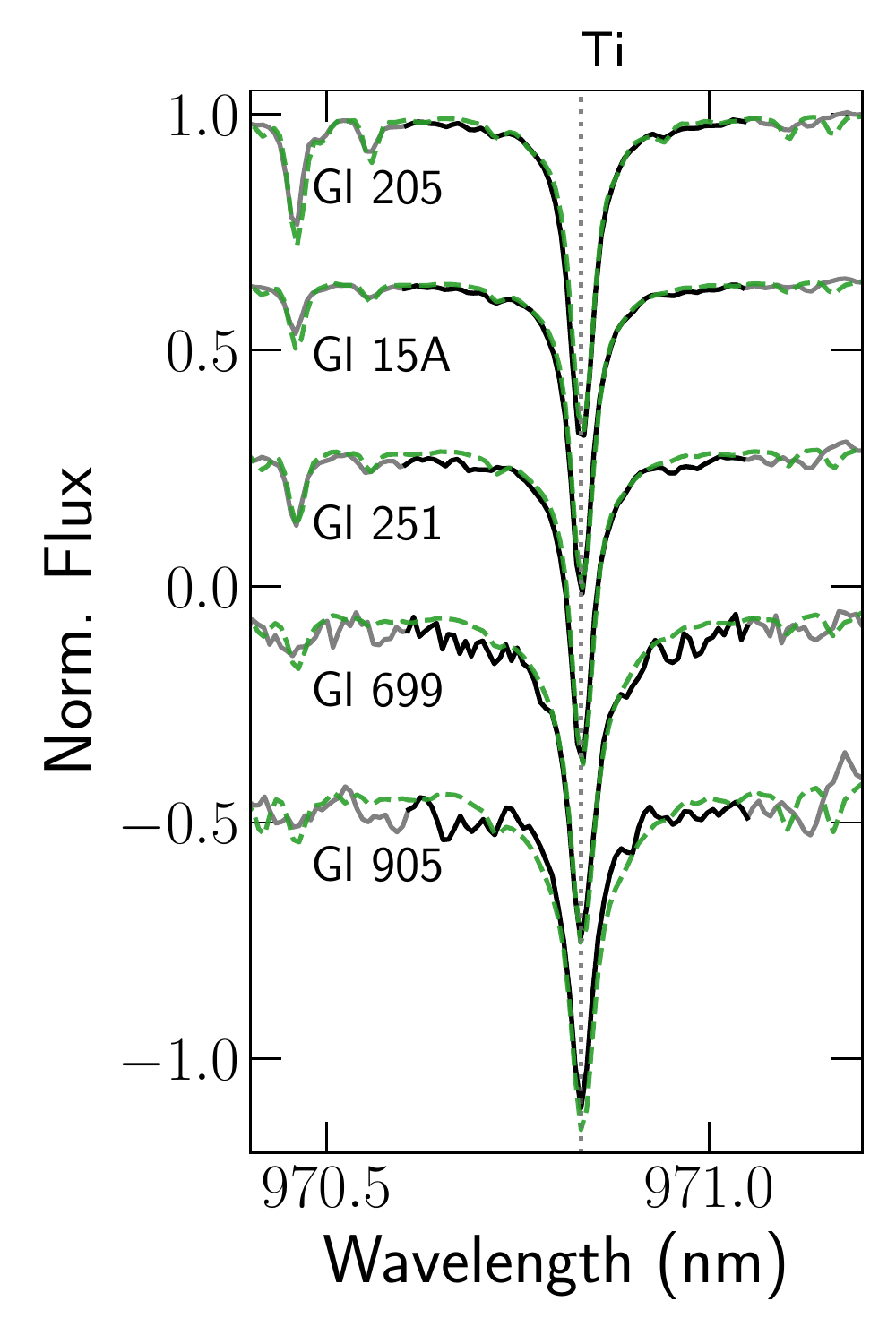}\includegraphics[trim={1cm 0  0cm 0}, clip,scale=0.5]{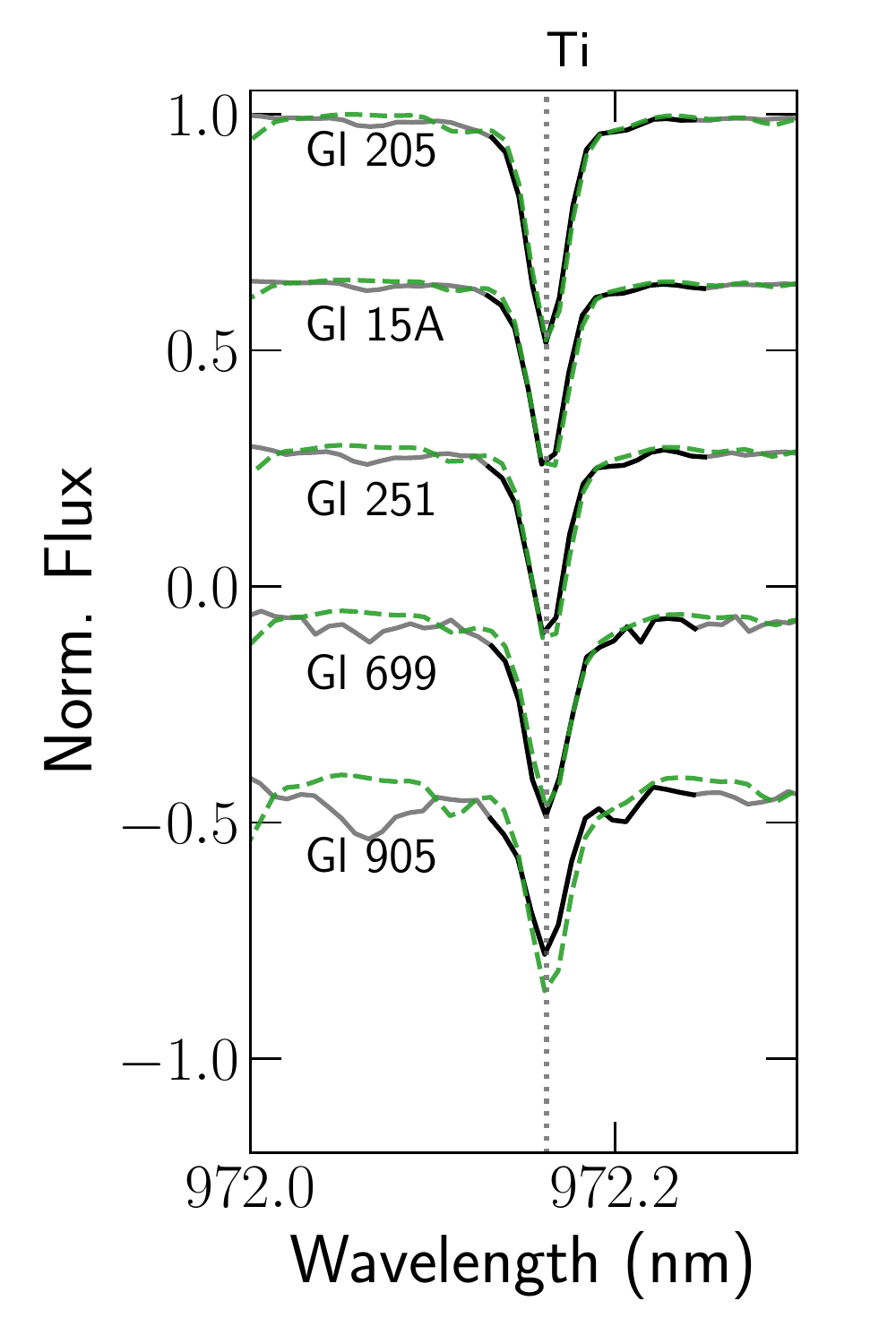}
	\includegraphics[scale=0.5]{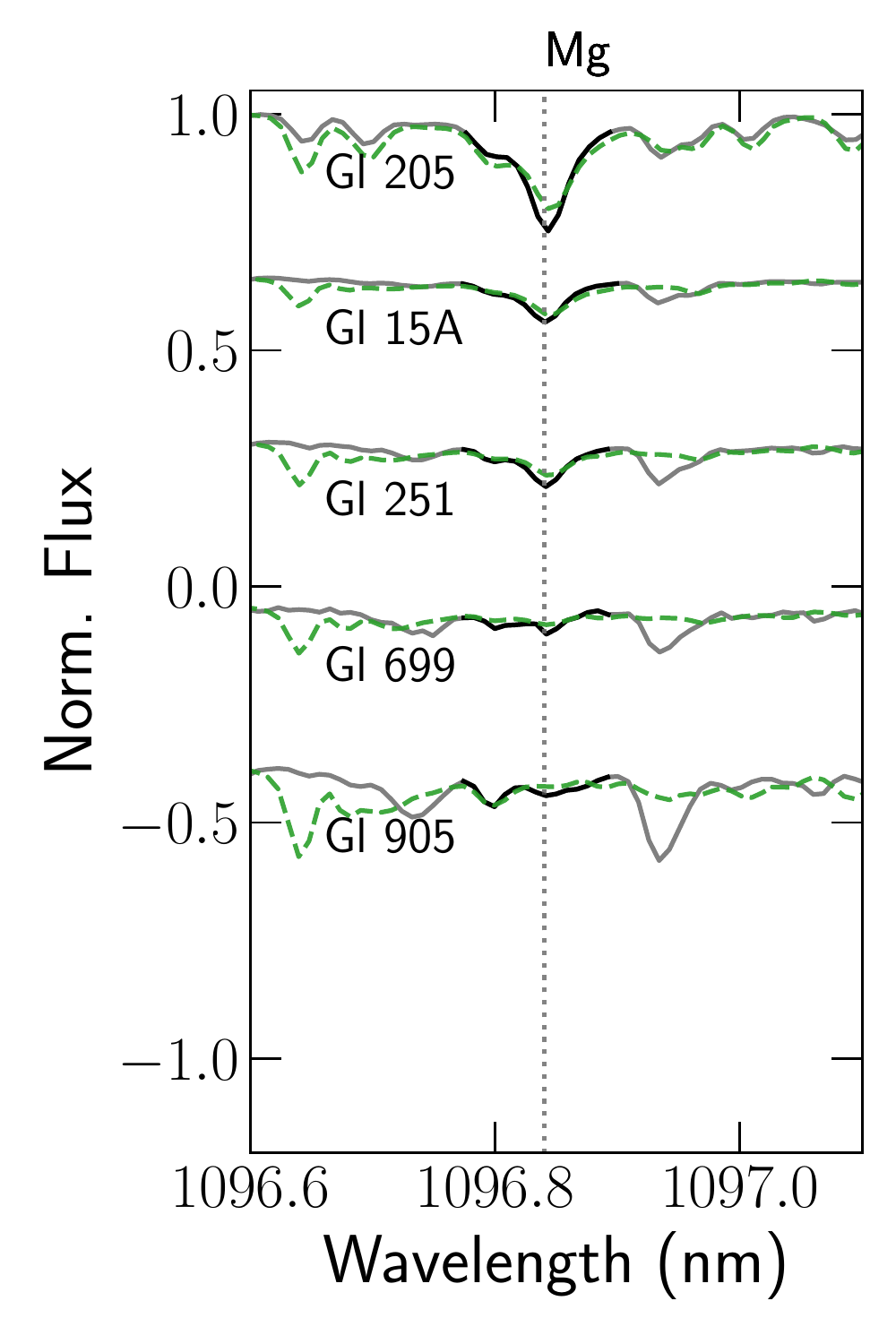}\includegraphics[trim={1cm 0  0cm 0}, clip, scale=0.5]{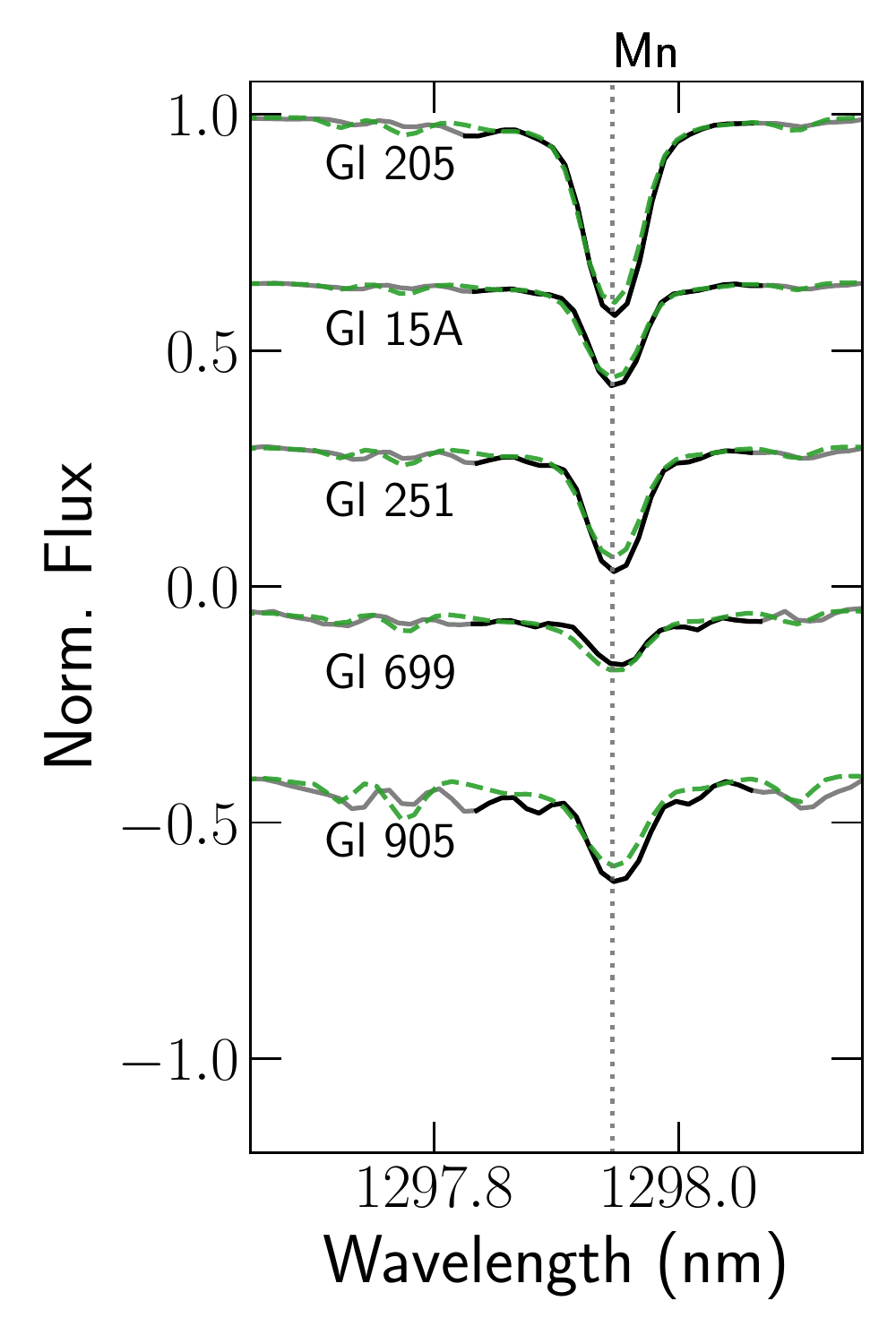}\includegraphics[trim={1cm 0  0cm 0}, clip,scale=0.5]{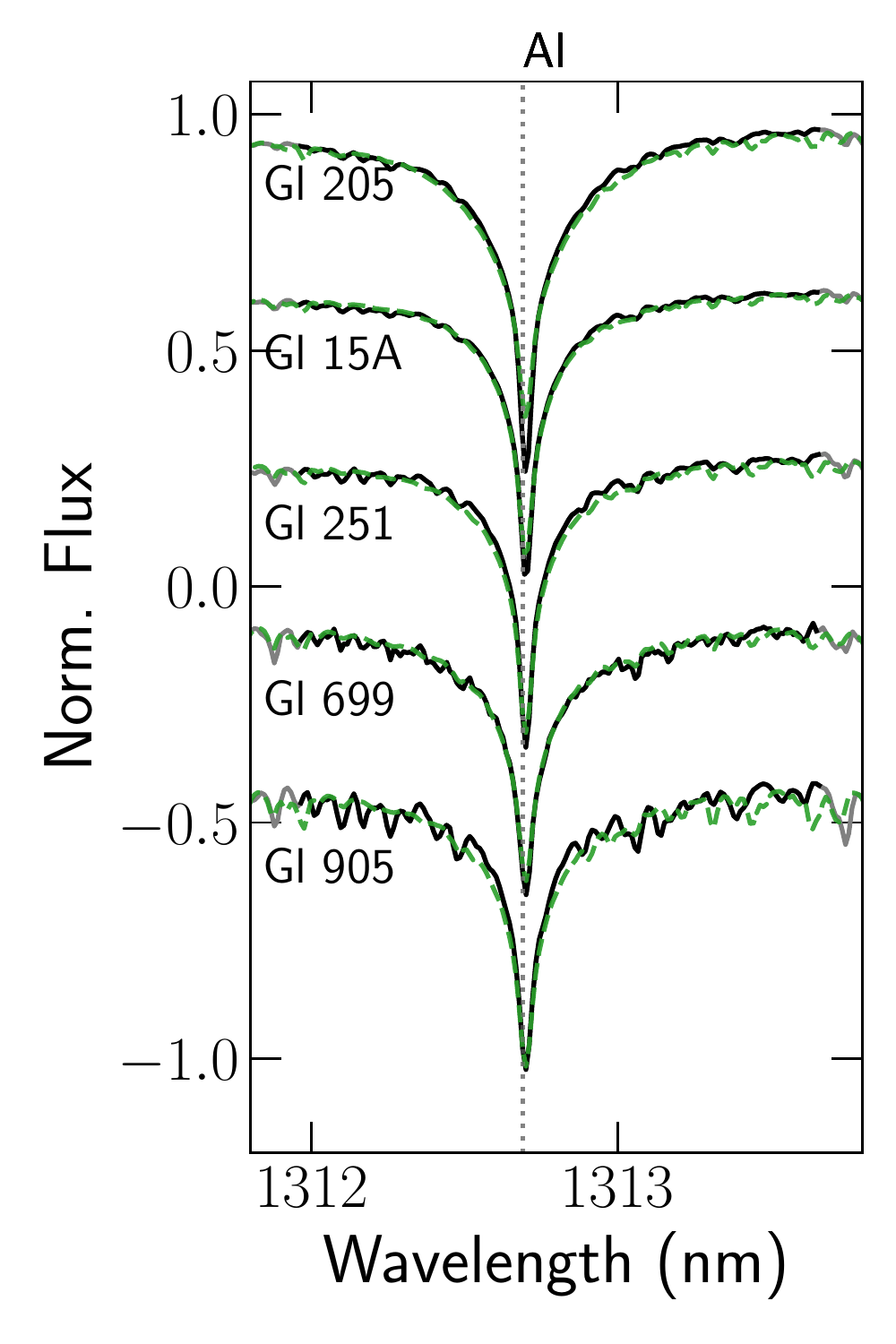}\includegraphics[trim={1cm 0  0cm 0}, clip,scale=0.5]{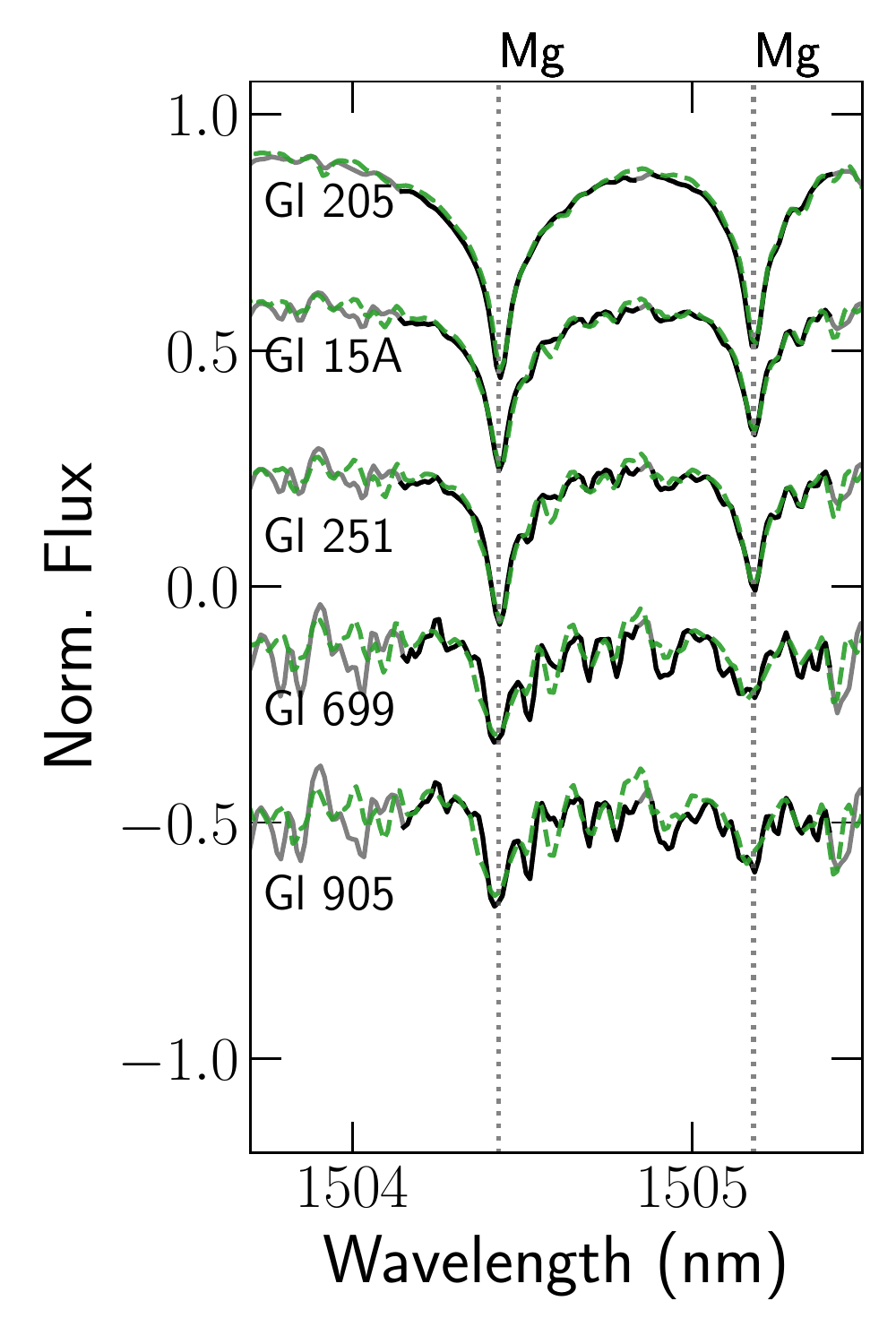}
	\caption*{Figure C1. Best fitted models obtained for 5 stars in our sample.}
	\label{fig:figure}
\end{figure}

\begin{figure}[h]
		\includegraphics[scale=0.5]{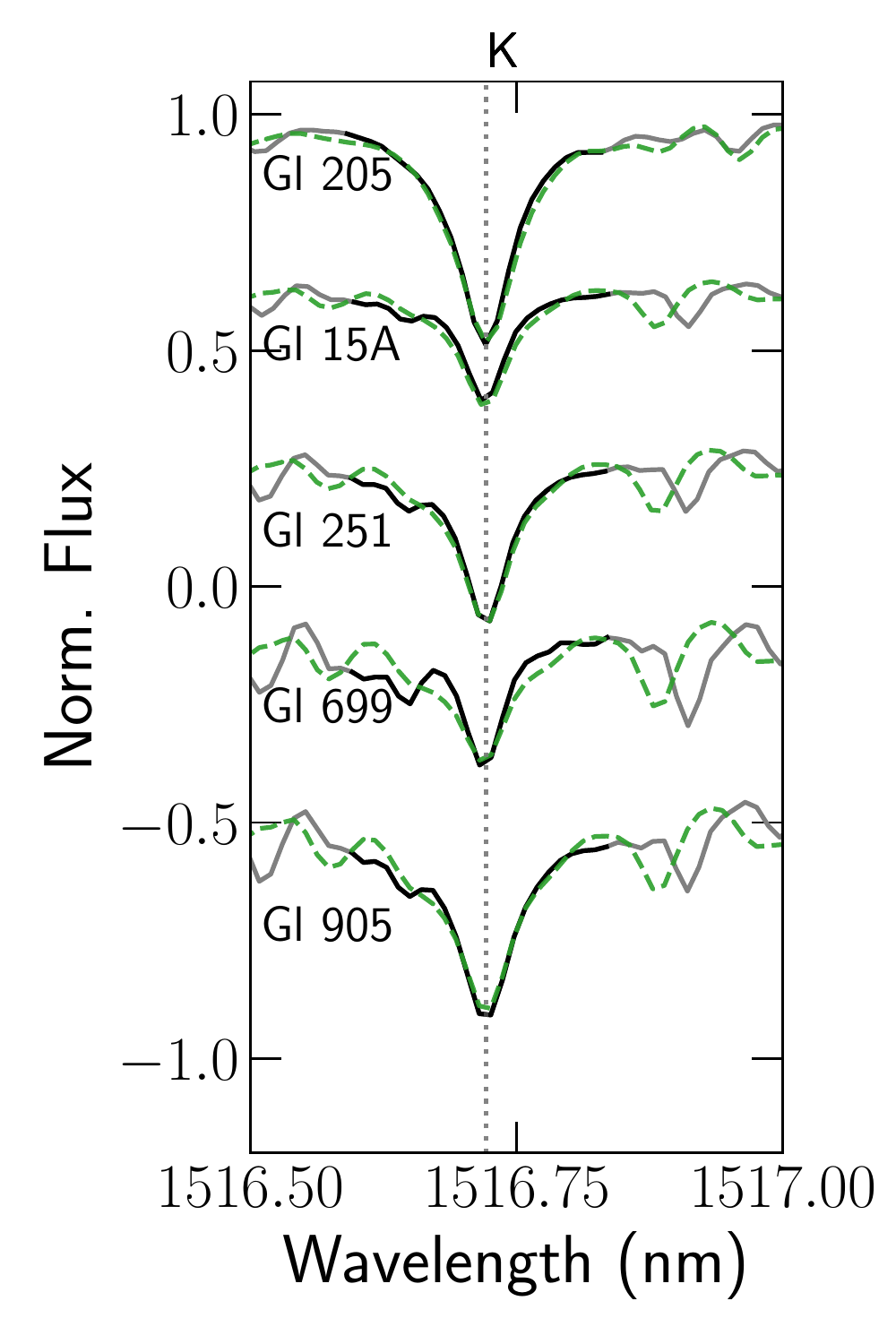}\includegraphics[trim={1cm 0  0cm 0}, clip, scale=0.5]{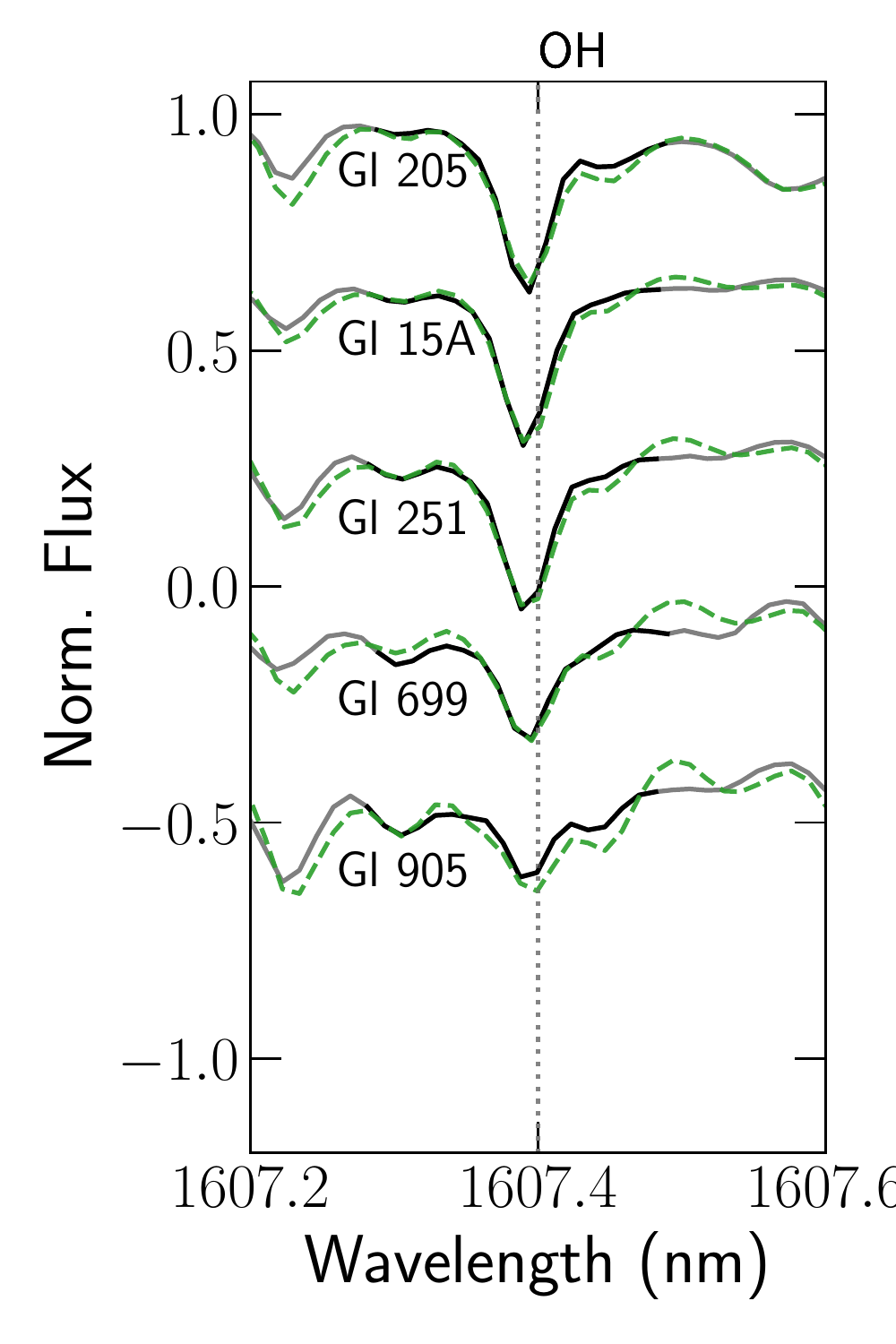}\includegraphics[trim={1cm 0  0cm 0}, clip,scale=0.5]{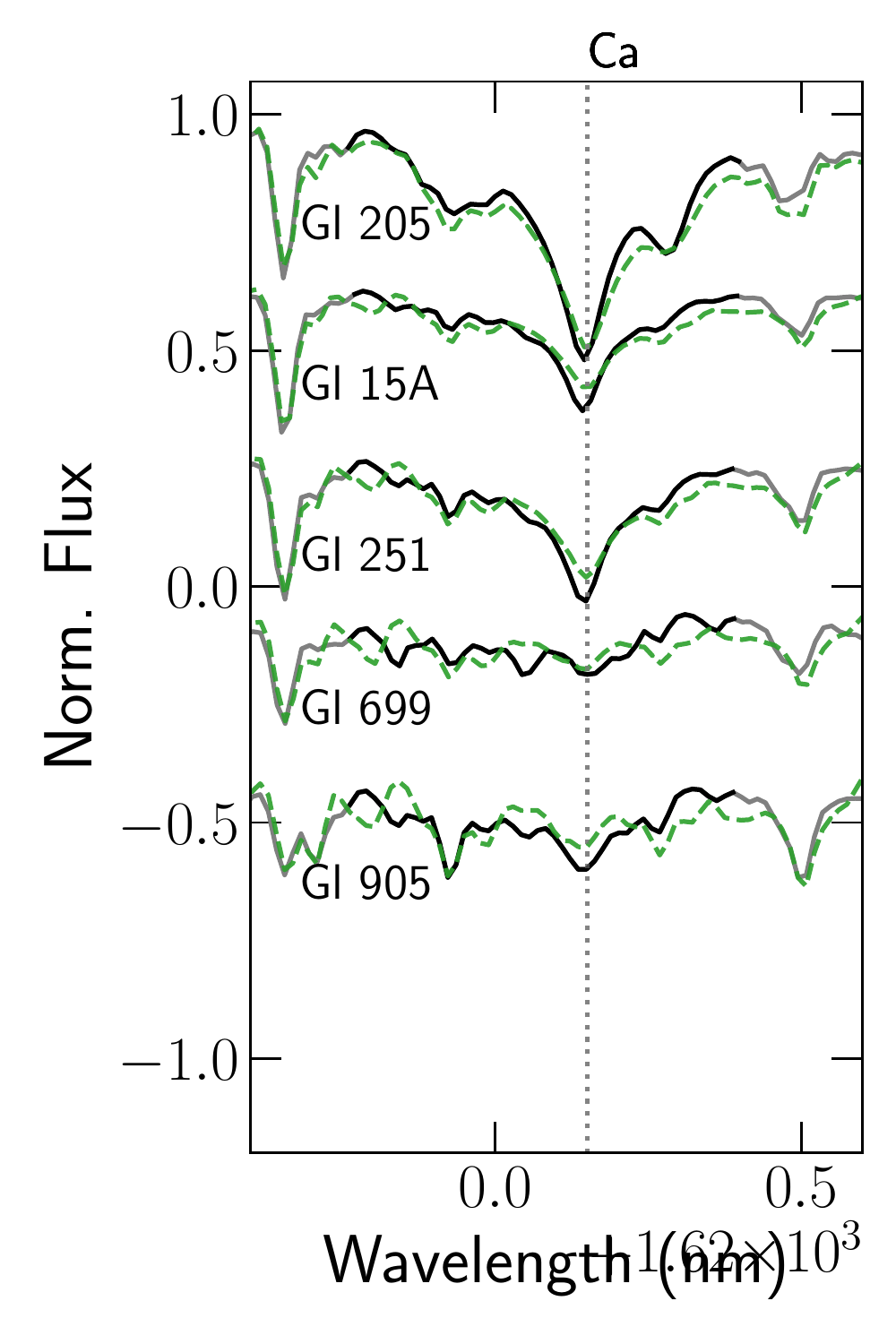}\includegraphics[trim={1cm 0  0cm 0}, clip,scale=0.5]{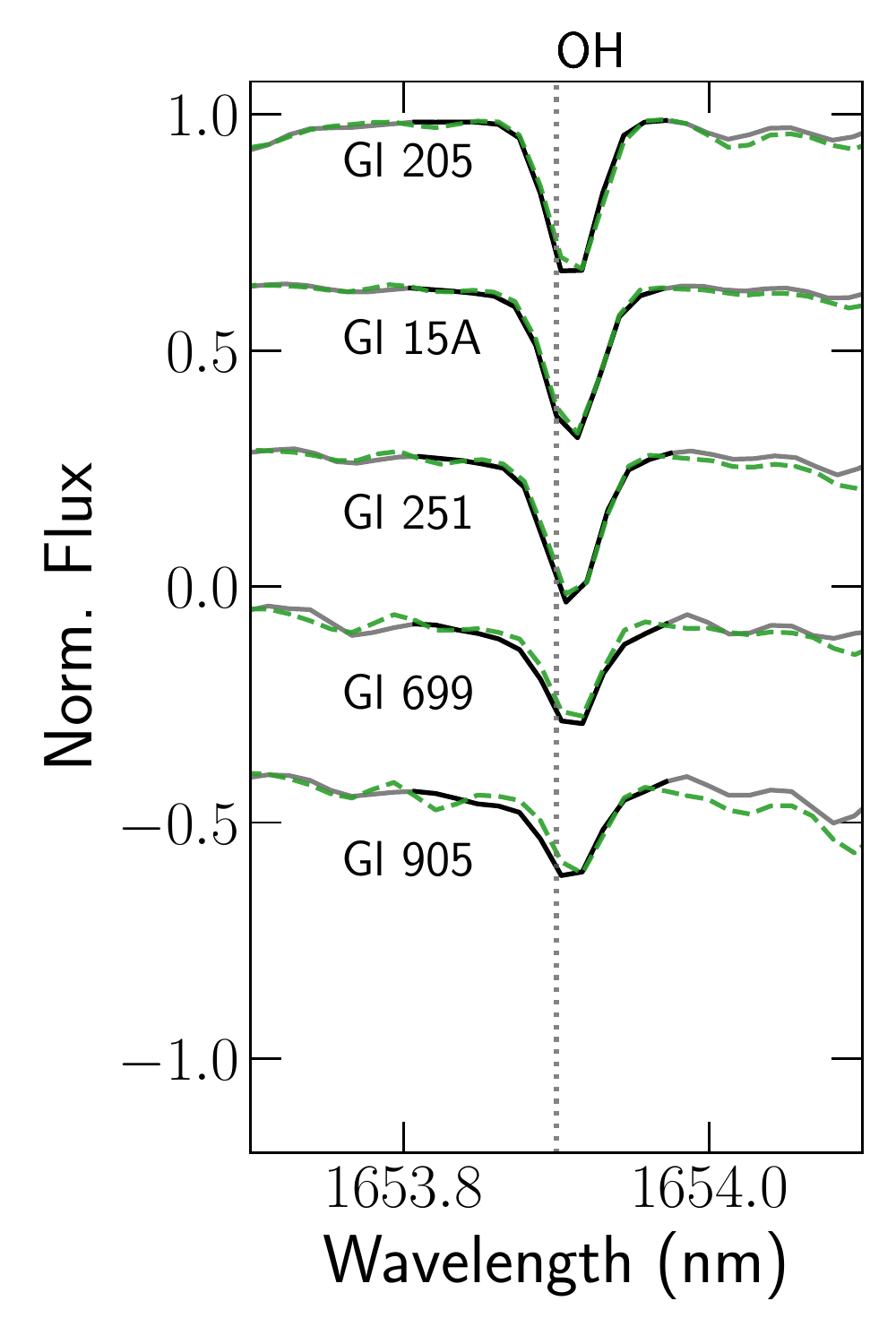}
	\includegraphics[scale=0.5]{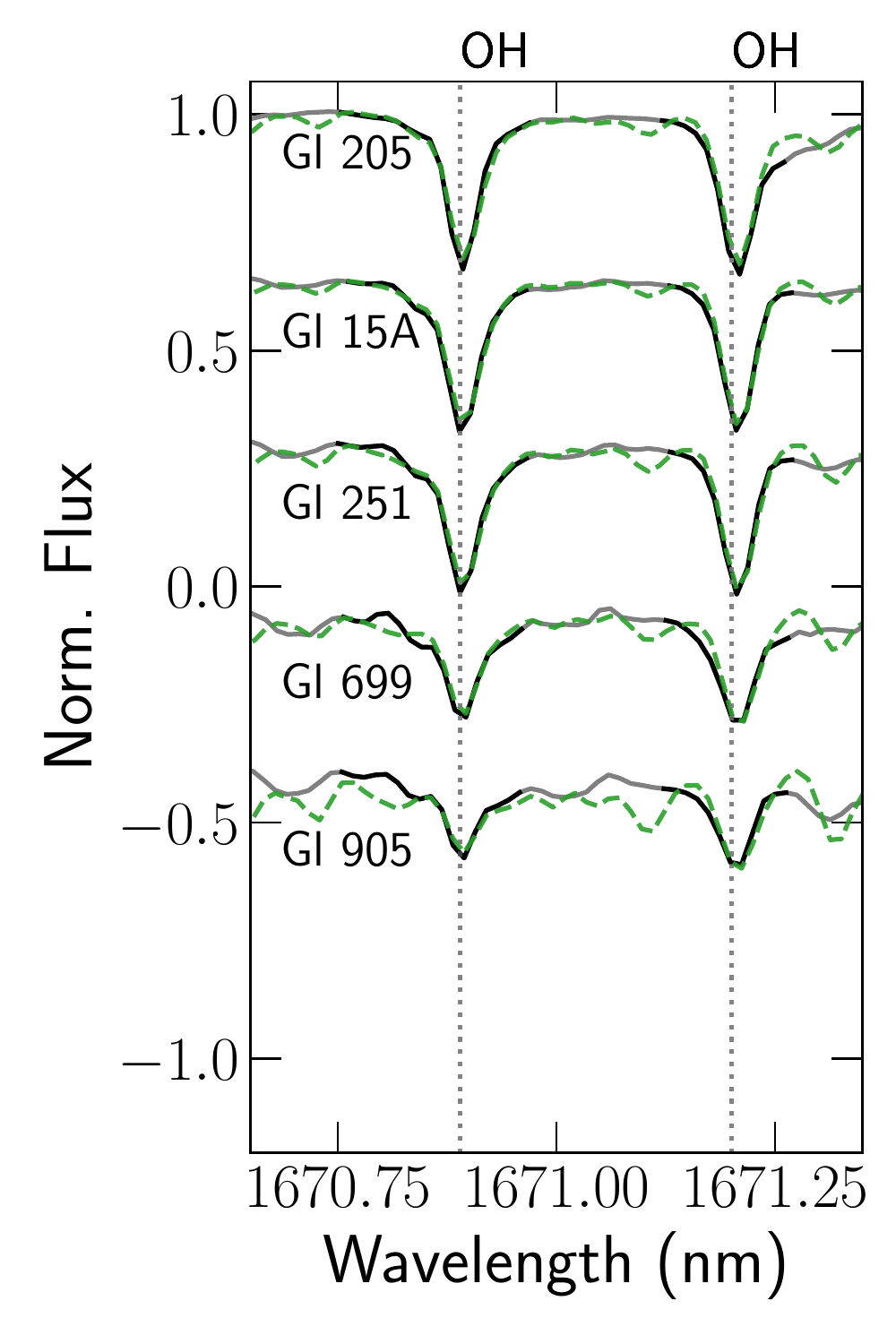}\includegraphics[trim={1cm 0  0cm 0}, clip, scale=0.5]{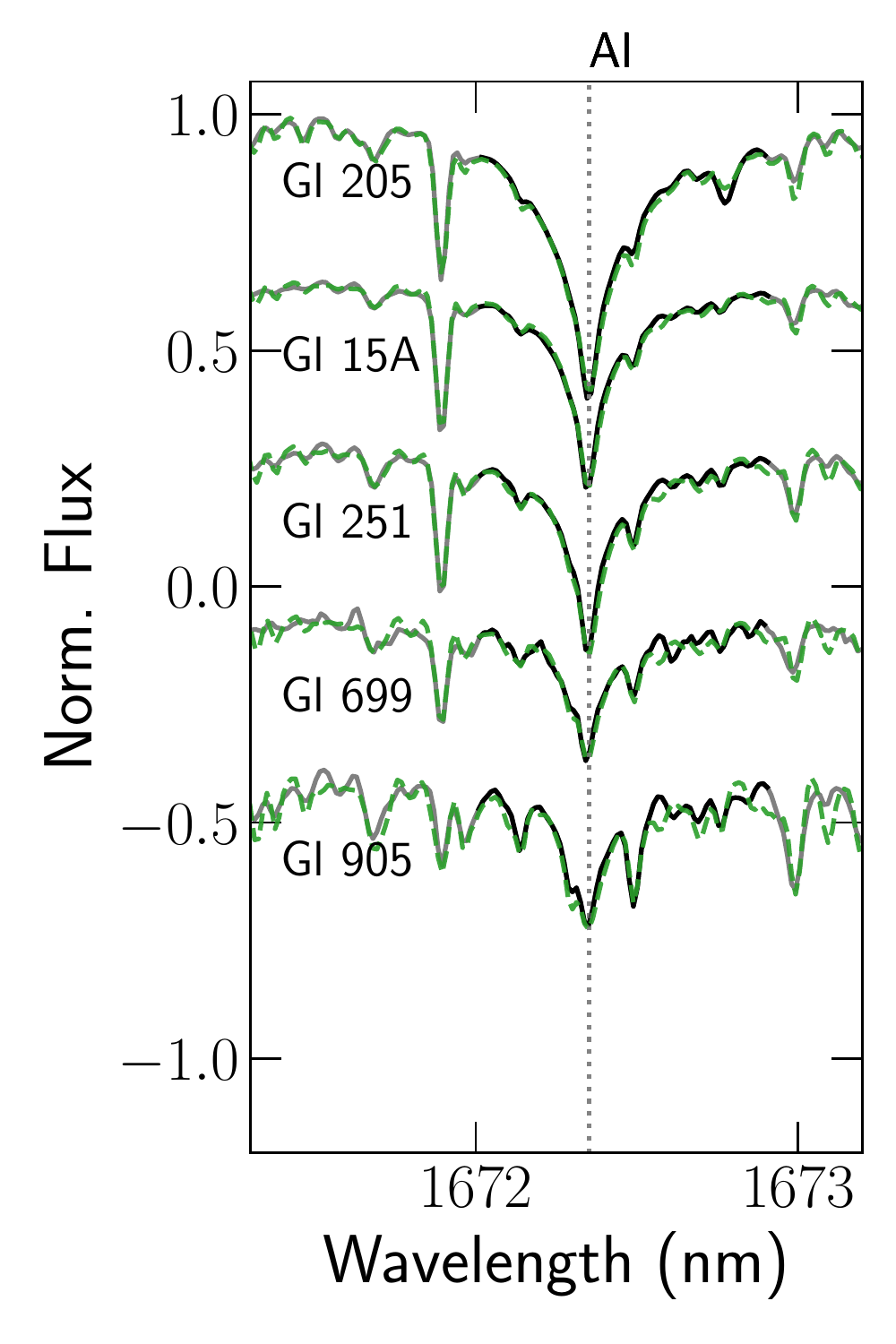}\includegraphics[trim={1cm 0  0cm 0}, clip,scale=0.5]{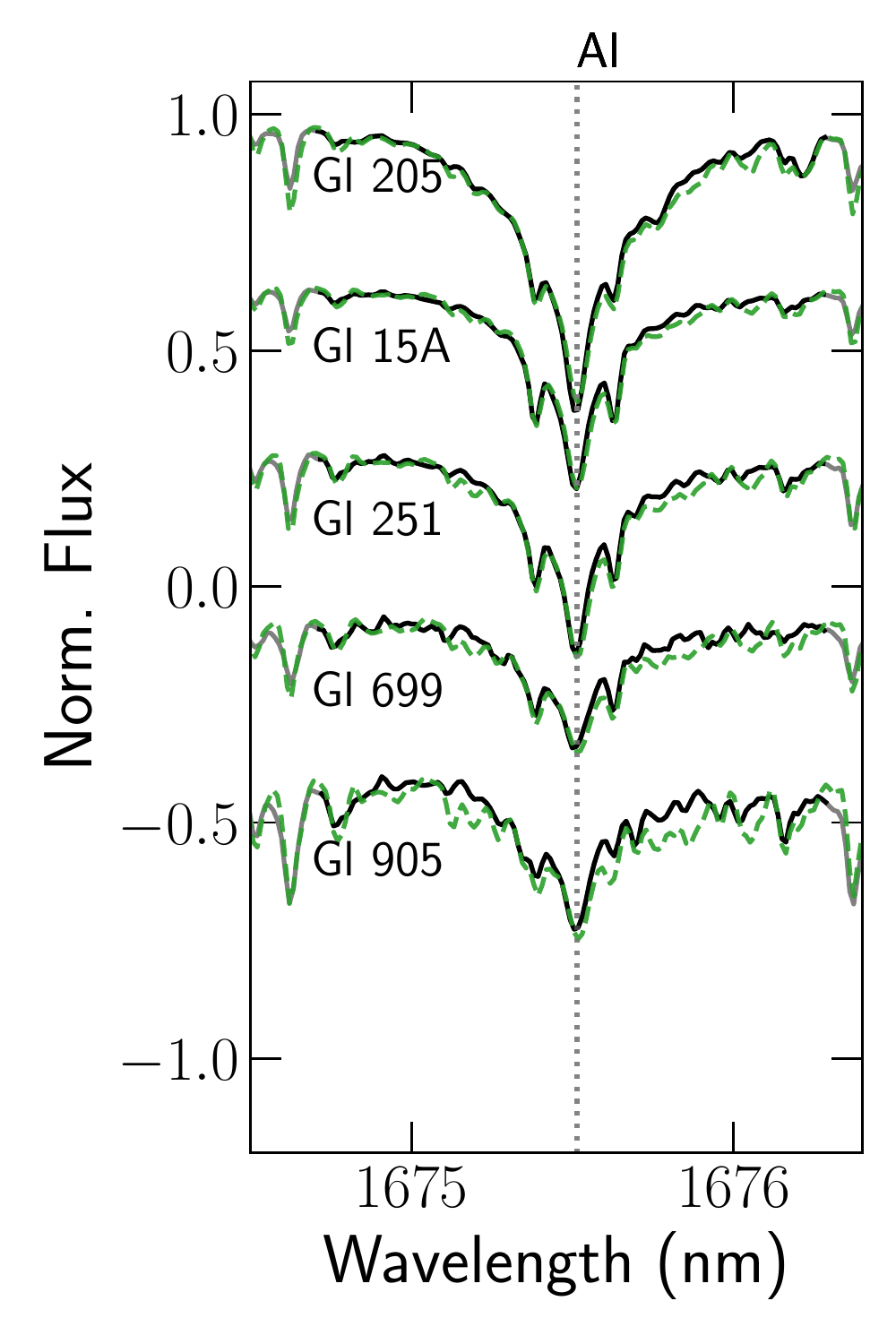}\includegraphics[trim={1cm 0  0cm 0}, clip,scale=0.5]{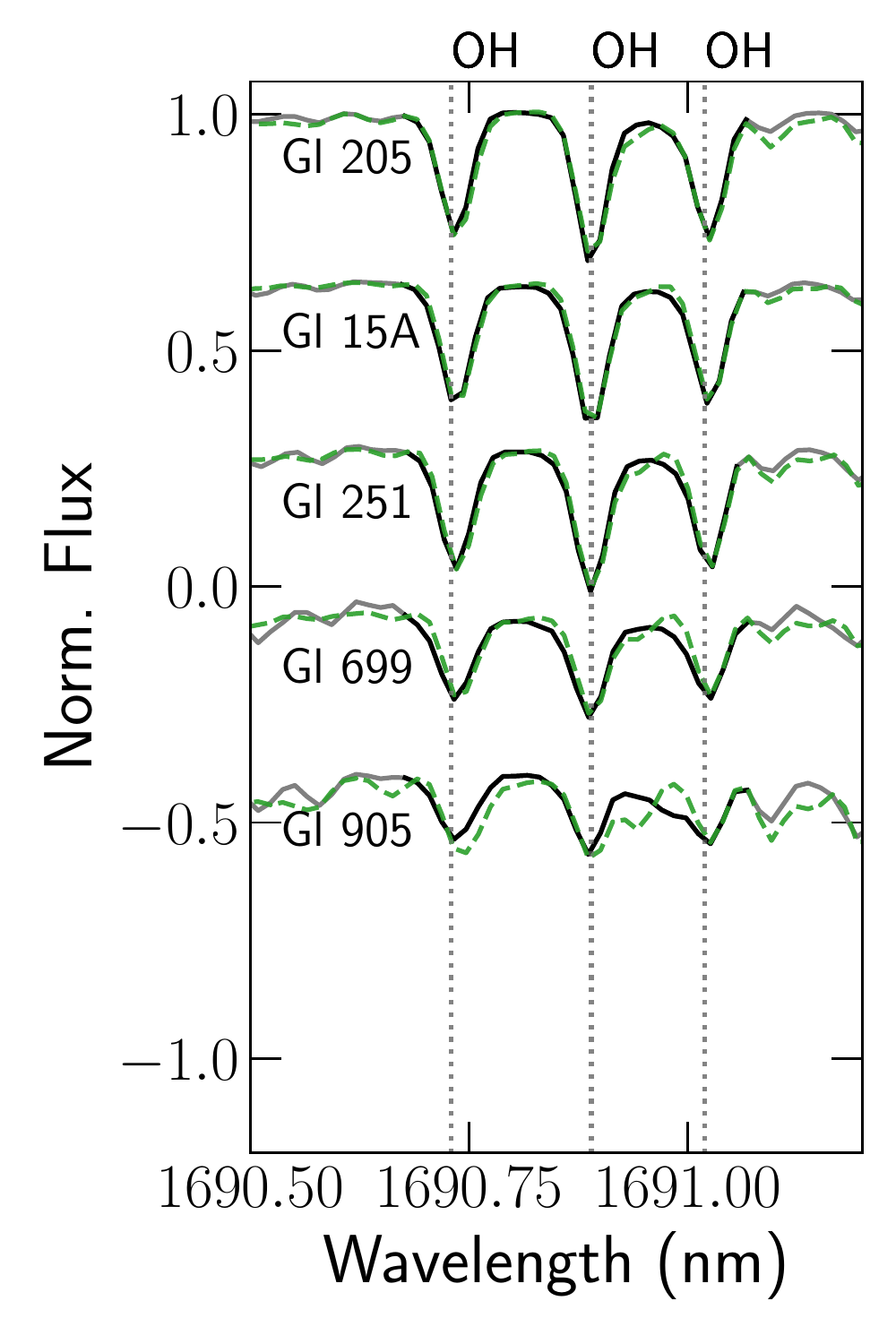}
	\caption*{Figure C1 -- continued}
\end{figure}

\begin{figure}[h]
	\includegraphics[scale=0.5]{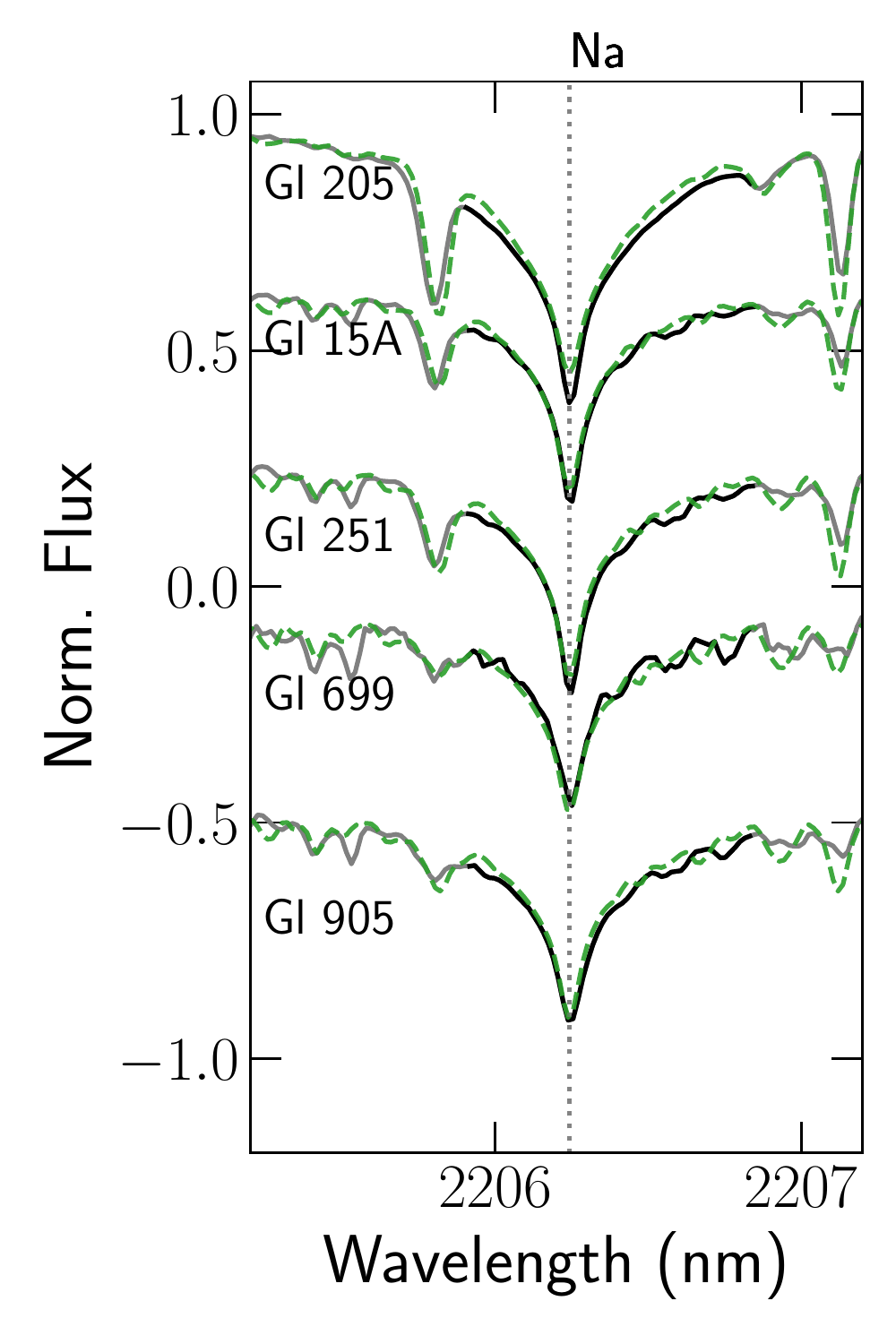}\includegraphics[trim={1cm 0  0cm 0}, clip, scale=0.5]{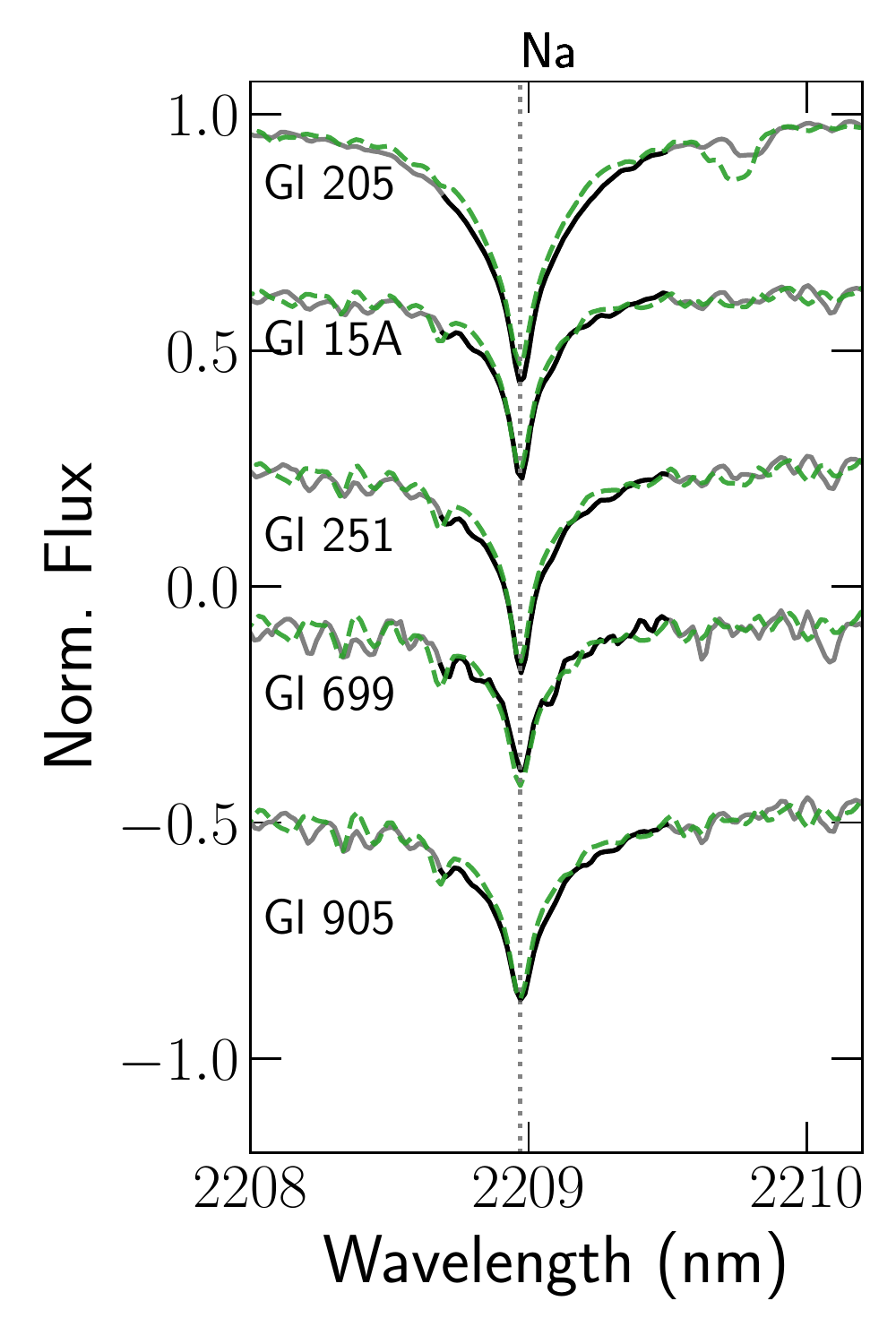}
	\caption*{Figure C1 -- continued}
\end{figure}

\begin{figure}[t!]
    \includegraphics[trim={0cm 0  0cm 0}, clip,scale=0.45]{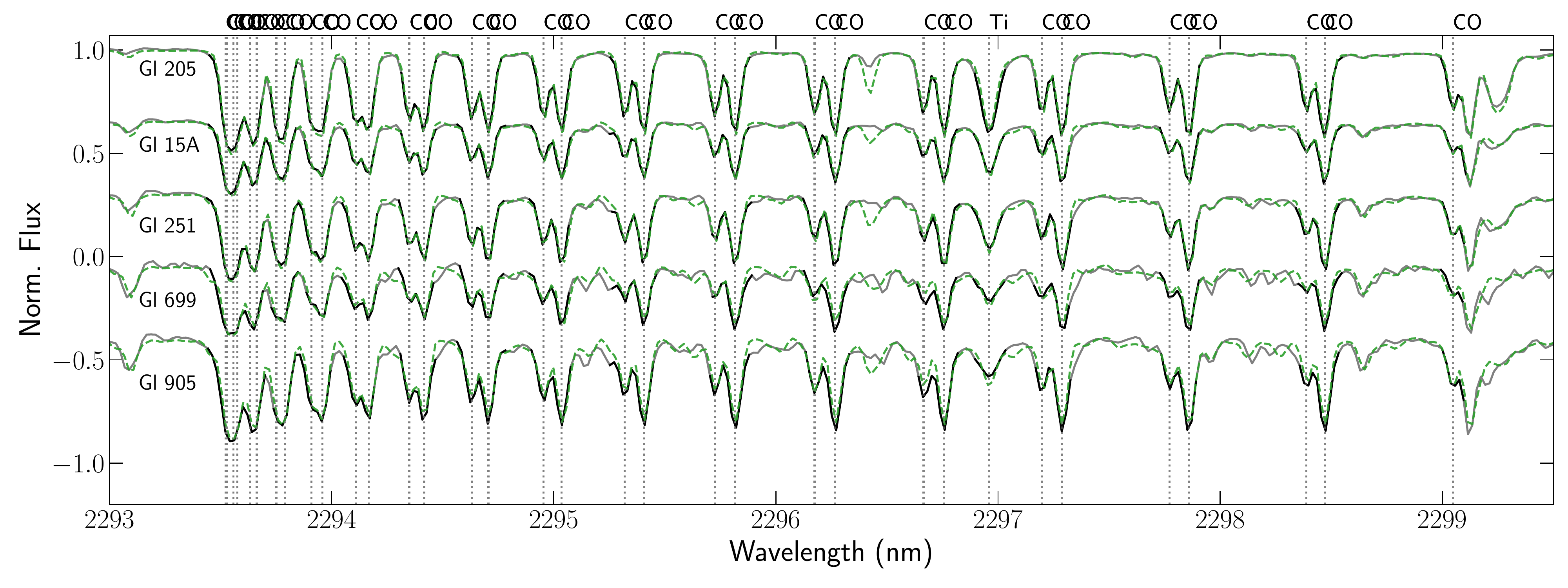}
	\caption*{Figure C1 -- continued}
\end{figure}